\documentclass[%
	pdftex,
	a4paper,
	oneside,
	12pt,
	headsepline,
	bibliography=totocnumbered,
	index=totoc
]{scrreprt}
\normalsize    

\usepackage{dsfont}
\usepackage[latin1]{inputenc}
\usepackage{amsmath}
\usepackage{amssymb}
\usepackage{graphicx}
\usepackage{array}
\usepackage{makeidx}
\usepackage[ngerman]{babel}
\usepackage[babel,german=quotes]{csquotes} 
\usepackage[%
	pdftitle={Masterarbeit},
	pdfauthor={Gunther Caspar},
	pdfcreator={MiKTeX, LaTeX with hyperref and KOMA-Script},
	pdfpagemode=UseOutlines,
	pdfdisplaydoctitle=true,
	pdflang=de,
  hyperfootnotes=false,
	plainpages=false,
	pdfpagelabels
]{hyperref}
\hypersetup{pdfsubject={Die Reissner-Nordström-Metrik in der pseudokomplexen Allgemeinen Relativitätstheorie}}
 \renewcommand\appendix{\par 
   \setcounter{section}{0}%
   \setcounter{subsection}{0}%
   \setcounter{figure}{0}%
   \renewcommand\thesection{\Alph{section}}%
   \renewcommand\thefigure{\Alph{section}.\arabic{figure}}}

\numberwithin{equation}{section}  

\renewcommand{\thefigure}{\arabic{chapter}.\arabic{figure}}
\makeatletter \@addtoreset{figure}{chapter} \makeatother

\makeindex

\newcommand{\fracpd}[2]{\frac{\partial #1}{\partial #2}} 

\newcommand{\crs}[3]{\genfrac{\{}{\}}{0pt}{}{ #1}{ #2 #3}}

\begin{document}

%
%
\begin{titlepage}
\pagenumbering{alph}
\begin{center}
\begin{huge}
\textbf{Die Reissner-Nordström-Metrik in der Pseudokomplexen Allgemeinen Relativitätstheorie}\\
\end{huge}
\vspace*{2 cm}
\begin{large}
Masterarbeit\\ von \\ Gunther Caspar\\
\vspace{18pt}
März, 2011\\
\vspace*{2 cm}
Betreuer:\\ Prof. Dr. Dr. h.c. mult Walter Greiner\\ Prof. Dr. Peter O. Hess\\
\vspace*{1.5 cm}
Frankfurt Institute for Advanced Studies (FIAS)\\
an der\\
Johann Wolfgang Goethe-Universität\\ Frankfurt am Main \\ Fachbereich Physik\\
\end{large}
\vspace*{2 cm}
\end{center}
Diese Masterarbeit erläutert die Änderungen der Reissner-Nordström-Metrik in der pseudokomplexen Allgemeinen Relativitätstheorie und die Auswirkungen der neuen Theorie auf bestimmte elektromagnetische Probleme in starken Feldern. Zudem werden Probleme der neuen Schwarzschildlösung diskutiert.\vspace{12pt}\\
Hierfür werden in den ersten drei Kapiteln die nötigen Grundlagen über Allgemeine Relativitätstheorie, pseudokomplexe Zahlen und die pseudokomplexe Allgemeine Relativitätstheorie erklärt, die mit Hilfe beider Betreuer und in Zusammenarbeit mit meinem Kommilitonen Thomas Schönenbach am Frankfurt Institute for Advanced Science(FIAS) erarbeitet und teilweise in vierköpfiger Zusammenarbeit weiterentwickelt wurden.
\end{titlepage}


\thispagestyle{empty}
\pagenumbering{roman}
\tableofcontents
\clearpage
\pagenumbering{arabic}
\setcounter{page}{1}

\chapter{Einleitung}
Schon vor Jahrtausenden blickten Menschen in den Himmel und versuchten die Bewegungen der Gestirne zu verstehen ohne zu ahnen, dass sie im Wesentlichen von nur einer Kraft bestimmt werden, der Gravitation.\vspace{12pt}\\
Lange Zeit waren die durch Beobachtungen gewonnenen Erkenntnisse auf empirische Befunde beschränkt, durch die zwar gewisse Ereignisse vorhergesagt werden konnten, aber es fehlte das Verständnis die Verbindung dieser Ereignisse zu erklären und eine mathematische Beschreibung war nicht möglich. Erst Johannes Kepler gelang mit den im Jahre 1609 und 1619 erschienen Werken ``Astronomia Nova'' und ``Harmonica Mundi'' ein erster Schritt in Richtung mathematischer Beschreibung, indem er die Daten von Tycho Brahe analysierte und daraus seine drei Gesetzmäßigkeiten ableitete. Zudem nahm er an, dass die Bahnen der Planeten durch eine kontaktlose Wirkung der Sonne bestimmt ist \cite{Greiner2003,Hiller1964}.\vspace{12pt}\\
Sir Isaak Newton gelang mit seinem 1687 erschienenen Werk ``Philosophiae Naturalis Principia Mathematica'' ein gewaltiger Durchbruch, indem er es schaffte sowohl Vorgänge auf der Erde als auch diese Wirkung mathematisch zu bestimmen. Er nannte sie Schwerkraft oder auch Gravitation\cite{Greiner2003}.\vspace{12pt}\\
Lange Zeit behielt sein Gravitationsgesetz und seine im gleichen Werk erschienene Theorie der Mechanik Gültigkeit und Abweichungen konnten entweder nicht gefunden werden oder wurden durch noch nicht gesichtete Objekte erklärt (Vulkanhypothese \cite{Adler1965, Adler1975,Turyshev:2008ur}). Das darauf aufbauende Theoriegebäude begann erst 1864 zu bröckeln, als James Clerk Maxwell die vollendeten Gleichungen der klassischen Elektrodynamik veröffentliche und es sich herausstellte, dass diese nur in einem bestimmten Bezugssystem gelten dürften. 1881 überprüfte Albert Abraham Michelson die daraus resultierende Vorhersage, dass die Lichtgeschwindigkeit vom Bezugssystem abhängig ist, und fand keine Abhängigkeit \cite{Hiller1964}.\vspace{12pt}\\
Dieses Phänomen konnte erst zufriedenstellend erklärt werden als Albert Einstein 1905 seine Spezielle Relativitätstheorie (SRT) veröffentlichte, die im Grenzfall langsamer Geschwindigkeiten in die Newtonsche Mechanik übergeht. Jedoch arbeitete Einstein weiter und es gelang ihm mit seiner im Jahre 1915 vorgetragenen Allgemeinen Relativitätstheorie (ART) erneut die Theorie der Mechanik zu revolutionieren und gleichzeitig eine bessere Theorie der Gravitation zu liefern, die bis heute ihre Gültigkeit behalten hat. Obgleich die ART außerordentlich erfolgreich ist, wird von vielen Physikern angenommen, dass sie nicht die endgültige Fassung der Theorie der Gravitation ist, da in ihr Singularitäten auftreten \cite{Adler1965, Adler1975}, die aus philosophischen Gründen äußerst unbeliebt sind, und bisher keine Möglichkeit gefunden wurde sie mit der Quantenmechanik oder mit einer der Theorien für die anderen Grundkräfte zu vereinigen.\vspace{12pt}\\
Folglich gibt und gab es zahlreiche Versuche eine allgemeinere Theorie der Gravitation zu formulieren, von denen jedoch bisher noch keine im Experiment erfolgreicher als die ART war.\vspace{12pt}\\ 
In dieser Arbeit werden Rechnungen in einer neuen Gravitationstheorie der beiden Betreuer \cite{Hess:2008wd} durchgeführt, deren Zweck das tiefere Verständnis der Theorie und das Hinarbeiten auf experimentell testbare Vorhersagen sind. 
\section{Allgemeine Relativitätstheorie}
Zum Verstehen dieser Masterarbeit sind Grundkenntnisse der Allgemeinen Relativitätstheorie notwendig, die in diesem Teilkapitel dem Leser in Erinnerung gerufen werden. Jedoch werden die notwenigen Kenntnisse der Tensoralgebra vorrausgesetzt und die Herleitung der Einsteinschen Feldgleichung nicht im Detail erklärt, sodass diese im Zweifelsfall in einem Lehrbuch wie \cite{Adler1965, Adler1975} nachgeschlagen werden müssen.\vspace{12pt}\\
In der Newtonschen Mechanik treten mehrere Kräfte proportional zur Masse auf, sobald die Bewegungen nicht in einem Inertialsystem berechnet werden. Ausgenommen der Gravitation wurden sie alle als Scheinkräfte bezeichnet, da sie alle durch die ``richtige'' Wahl des Koordinatensystems, d.h. der Geometrie, wegtransformiert werden können, indem man sie als Effekte der Trägheit erklärt.\vspace{12pt}\\ 
Diese Sonderstellung der Inertialsysteme und der Gravitation gaben Einstein den Anlass das Äquivalenzprinzip aufzustellen. Es ist ein Postulat, dass die Gravitation in einem geschlossenen lokalen System durch keine physikalische Messung von einem Trägheitseffekt unterschieden werden kann. Das übliche Bild ist, dass ein Passagier in einem Fahrstuhl nicht unterscheiden kann, ob er durch ein Seil nach oben beschleunigt wird oder sich in einem Gravitationsfeld befindet, da er in beiden Fällen eine zur Masse proportionale Kraft nach unten verspürt.\vspace{12pt}\\
Aus den naheliegenden Gründen versuchte Einstein das Erfolgsrezept der klassischen Mechanik beizubehalten und die Erklärung der Gravitation in der Geometrie des Raumes zu suchen. Da jedoch die Geschwindigkeit und somit die Zeit bei der Bahn eines Körpers durch ein Gravitationsfeld eine entscheidende Rolle spielen, aber räumlich in einer stationären Geometrie nur eine Geodäte (extremale Verbindung) existiert, ist dies in einem dreidimensionalen Raum nicht möglich. Aufbauend auf seine spezielle Relativitätstheorie versuchte Einstein deshalb eine Theorie in einer vierdimensionalen Raumzeit zu entwickeln.\vspace{12pt}\\
Aus theoretischen Überlegungen über mögliche reale Vorgänge entschied er sich nach einer Theorie in Riemann Räumen zu suchen, d.h.
\begin{align}
ds^2 = \sum_{i,j = 0}^{d} g_{ij} dx^idx^j
\end{align}
Wobei die Funktion g$_{ij}$ der metrische Tensor ist, der die Geometrie des Raums vollständig beschreibt.\vspace{12pt}\\
Weiterhin postulierte Einstein das Kovarianzprinzip, das besagt, dass die physikalischen Gesetze in allen Bezugssystemen die gleiche Form haben müssen. Deshalb formulierte er die Allgemeine Relativitätstheorie in kovarianter Tensorform.\\
Auf diesen Postulaten aufbauend lässt sich nun die Bedingung für das Auftreten eines Minkowskiraums bestimmen, dass der Riemannsche Krümmungstensor verschwinden muss. Da die Gesetze der SRT ziemlich genau zutreffen, muss der Raum im komplett masselosen Fall ein Minkowskiraum sein und somit sollte die Feldgleichung eine abgeschwächte Bedingung des verschwindenen Krümmungstensors sein. Weiterhin kann aus der Analogie zum klassischen Fall eine Verjüngung vermutet werden, sodass Einstein die Feldgleichungen für den energiefreien Raum aufstellen konnte:
\begin{align}
R_{\mu \nu} = 0
\end{align}
Wobei R$_{\mu \nu}$ der kontrahierte oder auch verjüngte Krümmungstensor ist.\vspace{12pt}\\
Alternativ kann sie auch mit dem divergenzfreien Riccitensor G$_{\mu \nu}$ geschrieben werden
\begin{align}
G_{\mu \nu} = R_{\mu \nu} - \frac{1}{2} g_{\mu \nu} R = 0
\end{align}  
Diese Gleichungen gelten jedoch wie schon erwähnt nur für den freien Raum und müssen für einen gefüllten Raum noch ergänzt werden.\vspace{12pt}\\
Die energetischen Verhältnisse im Raum werden durch den Energie-Impuls-Tensor widergespiegelt und deshalb sollte es einen Zusammenhang zwischen ihm und den Feldgleichungen geben. Einstein wählte die einfachst mögliche Lösung, indem er einen linearen Zusammenhang zwischen ihm und dem Riccitensor annahm und über den klassischen Grenzfall die Proportionalitätskonstante ermittelte. Somit ist die vollendete Feldgleichung gegeben durch
\begin{align}
G_{\mu \nu} = R_{\mu \nu} - \frac{1}{2} g_{\mu \nu} R = -\frac{8\pi\kappa}{c^2} T_{\mu \nu}
\end{align}
Theoretisch dürfte auch noch ein zur Metrik proportionaler Term auftreten, aber die Konstante müsste so klein sein, dass er erst auf Skalen weit größer als Sonnensysteme eine Rolle spielt.\vspace{12pt}\\ 
In der aktuellen Forschung wird häufig vermutet, dass es diesen Term tatsächlich gibt, aber seine Herkunft ist ungeklärt \cite{Barrow:2010xt,Bertolami:2009nr}. Zudem ist er für die in dieser Masterarbeit behandelten Probleme zweifellos irrelevant, da sich bei der Analyse der neuen Theorie herausstellt, dass der gleiche Effekt auch ohne ihn erreicht werden kann. Dementsprechend wird er in den Gleichungen vernachlässigt.
\subsection{Die Schwarzschild-Metrik} \label{secSMet}
Einstein selbst glaubte, dass es aufgrund ihrer Nichtlinearität Jahre oder Jahrzehnte dauern würde bis die ersten analytischen Lösungen zu seinen Gleichungen auftreten würden, aber schon 1916 fand Karl Schwarzschild eine Lösung. Er betrachtete den Außenraum einer statischen, sphärisch symmetrischen Masse und konnte mit Hilfe von Symmetrieargumenten die Gleichungen so weit vereinfachen, dass sie analytisch lösbar wurden.\vspace{12pt}\\
In der ART ist das infinitessimale Linienelement gegeben durch
\begin{align}
ds^2 = g_{\mu \nu} dx^\mu dx^\nu
\end{align}
Da das Problem statisch sein soll, muss es invariant unter der Vertauschung dx$^0$ = cdt $\rightarrow$ - dx$^0$ sein. Folglich verschwinden alle nicht diagonalen Metrikkomponenten, die den Index 0 besitzen.\vspace{12pt}\\
Außerdem ist die Masse spährisch symmetrisch, sodass in Kugelkoordinaten die Vertauschungen d$\vartheta~\rightarrow$ -d$\vartheta$ und d$\varphi~\rightarrow$ -d$\varphi$ das Linienelement nicht verändern dürfen. Dementsprechend fallen alle nicht diagonalen Metrikkomponenten weg, die den Index 2 oder 3 besitzen, und die Metrik muss diagonal und nur vom Radius abhängig sein.\vspace{12pt}\\
Durch die Wahl einer bestimmten Radialkoordinate ist es zudem möglich die Metrikkomponenten für die Winkel auf die in Kugelkoordinaten üblichen zu legen. Somit ergibt sich bei Wahl der Signatur als (+,-,-,-) folgender Ansatz für die Metrik:\\
\begin{align}
g_{\mu \nu} = \begin{pmatrix} e^{\nu(r)} & 0 & 0 & 0 \\ 0 & -e^{\lambda(r)} & 0 & 0 \\
0 & 0 & -r^2 & 0 \\ 0 & 0 & 0 & -r^2\sin^2(\vartheta) \end{pmatrix}\label{Ansmetradsym}\\
\notag
\end{align} 
Eingesetzt in die Einsteinschen Feldgleichugen ergibt sich daraus das Gleichungsystem\\
\begin{align}
R_{00} = e^{\nu-\lambda}\left[ - \frac{\nu''}{2} + \frac{\lambda'\nu'}{4} - \frac{\nu'^2}{4} - \frac{\nu'}{r} \right] &= 0 \label{ARTEinstschwarz0}\\
R_{11} = \frac{\nu''}{2} - \frac{\lambda'\nu'}{4} + \frac{\nu'^2}{4} - \frac{\lambda'}{r} &= 0 \label{ARTEinstschwarz1}\\
R_{22} = e^{-\lambda}\left[ 1 + \frac{r\nu'}{2} - \frac{r\lambda'}{2} \right] -1  &= 0 \label{ARTEinstschwarz2}\\
R_{33} = \sin^2(\vartheta) \left\{	e^{-\lambda}\left[ 1 + \frac{r\nu'}{2} - \frac{r\lambda'}{2} \right] -1 \right\} &= 0 \label{ARTEinstschwarz3}\\
\notag
\end{align}
Die Gleichungen für R$_{22}$ und R$_{33}$ sind redundant und insofern muss nur eine im Weiteren beachtet werden.\vspace{12pt}\\
Nun kann man \eqref{ARTEinstschwarz0} durch e$^{\nu-\lambda}$ teilen und mit \eqref{ARTEinstschwarz1} addieren und man erhält
\begin{align}
\nu' + \lambda' = 0
\end{align}
Diese Gleichung bedeutet zusammen mit der Forderung einer Minkowski-Metrik im Unendlichen, dass $\lambda = -\nu$ gilt.\vspace{12pt}\\
Wenn man diesen Zusammenhang nun in \eqref{ARTEinstschwarz2} einsetzt ergibt sich 
\begin{align}
\left (re^{-\lambda} \right )' &= 1 \label{ARTEinstschwarzl}\\
\Rightarrow e^{-\lambda} &= 1 - \frac{r_s}{r}
\end{align}
Wobei r$_s$ eine vorerst unbekannte Integrationskonstante ist, die als Schwarzschildradius bezeichnet wird und sich über den klassischen Limes als $\frac{2\kappa M}{c^2}$ bestimmen lässt.\vspace{12pt}\\ 
Zudem führt das Ableiten von \eqref{ARTEinstschwarzl} zu einer Gleichung, die unter der Vorraussetzung $\lambda = -\nu$ äquivalent zu \eqref{ARTEinstschwarz0} und \eqref{ARTEinstschwarz1} ist, sodass alle Einsteingleichungen erfüllt sind.\vspace{12pt}\\
Folglich ist das finale und als Schwarzschild-Metrik bezeichnete Ergebnis gegeben durch
\begin{align}
g_{\mu \nu} = \begin{pmatrix} 1 - \frac{r_s}{r} & 0 & 0 & 0 \\ 0 & \frac{1}{1 - \frac{r_s}{r}} & 0 & 0 \\
0 & 0 & -r^2 & 0 \\ 0 & 0 & 0 & -r^2\sin^2(\vartheta) \end{pmatrix}\\
\notag
\end{align}
\subsection{Die Reissner-Nordström-Metrik} \label{SecRMM}
Die Reissner-Nordström-Metrik beschreibt den Außenraum einer statischen, sphärisch symmetrischen und geladenen Masse und kann analog zum Schwarzschildfall angesetzt werden, da die Symmetrieargumente bei ihr ebenfalls gültig sind. Jedoch existiert das elektromagnetische Feld der Ladung auch außerhalt des Körpers, sodass der Außenraum nicht frei ist und folglich der Energie-Impuls-Tensor in die Einsteingleichungen mit eingehen muss.\vspace{12pt}\\
Dementsprechend muss zuerst der Energie-Impuls-Tensor des vorliegenden elektrostatischen Feldes bestimmt werden. Gemäß  \cite{Adler1965,Adler1975,Inverno2009} lässt er sich mit dem elektrischen Feldtensor F$_{\mu\nu}$ berechnen durch
\begin{align}
T_{\mu\nu} &= \frac{1}{c^2} \left [ F_{\mu\alpha} F^{\alpha}_{~\nu} + \frac 14 g_{\mu\nu} F_{\alpha\beta}F^{\alpha\beta} \right ]
\end{align}
Der elektrische Feldtensor widerrum ist nach \cite{Fliesbach2006} durch
\begin{align}
F^{\mu\nu} = \partial^\mu A^\nu - \partial^\nu A^\mu
\end{align} 
gegeben, wobei $A^\mu$ das elektromagnetische Potential ist.\\
Mit $\vec{E} = -\nabla \Phi - \partial_t \vec{A}$ und $\vec{B} = \nabla \times \vec{A}$ kann er in kartesischen Koordinaten einfach aufgeschrieben werden\\
\begin{align}
\bar{F}_{\mu\nu} = \begin{pmatrix} 0 & -E_x & -E_y & -E_z\\ E_x & 0 & H_x & -H_y \\ E_y & -H_x & 0 & H_z \\ E_z & H_y & -H_z & 0 \end{pmatrix}\\
\notag
\end{align}
Im Falle einer ruhenden geladenen Kugel oder eines ruhenden geladenen Punktes ist das elektrische Feld im Außenraum radialsymmetrisch, d.h. es zeigt in radiale Richtung und ist nur von r abhängig und das magnetische Feld ist im gesamten Raum 0. Folglich kann der Energie-Impuls-Tensor in kartesischen Koordinaten wie folgt umgeschrieben werden\\
\begin{align}
\bar{F}_{\mu\nu} = \begin{pmatrix} 0 & -E(r)\cos(\phi)\sin(\vartheta) & -E(r)\sin(\phi)\sin(\vartheta) & -E(r)\cos(\vartheta)\\ E(r)\cos(\phi)\sin(\vartheta) & 0 & 0 & 0\\ E(r)\sin(\phi)\sin(\vartheta) & 0 & 0 & 0\\ E(r)\cos(\vartheta) & 0 & 0 & 0 \end{pmatrix}
\end{align}
Nun kann der Tensor in Kugelkoordinaten transformiert werden durch:
\begin{align}
F_{\mu\nu} &= \fracpd{\bar{x}^{\alpha}}{x^\mu} \fracpd{\bar{x}^{\beta}}{x^\nu} \bar{F}_{\alpha \beta}\\
\Rightarrow F_{00} &= \delta^\alpha_0 \delta^\beta_0 \bar{F}_{\alpha\beta} = \bar{F}_{00}\\
F_{01} &= -F_{10} = \delta^\alpha_0 \fracpd{\bar{x}^\beta}{r} \bar{F}_{\alpha\beta} = \cos(\phi) \sin(\vartheta) \bar{F}_{01} + \sin{\phi} \sin(\vartheta) \bar{F}_{02} + \cos(\vartheta) \bar{F}_{03} = -E(r)\\
F_{02} &= -F_{20} = \fracpd{\bar{x}^\beta}{\vartheta} \bar{F}_{0\beta} = \cos(\phi) \cos(\vartheta) \bar{F}_{01} + \sin(\phi) \cos(\vartheta) \bar{F}_{02} - \sin(\vartheta) \bar{F}_{03} = 0\\
F_{03} &= F_{30} = \fracpd{\bar{x}^\beta}{\phi} \bar{F}_{0\beta} = -\sin(\phi) \sin(\vartheta) \bar{F}_{01} + \cos(\phi) \cos(\vartheta) \bar{F}_{02} = 0
\end{align}
Alle Terme ohne 0 verschwinden trivialerweise, da außer der Nullkoordinate keine Koordinate des ungestrichenen Systems von der Nullkoordinate des gestrichenen Systems abhängt, und da im ungestrichenen System alle Terme des Tensors verschwinden, solange keiner der Indizes 0 ist.\vspace{12pt}\\
Folglich gilt:
\begin{align}
\bar{F}_{\mu\nu} = E(r)\begin{pmatrix} 0 & -1 & 0 & 0\\ 1 & 0 & 0 & 0 \\ 0 & 0 & 0 & 0 \\ 0 & 0 & 0 & 0 \end{pmatrix}\\
\notag
\end{align}
Zusammen mit dem allgemeinen Ansatz für die Metrik ergeben sich daraus die gemischte und die kontravariante Form des elektrischen Feldtensors.\\
\begin{align}
F^\mu_{~\nu} &= E(r) \begin{pmatrix} 0 & -e^{-\nu} & 0 & 0\\ -e^{-\lambda} & 0 & 0 & 0 \\ 0 & 0 & 0 & 0 \\ 0 & 0 & 0 & 0 \end{pmatrix}\\
F^{\mu\nu} &= E(r) \begin{pmatrix} 0 & e^{-(\nu+\lambda)} & 0 & 0\\ -e^{-(\nu+\lambda)} & 0 & 0 & 0 \\ 0 & 0 & 0 & 0 \\ 0 & 0 & 0 & 0 \end{pmatrix}\\
\notag 
\end{align}
Wenn man zudem noch das Ergebnis der Maxwellgleichungen für das elektrische Feld nutzt  ( E(r) = e$^{\frac{\nu+\lambda}{2}} \frac{\epsilon}{r^2}$ ), erhält man den Energie-Impuls-Tensor\\ 
\begin{align}
T_{\mu\tau} &= \frac{1}{c^2} \left [ F_{\mu\alpha} F^{\alpha}_{~\tau} + \frac 14 g_{\mu\tau} F_{\alpha\beta}F^{\alpha\beta} \right ]\\
&= \frac{E^2}{c^2} \left [\begin{pmatrix} 0& -1 & 0 & 0 \\ 1 & 0 & 0 & 0 \\ 0& 0& 0& 0\\ 0&0&0&0 \end{pmatrix}\begin{pmatrix} 0& -e^{-\nu} & 0 & 0 \\ -e^{-\lambda} & 0 & 0 & 0 \\ 0& 0& 0& 0\\ 0&0&0&0 \end{pmatrix} + \frac 14 g_{\mu\tau} (-2e^{-(\nu+\lambda)}) \right]\\
&= \frac{E^2}{c^2} \left [\begin{pmatrix} e^{-\lambda}& 0 & 0 & 0 \\ 0 & -e^{-\nu} & 0 & 0 \\ 0& 0& 0& 0\\ 0&0&0&0 \end{pmatrix} - \frac 12 e^{-(\nu+\lambda)} \begin{pmatrix} e^{\nu}& 0 & 0 & 0 \\ 0 & -e^{\lambda} & 0 & 0 \\ 0& 0& r^2 & 0\\ 0&0&0& r^2\sin^2(\vartheta) \end{pmatrix} \right ]\\
&= \frac{E^2}{2c^2} \begin{pmatrix} e^{-\lambda}& 0 & 0 & 0 \\ 0 & -e^{-\nu} & 0 & 0 \\ 0& 0& -r^2e^{-(\nu+\lambda)} & 0\\ 0&0&0& -r^2\sin^2(\vartheta)e^{-(\nu+\lambda)} \end{pmatrix}\\
&= \frac{\epsilon^2}{2c^2r^4} \begin{pmatrix} e^{\nu}& 0 & 0 & 0 \\ 0 & -e^{\lambda} & 0 & 0 \\ 0& 0& r^2 & 0\\ 0&0&0& r^2\sin^2(\vartheta) \end{pmatrix}\\
\notag
\end{align}
Als Nächstes wird dieser Energie-Impuls-Tensor in die Einsteingleichungen eingestetzt und man erhält ein zu \eqref{ARTEinstschwarz0} bis \eqref{ARTEinstschwarz3} analoges Gleichungsystem mit Quelltermen auf der rechten Seite
\begin{align}
R_{00} = e^{\nu-\lambda}\left[ - \frac{\nu''}{2} + \frac{\lambda'\nu'}{4} - \frac{\nu'^2}{4} - \frac{\nu'}{r} \right] &= \frac{A}{r^4}e^\nu \label{ARTEinstReiss0}\\
R_{11} = \frac{\nu''}{2} - \frac{\lambda'\nu'}{4} + \frac{\nu'^2}{4} - \frac{\lambda'}{r} &= -\frac{A}{r^4}e^\lambda \label{ARTEinstReiss1}\\
R_{22} = e^{-\lambda}\left[ 1 + \frac{r\nu'}{2} - \frac{r\lambda'}{2} \right] -1  &= \frac{A}{r^2} \label{ARTEinstReiss2}\\
R_{33} = \sin^2(\vartheta) \left\{	e^{-\lambda}\left[ 1 + \frac{r\nu'}{2} - \frac{r\lambda'}{2} \right] -1 \right\} &= \frac{A}{r^2} \sin^2(\vartheta) \label{ARTEinstReiss3}
\end{align}
Wobei die Konstante $A := - \dfrac{8\pi \kappa \varepsilon^2}{2c^4}$ eingeführt wurde.\vspace{12pt}\\
Wie schon im Schwarzschildfall sind die Gleichungen für R$_{22}$ und R$_{33}$ redundant, sodass nur eine von beiden im Weiteren beachtet werden muss. Zudem erhält man durch vollkommen analoge Rechnung wie beim Schwarzschildfall wieder die Bedingung $\lambda = -\nu$, wobei die Bedingung einer Minkowskimetrik im Unendlichen erhalten bleibt, da das elektromagnetische Feld im Grenzfall verschwindet.\vspace{12pt}\\
Somit vereinfacht sich Gleichung \eqref{ARTEinstReiss2} zu
\begin{align}
\left (re^{-\lambda} \right )' &= 1 + \frac{A}{r^2} \label{ARTEinstReissl}\\
\Rightarrow e^{-\lambda} &= 1 - \frac{r_s}{r} - \frac{A}{r^2}
\end{align}
Durch Einsetzen der Lösung lässt sich dann noch zeigen, dass wieder alle Einsteingleichungen erfüllt sind und folglich ist das abschließende Ergebnis für die Reissner-Nordström-Metrik
\begin{align}
g_{\mu\nu} = \begin{pmatrix} 1 - \frac{r_s}{r} - \frac{A}{r^2} & 0 & 0 & 0 \\ 0 & \frac{1}{1 - \frac{r_s}{r} - \frac{A}{r^2}} & 0 & 0 \\ 0 & 0 & -r^2 & 0 \\ 0 & 0 & 0 & -r^2\sin^2(\vartheta) \end{pmatrix}
\end{align}
\section{Überprüfungen der Allgemeinen Relativitätstheorie}
Die Allgemeine Relativitätstheorie wurde mittlerweile vielfach geprüft und konnte bisher durch kein Experiment ausgeschlossen werden.\vspace{12pt}\\
In diesem Kapitel werden einige der bekanntesten Tests kurz diskutiert, um zu verdeutlichen für welche Effekte die Allgemeine Relativitätstheorie eine Rolle spielt.\\
\subsection{Rotverschiebung}
Nach Definition der Eigenzeit gilt
\begin{align}
c d\tau = ds = \sqrt{g_{\mu\nu}}dx^\mu dx^\nu
\end{align}
Daraus folgt für die Zeit, die eine Uhr in Ruhe anzeigt
\begin{align}
c d\tau = \sqrt{g_{00}}dt
\end{align}   
Wird nun ein statisches Gravitationsfeld betrachtet, gilt auch die aufintegrierte Form.\vspace{12pt}\\
Dementsprechend ergibt sich, wenn man die Zeitintervalle zwischen zwei Wellenbergen einer Lichtwelle von Punkt A nach Punkt B betrachtet 
\begin{align}
\frac{c}{\nu_A} = \sqrt{g_{00}(\vec{r}_A)} \Delta t && \frac{c}{\nu_B} = \sqrt{g_{00}(\vec{r}_B)} \Delta t
\end{align}
wobei die Zeitintervalle aufgrund der statischen Metrik gleich sind.\vspace{12pt}\\
Folglich erhält man 
\begin{align}
\frac{\nu_A}{\nu_B} = \sqrt{\frac{g_{00}(\vec{r}_B)}{g_{00}(\vec{r}_A)}}
\end{align}
Für schwache Felder folgt daraus nach einer Taylorentwicklung \cite{Adler1965,Adler1975,Fliesbach2006}
\begin{align}
\frac{\Delta \nu}{\nu} = \frac{\Delta \varphi}{c^2} \label{nahrs}
\end{align}
mit $\nu$ als der Frequenz beim Beobachter und $\Delta \varphi$ als die Differenz zwischen dem Newtonschen Gravitationsfeld beim Beobachter und dem beim Sender.\vspace{12pt}\\
Diese von der ART vorausgesagte Rotverschiebung des Lichts in Gravitationsfeldern ist durch erdnahe Versuche zu beweisen und kann somit quasi unter Laborbedingungen durchgeführt werden. Jedoch können dadurch die Feldgleichungen erst bei sehr präzisen Messungen getestet werden, da die genäherte Form \eqref{nahrs} auch allein über das Äquivalenzprinzip herleitbar ist \cite{Adler1965,Adler1975,Fliesbach2006}.\\
Die bisherigen Messungen bestätigen jedoch diesen Sachverhalt und die derzeit präziseste Messung ergab \cite{Turyshev:2008ur}
\begin{align}
| \frac{\Delta \nu}{\nu} - \frac{\Delta \varphi}{c^2} | < 2,1 \cdot 10^{-5}\frac{\Delta \varphi}{c^2}\\
\notag
\end{align}
\subsection{Der Parametrized Post Newtonian Formalismus}
Der Parametrized Post Newtonian (PPN) Formalismus ist eine Entwicklung der Metrik zur Erstellung genäherter Newtonscher Bewegungsgleichungen relativistischer Probleme zum Vergleich und Test verschiedener metrischer Gravitationstheorien. Somit eignet er sicht zum Auswerten experimenteller Tests im Sonnensystem.\vspace{12pt}\\
Zur Durchführung wird die Metrik gleichzeitig nach der Newtonschen Graviationsenergie durch die Ruhemasse (entspricht im Außenraum einer spährisch symmetrischen Masse $\frac{r_s}{r}$), der Geschwindigkeit durch die Lichtgeschwindigkeit, dem Quotienten der inneren und der Spannungsenergie mit einem Eichwert entwickelt. Danach werden die auftretenden Entwicklungskoeffizienten der ersten Ordnungen in den verschiedenen Theorien bestimmt und mit Experimenten verglichen, in denen alle Parameter viel kleiner als 1 sind, sodass höhere Ordnungen vernachlässigt werden können und eine Falsifizierung der Theorien möglich ist.\vspace{12pt}\\
In dieser Arbeit werden nur statische, sphärisch symmetrische Probleme im Außenraum der Massen betrachtet und die Betrachtung des PPN Formalismus findet nur für den Schwarzschildfall (d.h. $T_{\mu\nu} = 0$) statt, bei dem nur die Entwicklung nach $\frac{r_s}{r}$ relevant ist. Daher wird im Weiteren die Entwicklung  nur für diesen Spezialfall betrachtet und für den allgemeinen Fall wird auf \cite{GravMisner,Lammerzahl:2002tt} verwiesen.\vspace{12pt}\\
Im behandelten Fall ist der recht komplizierte PPN Formalismus äquivalent zur deutlich einfacheren Robertson Entwicklung, bei der einfach $g_{00}$ und $g_{11}$ in Kugelkoordinaten nach $\frac{r_s}{r}$ entwickelt und mit den PPN Parametern $\beta$, der angibt wieviel Krümmung pro Einheit ruhende Masse generiert wird, und $\gamma$, der die Nichtlinearität der Einsteingleichungen beschreibt, in Verbindung gebracht werden.\vspace{12pt}\\
Für das Linienelement
\begin{align}
ds^2 = B(r) c^2 dt^2 - A(r) dr^2 - r^2 \left ( d\vartheta^2 + \sin^2(\vartheta) d\varphi^2 \right )
\end{align}
ergibt sich nach \cite{Fliesbach2006} 
\begin{align}
B(r) &= 1 - \frac{r_s}{r} + \frac{\beta - \gamma}{2} \frac{r_s^2}{r^2} + ...\label{robertsonentw}\\
A(r) &= 1 + \gamma \frac{r_s}{r} + ...\\
\notag
\end{align} 
\subsection{Periheldrehung des Merkur}
Schon in der Mitte des 19. Jahrhunderts war bekannt, dass das Perihel des Merkur sich bei jedem Umlauf leicht (575 Bogensekunden pro Jahrhundert \cite{Fliesbach2006}) dreht. Der Großteil dieser Abweichung von der Newtonschen Zweikörper Vorhersage, nach der das Perihel stationär ist, ist über die Störungen der anderen Planeten erklärbar, aber bis zur Entwicklung der ART konnte keine wirklich zufriedenstellende Erklärung für einen Überschuss von etwa 43 Bogensekunden pro Jahrhundert gefunden werden.\vspace{12pt}\\
In Abhängigkeit von den PPN Parametern $\beta$, $\gamma$ und dem Quadrupolmoment der Sonne J$_2$ folgt aus der ART mit Hilfe der isotropen statischen Metrik eine Periheldrehung von \cite{lrr-2006-3}
\begin{align}
\dot{\Phi} = 42,98\frac{''}{Jh.} \left (\frac{1}{3} (2+ 2\gamma - \beta ) + 3 \cdot 10^{3} J_2 \right )
\end{align} 
Unter Vernachlässigung des Quadrupolmoments, dessen Effekt kleiner als der Messfehler ist, folgt mit dem aktuellen experimentellen Wert von $(42,98 \pm 0,04) \frac{''}{Jh.}$ \cite{Shapiro118}  
\begin{align}
|2\gamma - \beta - 1 | < 3 \cdot 10^{-3}
\end{align}
\subsection{Gravitationswellen}
Gravitationswellen sind über einen bestimten Zeitraum auftretende (annähernd) periodische Störungen der Raumzeit, d.h. der Metrik, die sich nach der ART wie elektromagnetische Wellen auf den Nullgeodäten der Raumzeit ausbreiten\cite{Adler1965,Adler1975}.\vspace{12pt}\\ 
Ihre Auswirkungen und Form sind aufgrund der Selbstwechselwirkungen des Gravitationsfeldes analytisch nicht bestimmbar, solange sie in starken Feldern betrachtet werden. Jedoch ist es möglich sie für schwache Felder mit Hilfe linearisierter Feldgleichungen in guter Näherung zu bestimmen.\vspace{12pt}\\
Dafür geht man davon aus, dass die Metrik als
\begin{align}
g_{\mu\nu} = \eta_{\mu\nu} + h_{\mu\nu} && |h_{\mu\nu}|<< 1
\end{align}
geschrieben werden kann. Wobei $\eta_{\mu\nu}$ die Minkowskimetrik ist, d.h. im vorliegenden Fall\\
\begin{align}
\eta_{\mu\nu} = \begin{pmatrix} 1 & 0 & 0 & 0 \\ 0 & -1 & 0 & 0 \\
0 & 0 & -1 & 0 \\ 0 & 0 & 0 & -1 \end{pmatrix}
\end{align}
Folglich können die Feldgleichungen in guter Näherung durch die in erster Ordnung in h$_{\mu\nu}$ entwickelten Gleichungen ersetzt werden.\vspace{12pt}\\
Da für den Fall einer Minkowskimetrik der Krümmungstensor verschwinden muss, verschwindet somit auch der verjüngte Krümmungstensor in nullter Ordnung. Für den verjüngten Krümmungstensor in erster Ordnung ergibt sich \cite{Fliesbach2006}\\
\begin{align}
R^{(1)}_{\mu\nu} = \frac{1}{2} \left ( \Box h_{\mu \nu} + h^\rho_{~\rho |\mu|\nu} - h^\rho_{~\mu |\rho|\nu} - h^\rho_{~\nu |\rho|\mu} \right )
\end{align}
Unter Verwendung der alternativen Form der Feldgleichungen erhält man nun
\begin{align}
\Box h_{\mu \nu} + h^\rho_{~\rho |\mu|\nu} - h^\rho_{~\mu |\rho|\nu} - h^\rho_{~\nu |\rho|\mu} = -\frac{16\pi G}{c^4} \left ( T_{\mu\nu} - \frac{T}{2} \eta_{\mu\nu} \right )
\end{align}
Wobei im letzten Term $\eta_{\mu\nu}$ steht, da T schon in erster Ordnung in h$_{\mu\nu}$ sein muss.\vspace{12pt}\\
Auf diese Gleichungen kann aufgrund der Freiheit des Koordinatensystems eine Eichtransformation angewandt werden \cite{Fliesbach2006} mit der man die entkoppelten linearisierten Feldgleichungen erhält:
\begin{align}
\Box h_{\mu \nu} = -\frac{16\pi G}{c^4} \left ( T_{\mu\nu} - \frac{T}{2} \eta_{\mu\nu} \right )
\end{align}
Somit erfüllen die Komponenten von h$_{\mu \nu}$ in linearer Näherung zu den elektromagnetischen Potentialen analoge Gleichung, deren Nulllösungen Wellen sind, die sich mit der Lichtgeschwindigkeit ausbreiten. Eine genauere Untersuchung zeigt, dass es sich um Quadrupolstrahlung handelt und das zugehörige hypotetische Austauschteilchen, das Graviton, den Spin zwei besitzt \cite{Fliesbach2006}. \vspace{12pt}\\
Man geht davon aus, dass eine experimentelle Auswertung von Gravitationswellen zu verblüffenden physikalischen Ergebnissen führen sollte, da sie einerseits schon zur Plankzeit von der restlichen Materie entkoppelten und andererseits aufgrund der schwächeren Kopplung an die interstellare Materie relativ ungestört durch das Weltall propagieren können.\vspace{12pt}\\ 
Bedauerlicherweise ist es bisher nicht gelungen Gravitationswellen direkt nachzuweisen und auszuwerten(siehe u.A. \cite{Collaboration:2009kk,Abadie:2010mt,Abbott:2005rt,Baggio:2005xi,Astone:2010mr}). Jedoch gelang es 1984 Weisberg und Taylor zu zeigen, dass die Verkürzung der Umlaufperiode des PSR 1913 + 16 Systems aus zwei sich umkreisenden Neutronensternen mit der von der ART vorhergesagten Energieabnahme (und somit Annäherung) durch Gravitationsstrahlung innerhalb der Messgenauigkeit übereinstimmt \cite{Taylor:1989sw}.

\chapter{Pseudokomplexe Zahlen}
Da die Pseudokomplexe Allgemeine Relativitätstheorie, wie der Name nahelegt, auf pseudokomplexen Zahlen basiert, sind zum weiteren Verständnis dieser Arbeit Grundkenntnisse über sie zwingend erforderlich.\vspace{12 pt}\\ 
Die Basis für dieses Kapitel bilden \cite{Antonuccio:1993et} und \cite{Hess:2008wd,Hess:2010wj}, wobei im Zweifelsfall die Nomenklatur der letzteren genutzt wird, um innerhalb der Arbeitsgruppe konsistent zu sein.\\
\section{Der Ring}
Dieses Unterkapitel beschäftigt sich mit der algebraischen Struktur des pseudokomplexen Zahlensystems und verdeutlicht den Unterschied zu den reellen und den komplexen Zahlen. Dafür ist es zweckmäßig zuerst einige grundlegende Definitionen zu wiederholen. Zum weiterführenden Nachlesen wird auf \cite{Fischer2005} verwiesen.\vspace{12pt}\\
\textbf{1. Definition}:\\
Eine Menge R zusammen mit zwei Verknüpfungen
\begin{align}
&+: R \times R \rightarrow R, ~ (a,b) \mapsto a+b\\
&\cdot: R \times R \rightarrow R, ~(a,b) \mapsto a \cdot b
\end{align}
heißt Ring, wenn folgendes gilt:
\begin{enumerate}
	\item R zusammen mit der Addition + ist eine abelsche Gruppe
	\item Die Multiplikation $\cdot$ ist assoziativ
	\item Es gelten die Distributivgesetze, d.h. für alle $a,b,c\in R$ gilt 
		\begin{align}
		a \cdot (b + c) = a\cdot b + a\cdot c && (a+b)\cdot c = a\cdot c + b\cdot c
		\end{align}
\end{enumerate}
Ein Ring heißt kommutativ, wenn $a\cdot b = b\cdot a$ für alle $a,b\in R$.\vspace{12pt}\\
\textbf{2. Definition}:\\
Ein Ring R heißt nullteilerfrei, wenn für alle $a,b\in R$ aus $a \cdot b=0$ stets $a=0$ oder $b=0$ folgt.\vspace{12pt}\\
\textbf{3. Definition}:\\
Ein nullteilerfreier kommutativer Ring mit einem neutralen Element der Multiplikation wird als Körper bezeichnet.\vspace{12pt}\\
Auf Grundlage dieser Definitionen wird nun das pseudokomplexe Zahlensystem betrachtet.\vspace{12pt}\\
Die pseudokomplexen Zahlen werden in Analogie zu den komplexen Zahlen mit einem Paar aus reellen Parametern (x$_1$,x$_2$) definiert über
\begin{align}
X = x_1 + Ix_2
\end{align}
mit den Regeln, dass I kommutiert und I$^2 = 1$ gilt (Dies mag zuerst befremdlich erscheinen, da die 1 dies als reelle Zahl schon liefert, ist aber beispielsweise in Matrixräumen nichts Ungewöhnliches. So existieren für 2x2 Matrizen mit der Einheitsmatrix und den Paulimatrizen vier linear unabhängige Elemente mit $M^2 = \mathds{1}$). Zudem wird in Analogie zu den komplexen Zahlen x$_1$ als Realteil und x$_2$ als Imaginärteil bezeichnet.\vspace{12pt}\\
Unter Annahme gewöhnlicher Rechenregeln ergibt sich für die Addition zweier pseudokomplexer Zahlen
\begin{align}
X + Y = x_1 + Ix_2 + y_1 + Iy_2 = (x_1 + y_1) + I(x_2 + y_2) 
\end{align}
und für die Multiplikation
\begin{align}
X \cdot Y &= (x_1 + Ix_2) \cdot (y_1 + Iy_2) = x_1y_1 + Ix_1y_2 + Ix_2y_1 + x_2 y_2\notag\\ 
&= (x_1y_1 + x_2y_2) + I(x_1y_2 + x_2y_1)
\end{align}
Da die reellen Zahlen die Eigenschaften eines Körpers erfüllen, ist es offensichtlich, dass die pseudokomplexen Zahlen mit der Addition eine abelsche Gruppe bilden und mit der Multiplikation assoziativ und kommutativ sind. Außerdem folgt aus
\begin{align}
X \cdot (Y + Z) &= (x_1 + Ix_2) [(y_1 + z_1) + I(y_2 + z_2)] \notag\\ 
&= x_1 (y_1 + z_1) + I x_1(y_2 + z_2) + I x_2(y_1+z_1) + x_2(y_2 + z_2) \notag\\
&= (x_1y_1 + x_2y_2) + I(x_1y_2 + x_2y_1) + (x_1z_1 + x_2z_2) + I(x_1z_2 + x_2z_1) \notag\\
&= XY + XZ
\end{align} 
dass das Distributivgesetz gilt.\vspace{12 pt}\\
Somit handelt es sich im mathematischen Sinn bei der Algebra der pseudokomplexen Zahlen D um einen kommutativen Ring. Jedoch liegt kein Körper vor, da er nicht nullteilerfrei ist, d.h.
\begin{align}
(1 + I) (1 - I) = 0
\end{align}
obwohl weder 1+I noch 1-I das neutrale Element der Addition ist, und somit handelt es sich bei Ihnen um zwei Nullteiler.\vspace{12pt}\\
Alle Nullteiler sind pseudokomplexe Zahlen mit einer Sonderstellung, sodass es nur dann möglich ist eine physikalische Theorie auf ihnen aufzubauen, falls allen Nullteilern wie in \cite{Antonuccio:1993et} geschehen, eine physikalische Interpretation zugewiesen werden kann.\vspace{12pt}\\
Um alle Nullteiler zu bestimmen, ist es zweckmäßig die pseudokomplexen Zahlen in die Basis aus $\sigma_+ = \frac{1+I}{2}$ und $\sigma_- = \frac{1-I}{2}$ umzuschreiben
\begin{align}
X = x_1 + Ix_2 &= X_+ \sigma_+ + X_- \sigma_-\\
\Rightarrow X_\pm &= x_1 \pm x_2 
\end{align}
und dann das Produkt zweier pseudokomplexer Zahlen 0 zu setzen
\begin{align}
XY &= (X_+ \sigma_++ X_- \sigma_-) (Y_+ \sigma_++ Y_- \sigma_-)\notag\\
&=X_+Y_+ \sigma_+^2 + (X_+Y_- + X_-Y_+)\sigma_+\sigma_- + X_-Y_- \sigma_-^2 \notag\\ 
&= X_+Y_+ \sigma_+ + X_-Y_- \sigma_- \stackrel{!}{=} 0
\end{align}
Da $\sigma_+$ und $\sigma_-$ linear unabhängig sind, folgt daraus
\begin{align}
X_+Y_+ = 0 \cap X_-Y_- = 0
\end{align}
Somit gibt es für X und Y folgende Möglichkeiten:
\begin{align}
X = 0 \cup Y = 0 \cup X = X_+ \sigma_+ \cap Y=Y_-\sigma_- \cup X = X_-\sigma_- \cap Y = Y_+\sigma_+
\end{align}
Die Fälle X,Y = 0 sind uninteressant, da dann ein Faktor das neutrale Element der Addition ist und alle Produkte 0 werden, und in den anderen beiden Fällen ist jeweils eine Zahl proportional zu $\sigma_+$ und die andere zu $\sigma_-$. Somit existieren nur $\sigma_+$ und $\sigma_-$ als linear unabhängige Nullteiler und die Theorie muss für die beiden Geraden mit einer physikalischen Interpretation aufwarten. Sobald dies möglich ist, wird dieser Nachteil des pseudokomplexen Zahlensystems sogar zu einer Stärke, da sich die $\sigma_\pm$-Nomenklatur als äußerst nützlich erweisen wird.
\section{Metrische Eigenschaften}
Da in der Realität und somit auch in der Physik Abstände eine entscheidende Rolle spielen, muss für die pseudkomplexen Zahlen eine Norm existieren, d.h. es wird eine Abbildung $||\cdot||$: R$\rightarrow \mathbb{R}^+$, $X \mapsto ||X||$ mit folgenden Eigenschaften benötigt
\begin{enumerate}
	\item $ ||X|| = 0 \Leftrightarrow X = 0 $
	\item $ ||XY|| = ||X|| ||Y||$
	\item $ ||X+Y|| \leq ||X|| + ||Y|| $
\end{enumerate}
Das Quadrat der Norm ist im Komplexen durch das Produkt aus einer Zahl mit ihrer komplex Konjugierten gegeben. Von daher liegt es nahe für die pseudokomplexen Zahlen auf dem gleichen Weg nach einer Norm zu suchen.\vspace{12pt}\\
Analog zu den komplexen Zahlen wird das komplex Konjugierte einer pseudokomplexen Zahl $X = x_1 + Ix_2$ definiert als $X^* = x_1 - Ix_2 = X_-\sigma_+ + X_+\sigma_-$ und das Betragsquadrat wird als 
\begin{align}
|X|^2 = XX^* = x_1^2-x_2^2 = X_+X_-
\end{align} 
definiert. Daraus ergibt sich zwar für alle Nichtnullteiler das Inverse
\begin{align}
X^{-1} = \frac{X^*}{|X|^2}
\end{align}
aber im Gegensatz zu den komplexen Zahlen kann der Nenner nicht die Norm sein, da er negativ und für $X \neq 0$ 0 werden kann und somit mindestens zwei grundlegende Vorraussetzungen der Definition verletzt.\vspace{12pt}\\
Es ist dennoch möglich eine Norm zu definieren, die zum komplexen Betrag analog ist, indem zwei Funktionen $|\cdot|_\pm$ wie folgt definiert werden
\begin{align}
|x_1 + Ix_2|_+ &= |x_1 + x_2| = |X_+|\\
|x_1 + Ix_2|_- &= |x_1 - x_2| = |X_-|
\end{align}   
Mit Hilfe der $\sigma_\pm$-Notation lässt sich problemlos zeigen, dass es sich bei $|\cdot|_\pm$ um zwei Seminormen handelt, d.h. Abbildungen, die alle Normaxiome bis auf die Definitheitseigenschaft $||X|| = 0 \Rightarrow X = 0$ erfüllen
\begin{align}
|X|_\pm &= |X_\pm| \geq 0\\
|X+Y|_\pm &= |X_\pm + Y_\pm| \leq |X_\pm| + |Y_\pm| = |X|_\pm + |Y|_\pm\\
|XY|_\pm &= |X_\pm Y_\pm| = |X_\pm| |Y_\pm| = |X|_\pm |Y|_\pm
\end{align}
Sobald jedoch verlangt wird, dass beide Seminormen verschwinden, ist die Definitheit erfüllt, da
\begin{align}
|X|_\pm &= 0 \Leftrightarrow X_+ = 0 \cap X_- = 0 \Leftrightarrow X = 0
\end{align}
Damit ist es nun auf Basis der beiden Seminormen möglich eine Norm auf dem Ring der pseudokomplexen Zahlen D zu definieren, die analog zur Norm im Komplexen ist
\begin{align}
||X|| = \sqrt{|X|_+^2 + |X|_-^2} = \sqrt{2(x_1^2 + x_2^2)}
\end{align}
Der Beweis der Normeigenschaften ergibt sich mit Hilfe der $\sigma_\pm$-Notation einfach aus der komplexen Norm.\\
\section{Differentation}
Jede beliebige Funktion, die nach D abbildet und auf einer offenen Umgebung U in der pseudokomplexen Ebene definiert ist, kann man durch zwei reellwertige Funktionen (u,v) wie folgt umschreiben
\begin{align}
f(x_1 + I x_2) = u(x_1,x_2) + Iv(x_1,x_2)
\end{align}
wobei aus physikalischen Überlegungen im Allgemeinen angenommen wird, dass u und v unendlich oft stetig differenzierbar sind.\vspace{12pt}\\
Alternativ kann sie auch in die Nullteilerbasis geschrieben werden
\begin{align}
f(x_1 + Ix_2) = (u+v)\sigma_+ + (u-v)\sigma_- = f_+(X_+,X_-)\sigma_+ + f_-(X_+X_-)\sigma_-
\end{align}
Die partiellen Ableitungen werden wie üblich über den Differenzenquotienten bestimmt. Jedoch ist für die Definition einer totalen Ableitung eine analoge Vorgehensweise wie im Komplexen nötig.\vspace{12 pt}\\
Die Ableitung einer pseudokomplexen Funktion f: U $\rightarrow$ D an der Stelle X wird definiert als die pseudokomplexe Zahl f'(X), für die gilt
\begin{align}
\lim_{||\Delta X||\rightarrow 0} \frac{f(X+\Delta X) - f(X) - f'(X) \Delta X}{||\Delta X||} = 0
\end{align}
Analog zum komplexen Fall ist eine Funktion dann holomorph, falls ihre Ableitung auf der ganzen Umgebung U existiert.\vspace{12pt}\\
Wird basierend auf dieser Definition die Nullteilerbasis betrachtet, ergibt sich
\begin{align}
\lim_{||\Delta X||\rightarrow 0} \frac{(f_\pm(X_+ + \Delta X_+, X_- + \Delta X_-) - f_\pm(X_+,X_-) - f_\pm '(X_+,X_-) \Delta X_\pm)\sigma_\pm}{||\Delta X||} = 0
\end{align}
Aufgrund der linearen Unabhängigkeit von $\sigma_\pm$ folgt daraus, dass für eine holomorphe Funktion
\begin{align}
\fracpd{f_+}{X_-} = \fracpd{f_-}{X_+} = 0 \label{Nullteilerpd}
\end{align}
gelten muss.\vspace{12pt}\\ 
Eine kurze Rechnung zeigt zudem, dass diese zwei Bedingungen äquivalent zu den pseudokomplexen Cauchy-Riemann-Gleichungen aus \cite{Antonuccio:1993et} sind
\begin{align}
\fracpd{f_+}{X_-} &= \fracpd{u+v}{x_1} \fracpd{x_1}{X_-} + \fracpd{u+v}{x_2} \fracpd{x_2}{X_-} = \frac{1}{2} \left (\fracpd{u+v}{x_1} - \fracpd{u+v}{x_2} \right ) = 0\\
\fracpd{f_-}{X_+} &= \fracpd{u-v}{x_1} \fracpd{x_1}{X_+} + \fracpd{u-v}{x_2} \fracpd{x_2}{X_+} = \frac{1}{2} \left (\fracpd{u-v}{x_1} + \fracpd{u-v}{x_2} \right ) = 0\\
&~~~~~~\Leftrightarrow \fracpd{u}{x_1} = \fracpd{v}{x_2} \cap  \fracpd{u}{x_2} = \fracpd{v}{x_1}
\end{align} 
Zusätzlich kann auch die Ableitung nach X und $\bar{X}$ in der Nullteilerbasis dargestellt werden
\begin{align}
\fracpd{}{X} &= \fracpd{}{X_+}\sigma_+ + \fracpd{}{X_-} \sigma_-\\
\fracpd{}{\bar{X}}&= \fracpd{}{X_-}\sigma_+ + \fracpd{}{X_+} \sigma_-
\end{align}
und aus \eqref{Nullteilerpd} ist sofort ersichtlich, dass
\begin{align}
\fracpd{f}{\bar{X}} = 0
\end{align}
eine äquivalente Bedingung für eine holomorphe Funktion ist.\vspace{12pt}\\
Somit existieren drei einfache gleichwertige Berechnungsmethoden der Ableitung für holomorphe Funktionen, von denen die für die jeweilige Aufgabenstellung geschickteste gewählt werden kann.
\begin{align}
f' = \fracpd{f}{X} = \fracpd{f_+}{X_+} \sigma_+ + \fracpd{f_-}{X_-}\sigma_- = \fracpd{u}{x_1} + I\fracpd{v}{x_1}
\end{align}
\section{Integration}
Die Integration einer pseudokomplexen Funktion f = u + Iv auf einer stückweise glatten Kurve $\gamma$ ist definiert als
\begin{align}
\int_\gamma f dX &= \int_\gamma (u+Iv) (dx_1 + Idx_2)\notag\\
&= \int_\gamma \left ( u dx_1 + vdx_2 \right ) + I \int_\gamma \left ( vdx_1 + udx_2 \right )\\
&= \int_\gamma \left ( f_+ dX_+\sigma_+ + f_-dX_- \sigma_- \right ) \label{intsig}
\end{align}
Zudem gilt, wenn f holomorph und $\gamma$ eine geschlossene Kurve ist, eine analoge Beziehung zum Cauchyschen Integralsatz, die mit \eqref{intsig} leicht ersichtlich wird.
\begin{align}
\oint_\gamma f dX = \oint_\gamma \left ( f_+(X_+) dX_+\sigma_+ + f_-(X_-)dX_- \sigma_- \right ) = 0
\end{align} 
   
\chapter{Pseudokomplexe Allgemeine Relativitätstheorie}
In diesem Kapitel werden der Grundformalismus der Theorie und die in der vierköpfigen Arbeitsgruppe gewonnen Erkenntnisse über die pseudokomplexe Schwarzschildmetrik und die Quellfunktionen vorgestellt. Einige der Sachverhalte können in \cite{Hess:2008wd} nachgelesen werden, aber vieles ist erst intern verfügbar.\vspace{12 pt}\\
Wie schon im oben genannten Paper beschrieben wird der pseudokomplexe Ansatz einer verallgemeinerten Relativitätstheorie durch mehrere Faktoren motiviert, die hier der Vollständigkeit halber in Grundzügen wiederholt werden.\vspace{12pt}\\ 
Da viele physikalische Theorien auf einer größeren Algebra als den reellen Zahlen definiert sind, kam schon recht früh die Idee auf, auch die Raumzeitkoordinaten auf eine größere Algebra zu erweitern, sodass es mittlerweile schon eine Vielzahl von Versuchen gab, die Allgemeine Relativitätstheorie dementsprechend auszubauen (zum Beispiel komplex \cite{Einstein:1945eu,Einstein:1948kr,Mantz:2008hm}, pseudokomplex \cite{Crumeyrolle:1964,Clerc:1972qd} und weitere \cite{Caianiello:1981jq,Brandt:1992qq,Beil:1992vh,Moffat:1978tr,Kunstatter:1982ix}). Ihnen ist gemein, dass eine maximale Beschleunigung und somit eine minimale Länge auftritt und folglich ist dies auch bei der hier vorgestellten Theorie der Fall.\vspace{12 pt}\\
Zudem erscheint die Betrachtung einer pseudokomplexen Erweiterung der Raumzeitkoordinaten vielversprechend, da in \cite{Kelly:1986mt} gezeigt wurde, dass nur reale und pseudokomplexe Koordinaten ausschließlich physikalische Lösungen erlauben und somit sinnvoll sind. Weiterhin sprechen die guten Ergebnisse in der Feldtheorie \cite{Hess:2007zzc,Hess:2007pp}, die durch eine Erweiterung der Koordinaten automatisch normalisiert wird, dafür einen solchen Ansatz zu verfolgen und auf testbare Vorhersagen hinzuarbeiten.
\section{Postulate und Formalismus}
Analog zur ART wird davon ausgegangen, dass der Raum eine Riemannschen Manigfaltigkeit ist, die jedoch nun auf dem Zahlensystem der Pseudokomplexen basiert. Somit beschreibt die Metrik den kompletten Raum und es wird postuliert, dass sie eine pseudokomplexe holomorphe Funktion ist. Dementsprechend gilt für die Metrik
\begin{align}
g_{\mu\nu} &= g^+_{\mu\nu}(X_+) \sigma_+ + g^-_{\mu\nu}(X_-) \sigma_-\\
&= \frac{1}{2} \left ( g^+_{\mu\nu}(X_+) + g^-_{\mu\nu}(X_-) \right ) + \frac{I}{2} \left ( g^+_{\mu\nu}(X_+) - g^-_{\mu\nu}(X_-) \right ) \label{gpholom}
\end{align}
wobei die Metrikkomponenten aufgrund der Holomorphiebedingung \eqref{Nullteilerpd} nur von der jeweiligen Koordinate abhängen.\vspace{12pt}\\
Aus \eqref{gpholom} lässt sich leicht erkennen, dass die Metrik unter diesen Voraussetzungen keine rein reelle Größe sein kann, da sonst 
\begin{align}
&g^+_{\mu\nu}(X_+) - g^-_{\mu\nu}(X_-) = 0\\
&\Leftrightarrow  g^+_{\mu\nu}(X_+) = g^-_{\mu\nu}(X_-)
\end{align}
folgt und die Metrik somit aufgrund der Abhängigkeit der Projektionen $g^+_{\mu\nu}(X_+)$ und $g^-_{\mu\nu}(X_-)$ von unterschiedlichen Koordinaten auf dem gesamten Raum konstant sein müsste.\vspace{12pt}\\ 
Da pseudokomplexe Zahlen und Funktionen ebenfalls kommutativ sind, kann analog zu \cite{Adler1965, Adler1975} über das Linienelement gezeigt werden, dass die Metrik symmetrisch sein muss.
\begin{align}
d\omega^2 = g_{\mu\nu} dX^\mu dX^\nu &= g_{\nu\mu} dX^\nu dX^\mu = g_{\nu\mu}dX^\mu dX^\nu\\
\Leftrightarrow g_{\mu\nu} &= g_{\nu\mu}  
\end{align}
Weil die beiden Nullteiler $\sigma_\pm$ linear unabhängig voneinander sind, ist es möglich beide Räume zu trennen und für jeden eine eigene Formulierung der Relativitätstheorie zu entwickeln. Somit verhalten sich beide wie ein realer Raum, sodass die Überlegungen zur Parallelverschiebung, den Christoffelsymbolen und der kovarianten Ableitung aus \cite{Adler1965,Adler1975} übernommen werden können.\vspace{12pt}\\
Es ist leicht zu zeigen, dass unter diesen Voraussetzungen die kovariante Ableitung der Metrik wie in der ART Null ist. 
\begin{align}
g^\pm_{\mu\nu||\lambda} &= g^\pm_{\mu\nu|\lambda} - \crs{\kappa}{\nu}{\lambda}_\pm g^\pm_{\mu\kappa} - \crs{\kappa}{\mu}{\lambda}_\pm g^\pm_{\kappa\nu}\notag\\
&= g^\pm_{\mu\nu|\lambda} - g^\pm_{\mu\kappa} g^{\kappa\sigma}_\pm [\nu\lambda,\sigma]_\pm - g^\pm_{\kappa\nu}  g^{\kappa\eta}_\pm [\mu\lambda,\eta]_\pm \notag\\
&= g^\pm_{\mu\nu|\lambda} - \frac 12 \left ( g^\pm_{\nu\mu|\lambda} + g^\pm_{\lambda\mu|\nu} - g^\pm_{\nu\lambda|\mu} \right ) - [\mu\lambda,\nu]_\pm\notag\\
&= \frac 12 \left ( g^\pm_{\mu\nu|\lambda} + g^\pm_{\nu\lambda|\mu} - g^\pm_{\lambda\mu|\nu} \right ) - [\mu\lambda,\nu]_\pm\notag\\
&= [\mu\lambda,\nu]_\pm - [\mu\lambda,\nu]_\pm = 0
\end{align}
Folglich ist die Metrik weiterhin eine intrinsische Eigenschaft des Raums und das Relativitätsprinzip wird erfüllt. Zudem besitzt sie weiterhin die Eigenschaft, dass man mit ihr ko- und kontravariante Koordinaten ineinander überführen kann, da 
\begin{align}
	g^{\mu \nu} g_{\nu \lambda} &= g^{\mu \nu}_+ g^+_{\nu\lambda} \sigma_+ + g^{\mu \nu}_- g^-_{\nu\lambda} \sigma_-\\
	&= \delta^\mu_\lambda (\sigma_+ + \sigma_-) = \delta^\mu_\lambda  
\end{align}
gilt.\vspace{12pt}\\
Offensichtlich sollte das Längenelement für makroskopische Probleme in guter Näherung die gleiche Form wie in der ART besitzen, sodass $d\omega^2 \approx g^{eff}_{\mu \nu} dx^\mu dx^\nu$ erfüllt ist und die reale effektive Metrik $g^{eff}_{\mu \nu}$ die Koordinaten näherungsweise ineinader überführt. Dafür wurde in \cite{Hess:2008wd} die naheliegendenste Möglichkeit vorgeschlagen die Realprojektion $g^{0}_{\mu \nu}$ als effektive Metrik zu benutzen.
\begin{align}
g^{0}_{\mu \nu} &:= \frac 12 \left ( g^+_{\mu\nu} + g^-_{\mu\nu} \right)\\
h_{\mu\nu} &:= \frac 12 \left ( g^+_{\mu\nu} - g^-_{\mu\nu} \right)
\end{align}
Jedoch führt das neben einer verkomplizierten Dispersionsrelation ($ h_{\mu\nu} \left (dx^\mu dx^\nu + l^2 du^\mu du^\nu \right ) + 2lg^0_{\mu\nu} dx^\mu du^\nu = 0$) dazu, dass Real- und Imaginärteil von ko- und kontravarianten pseudokomplexen Vektoren auch näherungsweise keine entsprechenden Vektoren mehr sind. Denn für $X_\mu = x_\mu + Ilu_\mu$ folgt
\begin{align}
x_\mu \pm lu_\mu = g^\pm_{\mu\nu} (x^\nu \pm lu^\nu)
\end{align}
und damit
\begin{align}
	x_\mu &= \frac 12 \left ( g^+_{\mu\nu} + g^-_{\mu\nu} \right ) x^\nu + \frac {l}{2} \left ( g^+_{\mu\nu} - g^-_{\mu\nu} \right ) 	u^\nu \notag\\
	&= g^0_{\mu\nu} x^\nu + lh_{\mu\nu} u^\nu\\
	lu_\mu &= \frac 12 \left ( g^+_{\mu\nu} - g^-_{\mu\nu} \right ) x^\nu + \frac {l}{2} \left ( g^+_{\mu\nu} + g^-_{\mu\nu} \right ) 	u^\nu \notag\\
	&= lg^0_{\mu\nu} u^\nu + h_{\mu\nu} x^\nu
\end{align}
Die Berechnung für kontravariante Vektoren ist äquivalent und führt zum selben Ergebnis.\vspace{12pt}\\
Aus diesen Gründen wurde in Analogie zur komplexen Elektrodynamik, bei der für Berechnungen der Realteil der Felder benötigt wird \cite{Greiner2008}, ein neues Abbildungsverfahren festgelegt, bei dem die effektive Metrik durch Realabbildung jeder einzelnen pseudokomplexen Größe von $g^{\pm}_{\mu\nu}$ gewonnen wird. Somit ist klar, dass sie ko- und kontravariante Größen ineinader überführt, da $g^{\pm}_{\mu\nu}$ diese Eigenschaft besitzen. \vspace{12pt}\\
Außerdem folgt daraus für das Längenelement
\begin{align}
d\omega^2 &= g_{\mu\nu} dX^\mu dX^\nu\notag\\
 &= g_{\mu\nu} \left [ \left ( dx^\mu dx^\nu + l^2 du^\mu du^\nu + 2lI dx^\mu du^\nu \right ) \right ]\\
 &\rightarrow g^{eff}_{\mu\nu} \left [ \left ( dx^\mu dx^\nu + l^2 du^\mu du^\nu + 2lI dx^\mu du^\nu \right ) \right ]
\end{align}
und somit ergibt sich aus der Forderung, dass es reell sein muss die Dispersionsrelation der ART
\begin{align}
g^{eff}_{\mu\nu}dx^\mu du^\nu = dx^\mu du_\mu = 0
\end{align}
Zudem ist die Korrektur des Längenelement in der Ordnung l$^2$ und somit für makroskopische Probleme zu vernachlässigen.\vspace{12pt}\\ 
Damit ist die Theorie bis auf die Bestimmungsgleichungen für $g^\pm_{\mu\nu}$ festgelegt. Soweit gibt es für ihre Berechnung keine entscheidenden Änderungen verglichen mit der ART, sodass es so aussieht als ob es sich um eine doppelte, parallele Fomulierung von ihr handelt. Dies ändert sich jedoch mit der Einführung eines neuen Variationsprinzips.\vspace{12pt}\\
Sei L die Lagrangefunktion in einem Integral, dann ist die Wirkung gegeben durch
\begin{align}
S = \int L d\tau
\end{align}
In der pseudokomplexen Allgemeinen Relativitätstheorie wird das Variationsprinzip zur Bestimmung des Gravitationsfeldes wie folgt definiert
\begin{align}
\delta S = \delta \int L d\tau \in \mathbf{P^0}
\end{align}
wobei \textbf{P$^0$} die Menge der Nullteiler sind.\vspace{12pt}\\ Diese Abänderung ist sinnvoll, da die Zahlen in der Nullteilermenge ein verschwindendes Betragsquadrat besitzen und somit eine verallgemeinerte Null darstellen. Zudem würde das Fordern einer Null für die Variation bedeuten, dass die Theorie äquivalent zu einer doppelten Formulierung der ART ist, sodass es erforderlich ist, die beiden Komponenten über das Variationsprinzip zu verknüpfen, um eine neue Theorie der Raumzeit zu erhalten.\vspace{12pt}\\
Über die Lagrangegleichungen folgt daraus
\begin{align}
\frac{D}{Ds} \frac{DL}{D\dot{X}^\mu} - \frac{DL}{DX^\mu}\in \mathbf{P^0}
\end{align}
mit einem Kurvenparameter s.\vspace{12pt}\\
Dieses Variationsprinzip bedeutet, dass sich zwei auf unterschiedlichen Kurven parallelverschobene Vektoren um ein Element aus dem Nullteilerraum unterscheiden können. Dementsprechend werden alle Vektoren als äquivalent bezeichnet, deren Differenz im Nullteilerraum liegt.\vspace{12pt}\\  
Daraus folgen wie in \cite{Adler1965,Adler1975} für eine vom Krümmungstensor abhängige Lagrangefunktion ($L = \sqrt{-g}R$ mit g als der Determinante der Metrik und dem Kümmungsskalar R) die umformulierten Einsteinschen Feldgleichungen für den freien Raum
\begin{align}
G_{\mu\nu} = R_{\mu\nu} - \frac 12 g_{\mu\nu} R \in \mathbf{P}^0
\end{align}
wobei G$_{\mu\nu}$ der pseudokomplexe Einsteintensor und R$_{\mu\nu}$ der zugehörige Riccitensor ist.\vspace{12pt}\\ 
Aufgrund der Symmetrie in $\sigma_\pm$ ist das reelle Ergebnis unabhängig davon in welchem Nullteilerraum sich der Term auf der rechten Seite der Einsteinschen Feldgleichungen befindet und man kann ohne Beschränkung der Allgemeinheit annehmen
\begin{align}
G_{\mu\nu} = R_{\mu\nu} - \frac 12 g_{\mu\nu} R = \Xi_{\mu\nu} \sigma_-
\end{align}
wobei $\Xi_{\mu\nu}$ ein vorerst unbekannter Tensor ist, der eine Quelle des Gravitationfeldes darstellt.\\
Weiterhin folgt daraus, dass für g$^+_{\mu\nu}$ die gewöhnlichen Einsteinschen Feldgleichungen gelten und dementsprechend
\begin{align}
g^+_{\mu\nu} = g^{ART}_{\mu\nu}
\end{align} 
ist.\vspace{12pt}\\
Sobald das Gravitationsfeld bestimmt ist, ergibt sich die Geodätengleichung aus dem unveränderten Variationsprinzip, da nach wie vor die kürzeste Verbindung gesucht wird. Also folgt für die Geodäte:
\begin{align}
\ddot{X}^\mu + \crs{\mu}{\nu}{\lambda} \dot{X}^\nu \dot{X}^\lambda = 0 \label{Geoglg}
\end{align}
\section{Bestimmung der Quellfunktionen}
Aktuell gibt es zwei Ansätze, wie die $\Xi$-Funktionen bestimmt werden können.\vspace{12pt}\\
Der erste Ansatz wurde im Originalpaper \cite{Hess:2008wd} verwendet und basiert auf der Überlegung, dass möglichst viele Eigenschaften der ART Lösungen erhalten werden sollen. Dafür wird angenommen, dass der Krümmungsskalar weiterhin für den ``leeren'' Raum immer 0 ist. Wie später zu sehen sein wird, führt zudem die Absicht die Symmetrie $g_{11} = -\frac{1}{g_{00}}$ der Schwarzschild-Metrik zu erhalten zur Annahme, dass $\Xi_0 = \Xi_1$ gilt. Zwar sind damit die Quellfunktionen allein dadurch noch nicht eindeutig bestimmt, aber zusammen mit den Feldgleichungen für sphärisch symmetrische Probleme sind sie es. \vspace{12pt}\\
Der zweite und neuere Ansatz ist die Quelle in unserem ``freien'' Raum als ideale Flüssigkeit zu betrachten. Dieses Bild ist mit der dunklen Energie verbunden, von der häufig ebenfalls diese Eigenschaft erwartet wird (siehe beispielsweise \cite{Carturan:2002si,Brevik:2007jt}).\vspace{12pt}\\
Der Energie-Impuls-Tensor einer idealen Flüssigkeit ist in \cite{Adler1965,Adler1975} zu finden.
\begin{align}
T^{frei}_{\mu\nu} = \rho u_\mu u_\nu + \frac{p}{c} (u_\mu u_\nu - g_{\mu\nu})
\end{align} 
Für statische Probleme ist zu erwarten, dass die Flüssigkeit ruht, sodass die räumlichen Geschwindigkeiten wegfallen. In diesem Fall ist u$_0$ gegeben durch
\begin{align}
&ds^2 = g_{00} (dx^0)^2 \Leftrightarrow 1 = g_{00}(u^0)^2\\ 
&\Rightarrow u^0 = \frac{1}{\sqrt{g_{00}}} \Rightarrow u_0 = \sqrt{g_{00}}
\end{align}
und damit folgt der Energie-Impuls-Tensor als \\
\begin{align}
T^{frei}_{\mu\nu} = \begin{pmatrix} \rho g_{00} & -\frac{p}{c^2} g_{01} & -\frac{p}{c^2} g_{02} & -\frac{p}{c^2} g_{03}\\ -\frac{p}{c^2} g_{10} & -\frac{p}{c^2} g_{11} & -\frac{p}{c^2} g_{12} & -\frac{p}{c^2} g_{13}\\ -\frac{p}{c^2} g_{20} & -\frac{p}{c^2} g_{21} & -\frac{p}{c^2} g_{22} & -\frac{p}{c^2} g_{23} \\ -\frac{p}{c^2} g_{30} & -\frac{p}{c^2} g_{31} & -\frac{p}{c^2} g_{32} & -\frac{p}{c^2} g_{33}  \end{pmatrix}\\
\notag
\end{align}
In dieser Arbeit spielen nur radialsymmetrische Probleme eine Rolle, sodass hier immer \\
\begin{align}
T^\text{frei}_{\mu\nu} &= \begin{pmatrix} \rho e^{\nu} & 0 & 0 &0 \\ 0 & pe^{\lambda} & 0 & 0 \\ 0 & 0 & pr^2 & 0 \\ 0 & 0 & 0 & pr^2\sin^2(\vartheta) \end{pmatrix}
\end{align}
gilt.\vspace{12pt}\\
Somit lässt sich $\Xi_{\mu\nu}$ bestimmen:
\begin{align}
\Xi_{\mu\nu} = - \frac{8\pi \kappa}{c^2} T^{frei}_{\mu\nu} = - \frac{8\pi \kappa}{c^2} \begin{pmatrix} \rho e^{\nu} & 0 & 0 &0 \\ 0 & pe^{\lambda} & 0 & 0 \\ 0 & 0 & pr^2 & 0 \\ 0 & 0 & 0 & pr^2\sin^2(\vartheta) \end{pmatrix}
\end{align}
Dementsprechend erhält man per Tensortransformation die $\Xi_\mu$
\begin{align}
\Xi_0 &= - \frac{8\pi \kappa}{c^2} \rho \notag\\
\Xi_1 &= \Xi_2 = \Xi_3 = - \frac{8\pi \kappa}{c^4} p \label{Xiradsym}
\end{align}

\section{Die Schwarzschildlösung}
Die erste und einfachste nichttriviale Lösung der Einsteinschen Feldgleichung war die Schwarzschild-Metrik für den Außenraum einer ruhenden Massekugel. Da die neuen Feldgleichungen relativ nah an denen von Einstein sind und die Metrik nach wie vor die Symmetrien der Energieverteilung übernimmt, ist davon auszugehen, dass dieses Problem ebenfalls die einfachste Lösung nach dem freien Raum ist. Dementsprechend wurde dieses Problem auch zuerst angegangen. \vspace{12pt}\\
Wie oben schon erwähnt ist $g^+_{\mu\nu}$ die Schwarzschild-Metrik der ART und ist somit ohne jegliche Rechnung bestimmt.\vspace{12pt}\\
Die Metrik insgeamt kann wie in \ref{secSMet} angesetzt werden, da die dort gelieferten Argumente weiterhin gültig sind. Folglich gilt:\\
\begin{align}
g_{\mu \nu} = \begin{pmatrix} e^{\nu(r)} & 0 & 0 & 0 \\ 0 & -e^{\lambda(r)} & 0 & 0 \\
0 & 0 & -r^2 & 0 \\ 0 & 0 & 0 & -r^2\sin^2(\vartheta) \end{pmatrix}\\
\notag
\end{align}
Hierbei ist natürlich zu beachten, dass es sich nun bei allen Variablen um pseudokomplexe Größen handelt (zum Beispiel $e^{\nu(r)} = e^{\nu_+(r_+)}\sigma_+ + e^{\nu_-(r_-)}\sigma_-$).\vspace{12pt}\\
Jedoch ist g$^+_{\mu\nu}$ schon bekannt, und es ist wenig sinnvoll die Minuszeichen und das $\sigma_-$ in der gesamten Rechnung mitzunehmen. Insofern lässt man sie wegfallen und merkt sich, dass die folgenden Rechnugen sich nur auf den $\sigma_-$-Anteil beziehen.\vspace{12pt}\\
Dementsprechend lauten die Feldgleichungen in gemischter Form
\begin{align}
R^\mu_{\nu} - \frac 12 g^\mu_{\nu} R = \Xi^\mu_{\nu}
\end{align}
Da die Nebendiagonalelemente der Metrik verschwinden, gilt dies auch für den Ricci-Tensor und somit auch zwangsläufig für $\Xi^\mu_{\nu}$. Daher bietet es sich an den Schreibaufwand etwas zu reduzieren, indem man $\Xi^\mu_{\mu} := \Xi_\mu$ definiert (ohne Einsteinschen Summenkonvention).\vspace{12pt}\\
Um mit \cite{Hess:2008wd} konsistent zu sein und da es sich später als nützlich erweisen wird, werden die Größen $\xi_\mu$ definiert über
\begin{align}
R_{00} &:= -\frac{1}{2} e^{\nu-\lambda}\xi_0 \label{PARTRic0}\\
R_{11} &:= \frac{1}{2} \xi_1 \label{PARTRic1}\\
R_{22} &:= \xi_2\label{PARTRic2}\\
R_{33} &:= \xi_3 = \xi_2 \sin^2(\vartheta) \label{PARTRic3}
\end{align}
Die kovarianten Komponenten des Ricci-Tensors für diesen Metrikansatz sind in \ref{secSMet} zu finden, sodass leicht zu ersehen ist, dass der Zusammenhang zwischen $\xi_2$ und $\xi_3$ die Lösung nicht einschränkt, sondern nur eine Eigenschaft des Ricci-Tensors widerspiegelt.\vspace{12pt}\\
Somit ersetzen die Gleichungen \eqref{PARTRic0},\eqref{PARTRic1},\eqref{PARTRic2} die Feldgleiungen. Jedoch stellt $\Xi^\mu_\nu$ bis auf eine Konstante einen unbekannten Energie-Impuls-Tensor dar, dessen Bestimmung deutlich einfacher sein dürfte als die einer abstrakten Größe wie dem Ricci-Tensor multipliziert mit einem von der Metrik abhängigen Term. \vspace{12pt}\\ 
Darum bietet es sich an vor dem Lösen der Gleichungen \eqref{PARTRic0},\eqref{PARTRic1},\eqref{PARTRic2},\eqref{PARTRic3} den Zusammenhang zwischen den $\xi_\mu$ und den $\Xi_\mu$ zu bestimmen. Dafür wird die gemischte Form des Ricci-Tensors und der Krümmungsskalar in Abhänigkeit der $\xi_\mu$ bestimmt
\begin{align}
R^0_0 &= g^{0\mu}R_{\mu 0} = g^{00}R_{00} = -\frac{1}{2} e^{-\lambda}\xi_0 \\
R^1_1 &= g^{1\mu}R_{\mu 1} = g^{11}R_{11} = -\frac{1}{2} e^{-\lambda}\xi_1 \\
R^2_2 &= g^{2\mu}R_{\mu 2} = g^{22}R_{22} = -\frac{\xi_2}{r^2} = R_{33}\\
\Rightarrow R &= -\frac{1}{2} e^{-\lambda} (\xi_0 + \xi_1) - \frac{2\xi_2}{r^2} \label{Krschwarz}
\end{align}
und dann in die Einsteingleichungen eingesetzt
\begin{align}
- \frac 14 e^{-\lambda} \xi_0 + \frac 14 e^{-\lambda} \xi_1  + \frac{\xi_2}{r^2} &= \Xi_0 \label{PARTgemsch0}\\
\frac 14 e^{-\lambda} \xi_0 - \frac 14 e^{-\lambda} \xi_1  + \frac{\xi_2}{r^2} &= \Xi_1 \label{PARTgemsch1}\\
\frac 14 e^{-\lambda} (\xi_0 + \xi_1) &= \Xi_2 \label{PARTgemsch2}\\
\Xi_3 = \Xi_2 \label{PARTgemsch3}
\end{align}
Durch Addieren der ersten beiden Gleichungen und Subtrahieren der ersten von der zweiten erhält man
\begin{align}
\frac{2\xi_2}{r^2} &= \Xi_0 +\Xi_1\\
\frac{1}{2} e^{-\lambda} (\xi_0 - \xi_1) &= \Xi_1 - \Xi_0 \label{GKXis}\\
\frac{1}{4} e^{-\lambda} (\xi_0 + \xi_1) &= \Xi_2
\end{align} 
Multipliziert man nun die dritte Gleichung mit 2 und addiert das Ergebnis zu der zweiten Gleichung und zieht es zudem von ihr ab, dann ergibt sich
\begin{align}
e^{-\lambda} \xi_0 &= 2 \Xi_2 + \Xi_1 - \Xi_0 \notag\\
e^{-\lambda} \xi_1 &= 2 \Xi_2 - \Xi_1 + \Xi_0 \notag\\
\frac{2\xi_2}{r^2} &= \Xi_0 +\Xi_1 \label{xiXi}
\end{align}
Nachdem dieser Zusammenhang geklärt ist, sind als Nächstes die Feldgleichungen zu lösen. Mit den Ausdrücken aus \ref{secSMet} sind diese gegeben durch:\\
\begin{align}
R_{00} = e^{\nu-\lambda}\left[ - \frac{\nu''}{2} + \frac{\lambda'\nu'}{4} - \frac{\nu'^2}{4} - \frac{\nu'}{r} \right] &= -\frac{1}{2} e^{\nu-\lambda}\xi_0 \label{PARTschwarz0}\\
R_{11} = \frac{\nu''}{2} - \frac{\lambda'\nu'}{4} + \frac{\nu'^2}{4} - \frac{\lambda'}{r} &= \frac{1}{2} \xi_1  \label{PARTschwarz1}\\
R_{22} = e^{-\lambda}\left[ 1 + \frac{r\nu'}{2} - \frac{r\lambda'}{2} \right] -1  &= \xi_2 \label{PARTschwarz2}\\
\notag
\end{align}
Durch Multiplikation von \eqref{PARTschwarz0} mit -e$^{-\nu+\lambda}$ erhält man, wenn man \eqref{PARTschwarz1} subtrahiert\\
\begin{align}
\frac{\nu' + \lambda'}{r} &= \frac 12 (\xi_0 -\xi_1)\\
\Rightarrow \nu' &= -\lambda' +  \frac{r}{2} (\xi_0 -\xi_1)\label{nupschwarz}\\
\Rightarrow \nu''&= -\lambda'' + \frac 12 (\xi_0 -\xi_1) + \frac{r}{2} (\xi_0' -\xi_1') \label{nuppschwarz}\\
\notag
\end{align}
Einsetzen von \eqref{nupschwarz} und \eqref{nuppschwarz} in das Doppelte von \eqref{PARTschwarz1} führt zu
\begin{align}
\xi_1 &= -\lambda'' + \frac 12 (\xi_0 -\xi_1) + \frac{r}{2} (\xi_0' -\xi_1') \notag\\
&~~- \frac{\lambda'(-\lambda' +  \frac{r}{2} (\xi_0 -\xi_1))}{2} + \frac{(-\lambda' +  \frac{r}{2} (\xi_0 -\xi_1))^2}{2} - \frac{2\lambda'}{r}\\
&= - \left (\lambda'' - \lambda'^2 + \frac{2\lambda'}{r} \right ) + \frac 12 (\xi_0 -\xi_1) + \frac{r}{2} (\xi_0' -\xi_1')\notag\\ 
&~~-\frac{3r}{4} (\xi_0 -\xi_1) + \frac{r^2}{8}(\xi_0 -\xi_1)^2
\end{align} 
Mit Hilfe der Relation
\begin{align}
\frac{e^\lambda}{r} \left ( re^{-\lambda} \right )'' = -\lambda'' + \lambda'^2 - \frac{2\lambda'}{r} 
\end{align}
und \eqref{PARTschwarz2} kann das noch umgeschrieben werden zu
\begin{align}
\frac{e^\lambda}{r} \xi_2'- \xi_1 = -\frac r4 (\xi_0' -\xi_1') +  \frac r2 \lambda' (\xi_0 -\xi_1) - \frac{r^2}{8} (\xi_0 -\xi_1)^2 \label{xiDGL}
\end{align}
Dies ist eine Bestimmungsgleichung für die $\xi$-Funktionen, die jedoch aufgrund zu vieler Unbekannter ohne Ansatz nicht gelöst werden kann.\vspace{12pt}\\
Es ist dennoch möglich Aussagen über die Metrikterme zu erhalten, die nun jedoch noch von den $\xi$-Funktionen abhängen. Dafür schaut man sich \eqref{PARTschwarz2} an und setzt \eqref{nupschwarz} ein.
\begin{align}
&e^{-\lambda}\left[ 1 + \frac{r}{2} (-\lambda' + \frac{r^2}{2} (\xi_0 - \xi_1) - \frac{r\lambda'}{2} \right] -1  = \xi_2\\
&\Rightarrow e^{-\lambda}\left[ 1 - r\lambda' \right ] =  1 + \xi_2 - \frac{r^2}{4} e^{-\lambda} (\xi_0 - \xi_1)
\end{align} 
Und wegen $e^{-\lambda}\left[ 1 - r\lambda' \right ] = (re^{-\lambda})'$ gilt:
\begin{align}
e^{-\lambda} &= 1 - \frac{2M}{r} + \frac{1}{r} \int \xi_2 dr - \frac {1}{4r} \int e^{-\lambda} r^2 (\xi_0 - \xi_1) dr\\
&= 1 - \frac{2M}{r} + \frac{1}{r} \int \xi_2 dr + \frac {1}{2r} \int  r^2 (\Xi_0 - \Xi_1) dr
\end{align}
Daraus ergibt sich mit\eqref{nupschwarz}
\begin{align}
e^\nu &= e^{-\lambda} e^{\frac 12 \int r (\xi_0 - \xi_1) dr}\\
&= \left (1 - \frac{2M}{r} + \frac{1}{r} \int \xi_2 dr + \frac {1}{2r} \int  r^2 (\Xi_0 - \Xi_1) dr \right )e^{\frac 12 \int r (\xi_0 - \xi_1) dr}\\
\notag
\end{align}
Folglich ist die allgemeine Schwarzschildmetrik gegeben durch\\
\begin{align}
g_{\mu \nu} = \begin{pmatrix} -\frac{e^{\frac 12 \int r (\xi_0 - \xi_1) dr}}{g_{11}} & 0 & 0 & 0 \\ 0 & -\frac{1}{1 - \frac{2M}{r} + \frac{1}{r}\int \xi_2 dr + \frac {1}{2r} \int  r^2 (\Xi_0 - \Xi_1) dr} & 0 & 0 \\
0 & 0 & -r^2 & 0 \\ 0 & 0 & 0 & -r^2\sin^2(\vartheta) \end{pmatrix}\\
\notag
\end{align}
Zusammen mit g$^+_{\mu\nu}$ erhält man daraus die effektive Schwarzschildmetrik\\
\begin{align}
g^{eff}_{\mu \nu} = \begin{pmatrix} -\frac{e^{\frac 14 \int r (\xi_0 - \xi_1) dr}}{g_{11}} & 0 & 0 & 0 \\ 0 & -\frac{1}{1 - \frac{2M}{r} + \frac{1}{2r}\int \xi_2 dr + \frac {1}{4r} \int  r^2 (\Xi_0 - \Xi_1) dr} & 0 & 0 \\ 0 & 0 & -r^2 & 0 \\ 0 & 0 & 0 & -r^2\sin^2(\vartheta) \end{pmatrix}\\
\notag
\end{align}
\subsection{Originalansatz}\label{origans}
Wie vorher schon erwähnt, wird beim Originalansatz angenommen, dass $\Xi_0 = \Xi_1$ gilt und der Krümmungsskalar verschwindet. Aus der ersten Bedingung folgt mit \eqref{GKXis}, dass auch $\xi_0 = \xi_1$ gilt und somit vereinfachen sich die allgemeinen Formeln erheblich.
\begin{align}
e^\nu = e^{-\lambda} = 1 - \frac{2M}{r} + \frac{1}{r} \int \xi_2 dr \label{origlsg}
\end{align}
Somit muss zur Lösung des Problems nur noch $\xi_2$ bestimmt werden. Dafür nutzen wir die Bedingung R = 0 mit \eqref{Krschwarz} und \eqref{xiDGL}, um eine Differentialgleichung für $\xi_2$ zu erhalten
\begin{align}
R &= 0 = -\frac{1}{2} e^{-\lambda} (\xi_0 + \xi_1) - 
\frac{2\xi_2}{r^2} = -e^{-\lambda} \xi_1 - \frac{2\xi_2}{r^2}\\
\Leftrightarrow \xi_1 &= -e^{\lambda} \frac{2\xi_2}{r^2}\\
0 &= \frac{e^\lambda}{r} \xi_2'- \xi_1 = \frac{e^\lambda}{r} \xi_2' + e^{\lambda} \frac{2\xi_2}{r^2}\\
\Rightarrow \xi_2' &= -\frac{2\xi_2}{r}\\
\notag 
\end{align}
Diese DGL wird durch $\xi_2 = \frac{-B}{r^2}$ gelöst, wobei B eine noch zu bestimmende Konstante ist. Eingesetzt in \eqref{origlsg} folgt daraus
\begin{align}
e^\nu = e^{-\lambda} = 1 - \frac{2M}{r} + \frac{B}{r^2}
\end{align}
Zudem können daraus die $\xi_\mu$ und $\Xi_\mu$ bestimmt werden\\
\begin{align}
\xi_0 &= \xi_1 = -e^{\lambda} \frac{2\xi_2}{r^2} = \frac{2B}{r^4\left ( 1 - \frac{2M}{r} + \frac{B}{r^2} \right )}\\
\xi_3 &= \xi_2 \sin^2 (\vartheta)\\
\Xi_0 &= \Xi_1 = \frac{\xi_2}{r^2} = -\frac{B}{r^4}\\
\Xi_2 &= \Xi_3 = \frac{1}{2} e^{-\lambda} \xi_0 = - \frac{\xi_2}{r^2} = \frac{B}{r^4} \\
\notag
\end{align}
Somit ergibt sich die effektive Metrik\\
\begin{align}
g_{\mu \nu} = \begin{pmatrix} 1 - \frac{2M}{r} + \frac{B}{2r^2} & 0 & 0 & 0 \\ 0 & -\frac{1}{1 - \frac{2M}{r} + \frac{B}{2r^2}} & 0 & 0 \\ 0 & 0 & -r^2 & 0 \\ 0 & 0 & 0 & -r^2\sin^2(\vartheta) \end{pmatrix}\\
\notag
\end{align}
Eine Korrektur der Metrik der Ordnung $\frac{1}{r^2}$, wie sie hier auftritt, lässt sich unter anderem mit Hilfe des PPN Formalismus testen. Obwohl es bisher kein theoretisches Prinzip zur Bestimmung von B gibt, ist es insofern möglich, die Theorie teilweise zu testen und B zu bestimmen, indem man versucht, experimentelle Daten an eine Korrektur der Ordnung $\frac{1}{r^2}$ anzupassen.
\subsection{Ideale Flüssigkeit} 
Für die Schwarzschild-Lösung im Flüssigkeitsmodell müssen zuerst die $\xi_\mu$ mit Hilfe von \eqref{Xiradsym} und \eqref{xiXi} bestimmt werden.
\begin{align}
e^{-\lambda} \xi_0 &= \frac{8\pi\kappa}{c^2} (\rho + 3\frac {p}{c^2}) \\
e^{-\lambda} \xi_1 &=  -\frac{8\pi\kappa}{c^2} (\rho - \frac {p}{c^2})\\
\frac{2\xi_2}{r^2} &= -\frac{8\pi\kappa}{c^2} (\rho - \frac {p}{c^2})
\end{align}
Daraus leitet sich direkt eine Relation zwischen $\xi_1$ und $\xi_2$ ab
\begin{align}
e^{-\lambda} \xi_1 = \frac{2\xi_2}{r^2}
\end{align}
Aus den allgemeinen Gleichungen folgt so für $\nu$ und $\lambda$
\begin{align}
e^{-\lambda} &= 1 - \frac{2M}{r} -\frac{8\pi\kappa}{c^2r} \int r^2\rho dr \\
e^\nu &=  e^{-\lambda} e^{\frac{8\pi\kappa}{c^2} \int r e^\lambda \left (\rho+ \frac{p}{c^2} \right ) dr}
\end{align}
und somit ist die Metrik über $\rho$ und p bestimmt.\vspace{12pt}\\ 
Das Einsetzen der $\xi$-Funktionen in \eqref{xiDGL} führt zu einer Differentialgleichung für die Dichte und den Druck. Die linke Seite der Gleichung ergibt\\
\begin{align}
\frac{e^\lambda}{r} \xi_2'- e^\lambda \frac{2\xi_2}{r^2} &= -\frac{e^\lambda}{r} \left (Cr (\rho - \frac{p}{c^2} + \frac{r^2}{2}C \left (\rho' + \frac{p'}{c^2} \right ) \right ) + e^\lambda C \left (\rho + \frac{p}{c^2} \right ) \notag\\
&= - e^\lambda C \left (\rho + \frac{p}{c^2} \right ) -\frac{1}{2}e^\lambda Cr \left (\rho' + \frac{p'}{c^2} \right ) + e^\lambda C \left (\rho + \frac{p}{c^2} \right )\notag\\
&= -\frac{1}{2}e^\lambda Cr \left (\rho' + \frac{p'}{c^2} \right )\\
\notag
\end{align}
und die rechte Seite\\
\begin{align}
&-\frac r4 (\xi_0' -\xi_1') +  \frac r2 \lambda' (\xi_0 -\xi_1) - \frac{r^2}{8} (\xi_0 -\xi_1)^2 \notag\\
&= -\frac {Cr}{2} e^\lambda \left ( \lambda' \left (\rho + \frac{p}{c^2} \right ) + \left (\rho' + \frac{p'}{c^2} \right ) \right ) + Cr e^\lambda \lambda' \left (\rho + \frac{p}{c^2} \right ) - \frac{C^2r^2}{2} e^{2\lambda} \left (\rho + \frac{p}{c^2} \right )^2 \notag\\
&= \frac{Cr}{2} \lambda' \left (\rho + \frac{p}{c^2} \right ) - \frac{C^2r^2}{2} e^{2\lambda} \left (\rho + \frac{p}{c^2} \right )^2 - \frac{Cr}{2} e^\lambda \left (\rho'+ \frac{p'}{c^2} \right )\\
\notag
\end{align}
wobei $C := \frac{8\pi\kappa}{c^2}$ gesetzt wurde.\vspace{12pt}\\
Beide Seiten zusammen ergeben die gesuchte DGL
\begin{align}
\frac{p'}{c^2} = \frac{\lambda'}{2} \left (\rho+ \frac{p}{c^2} \right ) - \frac{Cr}{2} e^\lambda \left (\rho+ \frac{p}{c^2} \right )^2 \label{Druckdgl}
\end{align}
Diese Differentialgleichung ist offensichtlich nichtlinear und deshalb nicht trivial zu lösen. Zudem können die von $\lambda$ abhängigen Terme noch in Abhängigkeit von $\rho$ und p ausgedrückt werden, sodass noch weitere Nichtlinearitäten zu Tage treten.\vspace{12pt}\\
Jedoch handelt es sich dabei um eine Bestimmungsgleichung für die Zustandsgleichung der Flüssigkeit und dementsprechend existiert nur noch ein Freiheitsgrad des Systems. Somit ist zumindest theoretisch die vollständige Lösung durch die Kenntnis der Dichte oder des Drucks der Flüssigkeit gegeben.\vspace{12pt}\\
Im einfachsten Fall könnte man annehmen, dass die Dichte und der Druck konstant sind. Dann wird \eqref{Druckdgl} durch $\frac{p}{c^2} = -\rho$ gelöst und es gilt wie bei der ART Lösung mit kosmologischer Konstante $g_{00} = -\frac{1}{g_{11}} = 1 - \frac{2M}{r} -\frac{8\pi\kappa}{3c^2} r^2\rho$, sofern man $\Lambda = \frac{8\pi\kappa}{3c^2}\rho$ setzt. Dieser Zusammenhang ist eine weitere Rechtfertigung für die Identifikation der idealen Flüssigkeit mit der Dunklen Energie.  

\chapter{Resultate}
\section{Reissner-Nordström-Lösung}
In diesem Unterkapitel wird das gleiche Problem wie in \ref{SecRMM} in der neuen Theorie behandelt, d.h. die Metrik einer statischen, sphärisch symmetrischen und geladenen Masseverteilung.\vspace{12pt}\\ Dabei können die Symmetrieargumente wieder übernommen werden und der Energie-Impuls-Tensor wird unverändert wie in der ART in die Feldgleichungen eingebracht. Zudem ist er real, da Energie und Impuls eine reale Bedeutung besitzen.\vspace{12pt}\\
Somit ergeben sich die Feldgleichungen als\\
\begin{align}
R_\mu^\nu - \frac 12 g_\mu^\nu R = \Xi_\mu^\nu \sigma_- - \frac{8\pi \kappa}{c^2} T_\mu^\nu
\end{align}
wobei die Metrik und der Energie-Impuls-Tensor erneut durch
\begin{align}
g_{\mu \nu} &= \begin{pmatrix} e^{\nu(r)} & 0 & 0 & 0 \\ 0 & -e^{\lambda(r)} & 0 & 0 \\
0 & 0 & -r^2 & 0 \\ 0 & 0 & 0 & -r^2\sin^2(\vartheta) \end{pmatrix}\\
T_{\mu}^\nu &=  \frac{\varepsilon^2}{2c^2r^4} \begin{pmatrix} 1& 0 & 0 & 0 \\ 0 & 1 & 0 & 0 \\ 0& 0& -1 & 0\\ 0&0&0& -1 \end{pmatrix}
\end{align}
gegeben sind.\vspace{12pt}\\
Wie schon im Schwarzschildfall ist $g^+_{\mu\nu}$ gleich $g^{ART}_{\mu\nu}$ und für die Berechnung von $g^-_{\mu\nu}$ werden die Minuszeichen und das $\sigma_-$ wieder weggelassen. Zudem wird die Definition der $\Xi_\mu$ übernommen, sodass man die vier Gleichungen
\begin{align}
\frac 12 \left ( R_0^0 - R_1^1 - R_2^2 - R_3^3 \right ) = \Xi_0 - \frac{8\pi \kappa}{c^2} T_0^0 \label{gemEinst0}\\
\frac 12 \left ( R_1^1 - R_0^0 - R_2^2 - R_3^3 \right ) = \Xi_1 - \frac{8\pi \kappa}{c^2} T_1^1 \label{gemEinst1}\\
\frac 12 \left ( R_2^2 - R_0^0 - R_1^1 - R_3^3 \right ) = \Xi_2 - \frac{8\pi \kappa}{c^2} T_2^2 \label{gemEinst2}\\
\frac 12 \left ( R_3^3 - R_0^0 - R_1^1 - R_2^2 \right ) = \Xi_3 - \frac{8\pi \kappa}{c^2} T_3^3 \label{gemEinst3}
\end{align}
erhält.\vspace{12pt}\\
Die Differenz zwischen der dritten Gleichung und der zweiten Gleichung ergibt
\begin{align}
R_2^2 - R_3^3 = \Xi_2 - \Xi_3 - \frac{8\pi \kappa}{c^2} \left ( T_2^2 - T_3^3 \right )
\end{align}
und da T$_2^2$ = T$_3^3$ und R$_2^2$ = R$_3^3$ gilt, folgt
\begin{align}
\Xi_3 = \Xi_2
\end{align}
Für weitere Aussagen müssen die R$_\mu^\nu$ berechnet werden. Dies ist jedoch nicht sehr aufwändig, da nur die kovarianten Komponenten aus \eqref{PARTRic0},\eqref{PARTRic1}, \eqref{PARTRic2} und \eqref{PARTRic3} transformiert werden müssen. Daraus ergibt sich
\begin{align}
R_\mu^\nu = \begin{pmatrix} e^{-\lambda} \left ( - \frac{\nu''}{2} + \frac{\lambda' \nu'}{4} - \frac{\nu'^2}{4} - \frac{\nu'}{r} \right ) & 0 & 0 & 0 \\ 0 & e^{-\lambda} \left ( - \frac{\nu''}{2} + \frac{\lambda' \nu'}{4} - \frac{\nu'^2}{4} + \frac{\lambda'}{r}  \right )& 0 & 0 \\ 0 & 0 & -\frac{e^{-\lambda}}{r^2} \left ( 1 + \frac{r\nu'}{2} + \frac{r\lambda'}{2} \right ) + \frac{1}{r^2} & 0 \\ 0 & 0 & 0 & R_2^2 \end{pmatrix}
\end{align}
Durch Abziehen von \eqref{gemEinst1} von \eqref{gemEinst0} erhält man den Zusammenhang
\begin{align}
\lambda' + \nu' = re^\lambda \left ( \Xi_1 - \Xi_0 \right ) 
\end{align} 
Analog zum Schwarzschildfall werden die $\xi_\mu$ mit der gleichen Beziehung zu den $\Xi_\mu$ definiert, sodass sich die zur Schwarzschildrechnung analoge Gleichung
\begin{align}
\lambda' + \nu' &= \frac 12 r  \left ( \xi_0 - \xi_1 \right )\\
\Rightarrow \nu'' &= -\lambda'' + \frac{1}{2} (\xi_0 - \xi_1 ) + \frac 12 r (\xi_0' - \xi_1')  
\end{align}
ergibt.\vspace{12pt}\\
Zieht man \eqref{gemEinst1} von \eqref{gemEinst0} ab, addiert \eqref{gemEinst2} zweifach und multipliziert das Ergebnis mit 2$e^\lambda$, folgt die Gleichung
\begin{align}
\nu'' - \frac{\lambda' \nu'}{2} + \frac{\nu'^2}{2} - \frac{2\lambda'}{r} = \xi_1 - \frac{2A}{r^4}e^\lambda \label{bestDGL}
\end{align}
Wobei erneut $A := - \dfrac{4\pi \kappa \varepsilon^2}{c^4}$ gesetzt wurde.\\
Zudem ergibt sich aus der Addition von \eqref{gemEinst0} und \eqref{gemEinst1} 
\begin{align}
e^{-\lambda} \left ( 1 + \frac{r\nu'}{2} - \frac{r\lambda'}{2} \right ) - 1 &= \xi_2 + \frac{A}{r^2}\\
\Rightarrow \left (re^{-\lambda}\right )' &= 1 + \xi_2 - \frac 14 r^2e^{-\lambda} \left ( \xi_0 - \xi_1 \right ) + \frac{A}{r^2} \label{bestg11}
\end{align}
Wenn man nun diese Gleichung benutzt und sie zusammen mit den Gleichungen für $\nu'$ und $\nu''$ in \eqref{bestDGL} einsetzt, führt dies auf die schon aus dem Schwarzschildfall bekannte DGL zur Bestimmung der $\xi_\mu$.
\begin{align}
\frac{e^\lambda}{r}\xi_2' - \xi_1 = \frac 12 \lambda' r \left ( \xi_0 - \xi_1 \right) - \frac 14 r \left ( \xi_0' - \xi_1' \right ) - \frac 18 r^2 \left ( \xi_0 - \xi_1 \right)^2
\end{align}
Somit sind sie die gleichen Funktionen wie bei der Schwarzschildlösung und es ist möglich die Reissner-Nordström-Lösung in Abhänigkeit von der Schwarzschildlösung aufzuschreiben.\\
Aus \eqref{bestg11} erhält man nun
\begin{align}
e^{-\lambda} = 1 - \frac{r_s}{r} + \frac{1}{r} \int \xi_2 dr - \frac {1}{4r} \int e^{-\lambda} r^2 \left ( \xi_0 - \xi_1 \right ) dr - \frac{A}{r^2} = e^{-\lambda_s} - \frac{A}{r^2}
\end{align}
Und über den Zusammenhang von $\nu$ und $\lambda$ wird daraus die g$_{00}$-Komponente berechnet
\begin{align}
e^\nu = e^{-\lambda} e^{\frac 12 \int r (\xi_0 -\xi_1) dr} = e^{\nu_s} - \frac{A}{r^2} e^{\frac 12 \int r (\xi_0 -\xi_1) dr}
\end{align} 
Insofern weist $e^{-\lambda}$ exakt die Änderung von der Schwarzschildlösung zur Reissner-Nordström-Lösung auf, die aus der ART zu erwarten wäre. Jedoch gibt es für die g$_{00}$-Komponente  noch eine abgeänderte Korrektur, sobald $\xi_0$ und $\xi_1$ verschieden sind.\vspace{12pt}\\
In Anbetracht der Tatsache, dass die Quellfunktionen noch nicht endgültig bestimmt sind, ist dies das optimale Ergebnis, da somit die Reissner-Nordströmlösung sofort bestimmt ist, sobald die Schwarzschildlösung bekannt ist. So erhält man für die Anfangshypothese, dass die lokale Krümmung verschwindet (R = 0) und $\xi_0 = \xi_1$ gilt, die Lösung
\begin{align}
e^{-\lambda} = e^\nu = 1 - \frac{r_s}{r} + \frac{B-A}{r^2}
\end{align}
die die zu erwartende Korrektur aus der ART aufweist. 
\section{Überprüfung und Interpretation des Originalansatzes} \label{testorig}
Für den Originalansatz wurde in \ref{origans} eine analytische Lösung des Schwarzschildproblems gefunden, die nur noch von dem unbekannten Parameter B abhängt.
\begin{align}
g_{\mu \nu} = \begin{pmatrix} 1 - \frac{2M}{r} + \frac{B}{2r^2} & 0 & 0 & 0 \\ 0 & -\frac{1}{1 - \frac{2M}{r} + \frac{B}{2r^2}} & 0 & 0 \\ 0 & 0 & -r^2 & 0 \\ 0 & 0 & 0 & -r^2\sin^2(\vartheta) \end{pmatrix}\\
\notag
\end{align}
Für eine rein analytische Theorie sollte es möglich sein diesen aus grundlegenden Prinzipien zu bestimmen. Im urspünglichen Paper \cite{Hess:2008wd} wurde postuliert, dass $g_{00}$ ím gesamten Raum positiv sein muss, um zumindest eine untere Grenze zu erhalten. Da dies unter der Annahme $B>0$ für $r\rightarrow 0$ und $r\rightarrow \infty$ offensichtlich der Fall ist, ist es möglich B über die Bedingung, dass das lokale Minimum größer als 0 sein muss, nach unten abzuschätzen.
\begin{align}
\frac{dg_{00}}{dr} &= \frac{r_s}{r^2} - \frac{B}{r^3} \stackrel{!}{=} 0\notag\\
\Rightarrow r &= \frac{B}{r_s} \label{beschns}
\end{align}
Eingesetzt in $g_{00}$ folgt daraus mit der Bedingung
\begin{align}
g_{00}(\frac{B}{r_s}) &= 1 - \frac{r_s^2}{B} + \frac{r_s^2}{2B} > 0\\
\Leftrightarrow B &> \frac{r_s^2}{2} 
\end{align} 
Diese Bedingung hätte zur Folge, dass sowohl der Ereignishorizont als auch die Singularität einer sehr dichten Masseansammlung, die in der ART auftreten, nicht mehr existieren und somit ein schwarzes Loch grau würde. \vspace{12pt}\\
Wie in \ref{origans} schon erwähnt ist es jedoch auch möglich die Korrektur mittels des PPN-Formalismus' zu testen. Durch Vergleich mit der Robertson Entwicklung \eqref{robertsonentw} erhält man
\begin{align}
B = (\beta - \gamma) r_s^2
\end{align}
Aus Experimenten ist bekannt, dass $|\gamma - 1| < 2,3 \cdot 10^{-5}$ und $|\beta - 1| < 2,3 \cdot 10^{-4}$ gilt \cite{lrr-2006-3}, und daraus folgt
\begin{align}
|\beta - \gamma| = |\beta - 1 - (\gamma - 1)| \leq |\beta - 1| + |\gamma - 1| < 2,6 \cdot 10^{-4}
\end{align}
und somit
\begin{align}
B < 2,6 \cdot 10^{-4} r_s^2
\end{align}
Dies ist ein Widerspruch zur vorherigen Annahme, die dementsprechend falsifiziert ist.\vspace{12pt}\\
Da bisher keine anderen Ideen zur Bestimmung von B existieren, bleibt vorerst also nichts anderes übrig, als die Theorie ab diesem Punkt als phänomenologisch anzusehen. \vspace{12pt}\\
Daher ist es sinnvoll die Konsequenzen der Annahme, dass die Korrektur der Metrik proportional zu $\frac{B}{r^2}$ mit $B << r_s^2$ ist, zu untersuchen.\vspace{12pt}\\  Dafür werden zuerst die Nullstellen von $g_{00}$ bestimmt:
\begin{align}
g_{00}(r_0) &= 1 - \frac{r_s}{r_0} + \frac{B}{2r_0^2} \stackrel{!}{=} 0\\
\Rightarrow r_{01} &= \frac{r_s}{2} \left (1 + \sqrt{1-\frac{2B}{r_s^2}}\right ) &
r_{02} &= \frac{r_s}{2} \left (1 - \sqrt{1-\frac{2B}{r_s^2}}\right )
\end{align}
Für $ 0 < B < 2,6 \cdot 10^{-4}$ liegen demnach immer zwei Nullstellen vor, zwischen denen $g_{00}$ negativ und $g_{11}$ positiv ist. Dieser Bereich ist der Ergosphäre der Kerr-Metrik in der ART ähnlich \cite{Adler1965,Adler1975,Inverno2009}, da aufgrund des Postulats $ds^2 \geq 0$ in ihm kein ruhendes Teilchen existieren kann. Im vorliegenden Fall ist es jedoch nicht möglich in diesem Bereich umzukehren, da nur der Metrikkoeffizient der radialen Bewegung positiv ist und daher die radiale Geschwindigkeit ihr Vorzeichen nicht ändern kann. Diese Bedingung verliert aber für $r < r_{02}$ und $r>r_{01}$ ihre Gültigkeit, sodass a priori eine Rückkehr aus dem Inneren theoretisch möglich ist. \vspace{12pt}\\
Um dies zu veranschaulichen wird die Bewegung eines anfangs bei $r_A > r_{01}$ ruhenden Testteilchens betrachtet. Die für die Geodätengleichung zu variierende Lagrangefunktion ist
\begin{align}
L = \frac{ds^2}{ds^2} = 1 = g_{00} c^2t'^2 - \frac{1}{g_{00}} r'^2 - r^2 \vartheta'^2 - r^2 \sin^2(\vartheta) \varphi'^2
\end{align} 
wobei $x_\mu' = \frac{dx_\mu}{ds}$ definiert wurde.\vspace{12pt}\\ 
Aus L und den Lagrangegleichungen für t,$\vartheta$ und $\varphi$ folgt das Gleichungssystem
\begin{align}
g_{00} c^2t'^2 - \frac{1}{g_{00}} r'^2 - r^2 \vartheta'^2 - r^2 \sin^2(\vartheta) \varphi'^2 &= 1\\
\frac{d}{ds} ( 2g_{00}c^2t') &= 0\\
\frac{d}{ds} ( - 2r^2 \vartheta')  + 2r^2 \sin(\vartheta)\cos(\vartheta) \varphi'^2 &= 0\label{Geothe}\\
\frac{d}{ds} (-2r^2\sin^2(\vartheta) \varphi') &= 0\label{Geophi}
\end{align}
Aus \eqref{Geophi} und der Bedingung, dass das Testteilchen anfangs in Ruhe sein soll, folgt $r^2\sin^2(\vartheta) \varphi' = 0$. Dementsprechend muss $\vartheta$ oder $\varphi$ konstant sein, sodass $\sin(\vartheta) = 0$ oder $\varphi' = 0$ gilt.\vspace{12pt}\\
Unabhängig davon welche der beiden Bedingungen in \eqref{Geothe} eingesetzt wird, ergibt sich die Gleichung
\begin{align}
\frac{d}{ds} ( - 2r^2 \vartheta') = 0
\end{align}
aus der durch analoge Betrachtungen folgt, dass die räumliche Bewegung rein radial ist.\vspace{12pt}\\
Insofern kann das Gleichungssystem vereinfacht werden
\begin{align}
g_{00} c^2t'^2 - \frac{1}{g_{00}} r'^2 &= 1\label{infL}\\
\frac{d}{ds} ( 2g_{00}c^2t') &= 0\label{Geot}\\
\vartheta(s) = \varphi(s) = 0\notag
\end{align}
wobei $\vartheta$ und $\varphi$ gewählt wurden.\vspace{12pt}\\
Aus \eqref{Geot} lässt sich durch Integration t' als
\begin{align}
t' = \frac{A}{2g_{00}c^2}
\end{align}
bestimmen, wobei A eine vorerst unbekannte Konstante ist.\vspace{12pt}\\
Dieses Ergebnis in \eqref{infL} eingesetzt führt zu
\begin{align}
r'^2 = \frac{A^2}{4c^2} - g_{00}
\end{align}
Nun kann A durch Verwendung der Randbedingungen bestimmt werden
\begin{align}
r(0) &= r_A & r'(0) &= 0\\
&&\Rightarrow \frac{A^2}{4c^2} = g_{00}(r_A) \Rightarrow A = \pm 2c \sqrt{g_{00}(r_A)}
\end{align}
Da für $r\rightarrow \infty$ $t' \geq 0$ gilt, muss das positive Vorzeichen verwendet werden und A ist duch $2c \sqrt{g_{00}(r_A)}$ gegeben. Folglich gilt
\begin{align}
t' &= \frac{\sqrt{g_{00}(r_A)}}{g_{00}c}\\
r'^2 &= g_{00} (r_A) - g_{00} \label{squarevr} 
\end{align}
Bei den gegebenen Anfangsbedingungen existiert unabhängig von $r_A$ ein weiterer Punkt r$_U$ mit $g_{00}(r_U) = g_{00}(r_A)$. Somit hat $r'$ zwei Nullstellen. Das Verhalten an diesen Stellen kann durch die Betrachtung von $r''$ bestimmt werden. Um dieses zu erhalten wird \eqref{squarevr} abgeleitet.
\begin{align}
\frac{d}{ds} r'^2 = 2r'r'' &= - \frac{d}{ds} g_{00}(r(s)) = - r' \frac{dg_{00}}{dr}\\
\Rightarrow r'' = - \frac{1}{2} \frac{dg_{00}}{dr}  
\end{align}
Das Einsetzen der Originallösung $(g_{00} = 1 -\frac{r_s}{r} + \frac{B}{2r^2})$ ergibt
\begin{align}
r'' = -\frac 12 \left ( \frac{r_s}{r^2} - \frac{B}{r^3} \right ) = \frac{1}{2r^2} \left ( \frac{B}{r} - r_s \right)
\end{align}
Die letzte Darstellung ist ein Produkt aus zwei streng monoton fallenden Termen, von denen zudem der erste immer positiv ist. Folglich ist $r''$ ebenfalls streng monton fallend, das heißt es gilt für beliebige $r_1 > r_2$
\begin{align}
r''(r_2) > r''(r_1)
\end{align}
Aus \eqref{beschns} folgt zudem, dass $r''$ für $r_B = \frac{B}{r_s}$ verschwindet. Somit wird ein Testteilchen für alle $r < r_B$ nach außen gedrückt und für alle $r > r_B$ nach innen gezogen. Zusammen mit der Tatsache, dass $r_A > r_B$ und $r_U < r_B$ folgt daher, dass das Testteilchen in dieser Metrik aus seiner Sicht oszillieren wird. Damit ist vom Standpunkt des mitbewegten Beobachters eine Rückkehr aus dem Inneren eindeutig möglich, sodass auch für diese Lösung weder eine Singularität noch ein Ereignishorizont existiert.\vspace{12pt}\\
Allerdings ist diese Rechnung keinesfalls realistisch, da spätestens für $r \leq r_{02}$ die Annahme eines leeren Raumes widersprüchlich ist und die Metrik daher zumindest dort wenn nicht sogar für größere r modifiziert werden muss und die daher berechnete Trajektorie keine reale Situation widerspiegeln kann.\vspace{12pt}\\
Unter der Annahme, dass erst für $r \leq r_{02}$ ein nichtleerer Raum vorliegt, kann die mittlere Dichte im Inneren in Abhängigkeit von der Gesamtmasse abgeschätzt werden:
\begin{align}
\langle \rho \rangle = \frac{M}{V} = \frac{M}{\frac{4}{3}\pi r_{02}^3}
\end{align}
Das Volumen ist, solange $r_{02}$ existiert, streng monoton wachsend mit B, und die Dichte ist somit minimal für maximales B ($B = 2,6 \cdot 10^{-4} r_s^2$). Dementsprechend gilt
\begin{align}
\langle \rho \rangle >& \frac{3c^6}{4G^3M^2\left ( 1 - \sqrt{0,99974} \right )^3}\\
&= \frac{3c^6}{4G^3M_S^2\left ( 1 - \sqrt{0,99974} \right )^3} \frac{M_S^2}{M^2}
\end{align}
wobei $M_S$ die Sonnenmasse ist, auf die mittlere Dichte nun normiert ist. Die Werte für die Konstanten können in \cite{TBphys2005} nachgeschlagen werden und eingesetzt ergibt sich nach Rundung auf die erste Nachkommastelle
\begin{align}
\langle \rho \rangle > 2,1 \cdot 10^{32} \frac{kg}{m^3} \frac{M_S^2}{M^2} \label{density}
\end{align}
Um ein Gefühl für die Größenordnung der Dichte zu bekommen, wird sie für die zentrale Massen unserer Galaxie $M \approx 4 \cdot 10^6 M_S$ \cite{Gillessen:2008qv} betrachtet. Daraus folgt $\langle \rho \rangle > 1,3 \cdot 10^{19} \frac{kg}{m^3}$. Dies ist in etwa die gleiche Größenordnung wie die Dichte eines Neutronensterns \cite{Fliesbach2006} und somit keinesfalls vernachlässigbar. \vspace{12pt}\\
Betrachtet man erneut t' fällt auf, dass es für verschwindendes g$_{00}$ divergiert und damit die dem Teilchen von außen zugeordnete Zeit springt, wenn nicht divergiert. Unabhängig davon bedeutet das, dass t für den Bereich zwischen den beiden Nullstellen keine sinnvolle Koordinate ist.\vspace{12pt}\\
Das genaue Verhalten von t für die Nullstellen von g$_{00}$ könnte einfach mit Hilfe von r(s) geklärt werden. Unglücklicherweise ist \eqref{squarevr} zwar durch Separation der Variablen prinzipiell analytisch lösbar, aber die Komplexität des Integrals macht ein Umstellen nach r unmöglich.\vspace{12pt}\\
Deshalb muss man eine andere Herangehensweise wählen und betrachtet $\frac{dr}{dt}$ für ein von außen auf $r=r_{01}+\epsilon$ zufallendes Teilchen.
\begin{align}
 \dot{r} := \frac{dr}{dt} = \frac{r'}{t'} = -c\frac{\sqrt{g_{00}(r_A)-g_{00}}g_{00}}{\sqrt{g_{00}(r_A)}}
\end{align}
Über Trennung der Variablen erhält man
\begin{align}
 c(t_i - t) &= \int_{r_i}^{r_{01}+\epsilon} \frac{\sqrt{g_{00}(r_A)}}{\sqrt{g_{00}(r_A)-g_{00}}g_{00}} d\bar{r}\\
&= \int_{r_i}^{r_{01}+\epsilon} \frac{\bar{r}^3\sqrt{g_{00}(r_A)}}{\sqrt{g_{00}(r_A)\bar{r}^2-g_{00}\bar{r}^2}g_{00}\bar{r}^2} d\bar{r}
\end{align}
Nun wird $\bar{r} = r_{01}(1 - \bar{\epsilon})$ substituiert und nur der Unterschied für einen Ort $r_i = r_{01} + a$ nahe $r_{01}$ betrachtet, sodass Zähler und Nenner näherungsweise in die erste Ordnung von $\bar{\epsilon}$ entwickelt werden können. Dafür werden die einzelnen Terme betrachtetet
\begin{align}
 g_{00}\bar{r}^2 &= \bar{r}^2 - r_s\bar{r} + \frac{B}{2} = r_{01}^2 (1-2\bar{\epsilon} + \bar{\epsilon}^2) - r_sr_{01} (1-\bar{\epsilon} + \frac{B}{2})\\
&\approx r_{01}^2 - r_sr_{01} + \frac{B}{2} - \bar{\epsilon}r_{01} (2r_{01} - r_s)  = -\bar{\epsilon}r_{01} (2r_{01} - r_s)\\
\bar{r}^3 &\approx  r_{01}^3 (1-3\bar{\epsilon})\\
\bar{r}^2 &\approx  r_{01}^2 (1-2\bar{\epsilon})
\end{align}
Eingesetzt ergibt sich
\begin{align}
 c(t_i - t) &\approx \int_{-a}^{-\epsilon} \frac{\sqrt{g_{00}(r_A)}r_{01}^4 (1-3\bar{\epsilon)}}{\sqrt{g_{00}(r_A)r_{01}^2(1-2\bar{\epsilon})+\bar{\epsilon}r_{01} (2r_{01} - r_s)}\bar{\epsilon}r_{01} (2r_{01} - r_s)} d\bar{\epsilon}\\
&\approx \int_{-a}^{-\epsilon} \frac{r_{01}^2(1-3\bar{\epsilon})}{\bar{\epsilon}(2r_{01}-r_s)}\\
&= \frac{r_{01}^2}{2r_{01}-r_s} \left [ \ln \left ( \frac{\epsilon}{a} \right ) + 3 (a-\epsilon)\right ] 
\end{align}
Nun ist offensichtlich, dass t für den Grenzfall $\epsilon \rightarrow 0$ divergiert, das heißt für die Nullstellen von g$_{00}$ divergiert die Zeitkoordinate t. Insofern ist davon auszugehen, dass ein Ereignishorizont vorliegt.\vspace{12pt}\\
Um diese Problematik besser zu durchschauen, kann sie noch über eine andere Herangehensweise betrachtet werden. Analog zu Kapitel 7.8 in \cite{Adler1975} werden Ereignishorizonte gesucht. Dafür werden die Oberflächen mit konstantem Radius analysiert, deren gemeinsamer Normalvektor durch 
\begin{align}
n_\alpha = \begin{pmatrix} 0 & 1 & 0 & 0 \end{pmatrix}
\end{align}
gegeben ist. Der ``Betrag'' diese Vektors ist dann
\begin{align}
n^\alpha n_\alpha = g^{\alpha\mu}n_\mu n_\alpha = g^{11} = - \left ( 1 - \frac{r_s}{r} + \frac{B}{2r^2} \right )
\end{align}
Er verschwindet somit für $r_{01}$ und $r_{02}$, sodass bei diesen Radien Ereignishorizonte vorliegen, das heißt die Oberflächen können aus Sicht des weit entfernten Beobachters nur in eine Richtung durchschritten werden.\vspace{12pt}\\
Insofern tritt bei der Originallösung zwar die Singularität der Schwarzschildmetrik nicht mehr auf, aber der Ereignishorizont bleibt leicht verschoben erhalten. Dementsprechend ist das Innere nach wie vor vom Äußeren abgeschnürt.\vspace{12pt}\\ 
Dieses Problem tritt allgemein auf, sobald eine sphärisch symmetrische Metrik Nullstellen besitzt. Allerdings ist die Erklärung äußerst dunkler Objekte im Universum auch anders möglich. Materie mit einer Dichte wie in \eqref{density} für eine Sonnenmasse dürfte optisch dicht sein, sodass jegliche Strahlung von der Oberfläche ausgesandt wird. Liegt diese Oberfläche dann bei sehr hoher Rotverschiebung ist das Objekt äußerst dunkel und somit ist der Unterschied zu einem Schwarzen Loch experimentell schwer feststellbar.\vspace{12pt}\\
Zusammengefasst lässt sich sagen, dass die Korrektur proportional zu $\frac{1}{r^2}$ nicht auszuschließen ist. Jedoch erscheint sie aus philosophischen Gründen wenig attraktiv, da bisherige Beobachtungen auch ohne das Auftreten eines Ereignishorizontes erklärbar sind. 

\section{Semiklassische Beschreibung von Dirac-Teilchen in Gravitationsfeldern} \label{semiGrav}
Wie vorher schon erwähnt ist es bisher nicht gelungen eine Quantentheorie der Gravitation zu finden, sodass Quantenphänomene (insbesondere in Gravitationsfeldern) nicht vollkommen exakt berechnet werden können, weil jedes reale Teilchen Energie trägt und somit gravitativ wechselwirkt.\vspace{12pt}\\
Die Erfahrung zeigt jedoch, dass quantenmechanische Rechnungen im Allgemeinen sehr gut mit den zugänglichen experimentellen Befunden übereinstimmen. Dies liegt daran, dass die Gravitation die schwächste Wechselwirkung ist und somit äußerst hohe Energien benötigt werden, damit sie relevante Beiträge leistet. Folglich können durch Vakuumfluktuationen und durch die betrachteten Teilchen hervorgerufene Metrikveränderungen in sehr guter Näherung vernachlässigt werden.\vspace{12pt}\\
Jedoch sind die Veränderungen der Metrik sehr massiver nahegelegener Objekte durchaus relevant (wie man an den klassischen Metriken sieht), sodass diese in die Rechnungen mit einbezogen werden müssen. Dies führt zu einer semiklassischen Beschreibung von Teilchen in Gravitationsfeldern, da man ihre Quantennatur in einem gekrümmten Hintergrundraum untersucht.\vspace{12pt}\\
Da an dieser Stelle Dirac-Teilchen betrachtet werden sollen, muss dementsprechend die Dirac-Gleichung auf einen gekrümmten Hintergrundraum verallgemeinert werden. Die dafür nötigen Schritte können in \cite{GMRbuch} nachgelesen werden und werden hier der Vollständigkeit halber wiederholt.\vspace{12pt}\\
Die freie Dirac-Gleichung im Minkowskiraum ist gegeben durch
\begin{align}
\left (i\hbar\gamma^\mu \fracpd{}{x^\mu} - mc \right ) \Psi = 0 \label{MinDi}
\end{align}
mit der Anitkommutationsrelation
\begin{align}
\left \{ \gamma^\mu , \gamma^\nu \right \} = 2\eta^{\mu\nu} \label{Minkom}
\end{align}
Ein gekrümmter Raum zeichnet sich dadurch aus, dass der Riemannsche Krümmungstensor nicht verschwindet. Dies hat unter anderem zur Folge, dass die partielle Ableitung nicht mehr wie ein ko- bzw kontravarianter Vektor transformiert, da die Parallelverschiebung die Komponenten eines Vektors nicht konstant lässt. Da jedoch alle Gleichungen form- invariant unter Koordinatentransformationen sein sollen, ist klar, dass die Ableitung in der Dirac-Gleichung modifiziert werden muss. \vspace{12pt}\\
Die Antikommutationsrelation der Dirac-Matirzen wird indirekt über die Bedingung gegeben, dass jede Komponente des Spinors eine Klein-Gordon-Gleichung erfüllen muss \cite{GMRbuch}. Da in dieser ebenfalls die Ableitungen modifiziert werden müssen, wird auch die Relation für einen gekrümmten Raum abgeändert. Es zeigt sich, dass die naheliegendste Möglichkeit erfüllt ist, und \eqref{Minkom} wird zu
\begin{align}
\left \{ \gamma^\mu , \gamma^\nu \right \} = 2g^{\mu\nu} \label{Riekom}
\end{align}
Somit sind die $\gamma^\mu$ abhängig vom Raumzeitpunkt und kommutieren folglich nicht mehr mit den Ableitungsoperatoren $\fracpd{}{x^\mu}$. Zusammengenommen wird \eqref{MinDi} somit zu
\begin{align}
\left [ i\hbar \gamma^\mu \left ( \fracpd{}{x^\mu} + \Gamma_\mu \right ) - mc \right ]\Psi \equiv \left (i\hbar\gamma^\mu D_\mu - mc \right )\Psi = 0
\end{align}
wobei $\Gamma_\mu$ von Weyl, Fock und Iwanenko aus der Bedingung, dass der Hamiltonian hermitesch sein muss, berechnet wurde \cite{Weyl:1929fm,Fock:1929vt,Iwa:1929vt} als
\begin{align}
\Gamma_\mu = \frac{1}{4} \gamma_\nu \left ( \fracpd{\gamma^\nu}{x^\mu} + \crs{\nu}{\lambda}{\mu} \gamma^\lambda \right ) \label{GroGA}
\end{align}
\subsection{Dirac-Gleichung in sphärischer Symmetrie}
In der gesamten Arbeit werden ausschließlich sphärisch symmetrische Problemstellungen betrachtet, sodass das Ziel dieser Sektion ist, die Dirac-Gleichung für diesen Fall möglichst weit zu vereinfachen. Jedoch ist die Herleitung der radialen Gleichungen, die die weitreichenste allgemeine Vereinfachung darstellen, äußerst aufwändig und befindet sich deshalb nicht im Hauptteil der Arbeit. Allerdings befinden sich alle Rechnungen äußerst detailliert im Anhang und können dort nachvollzogen werden. Hier wird eine deutlich verkürzte Form präsentiert.\vspace{12pt}\\ 
Aufgrund der sphärischen Symmetrie nimmt die zu betrachtende Metrik wieder die Form \eqref{Ansmetradsym} an. Aber es bietet sich aus rechentechnischer Sicht an, die  Berechnungen in isotropen Koordinaten durchzuführen und dann auf die gewöhnlichen Koordinaten zurückzurechnen. Somit ist das Linienelement
\begin{align}
 ds^2 = \omega(\rho) c^2dt^2 - \delta(\rho) \left ( (dx^1)^2 + (dx^2)^2 + (dx^3)^2 \right ) \label{Isolin}
\end{align}
Folglich ist die Metrik diagonal, und es ist möglich die Dirac-Matrizen als Produkt der jeweiligen alten Matrix und einer vom zugehörigen Metrikkoeffizienten abhängigen Funktion auszudrücken:
\begin{align}
\gamma_0 &= \sqrt{\omega} \tilde{\gamma}_0 & \gamma^0 &= \frac{1}{\sqrt{\omega}} \tilde{\gamma}^0\notag\\
\gamma_i &= \sqrt{\delta} \tilde{\gamma}_i & \gamma^i &=  \frac{1}{\sqrt{\delta}} \tilde{\gamma}^i
\end{align}
wobei $\tilde{\gamma_\mu}$ wie in \cite{GMRbuch} die jeweilige Dirac-Matrix im Minkowskiraum bezeichnet.\vspace{12pt}\\
Die nichtverschwindenden Christoffelsymbole für die isotrope Metrik können mittels der Geodätengleichung einfach berechnet oder aus \cite{Papapetrou:1956} übernommen werden
\begin{align}
 \crs{0}{0}{i} &= \crs{0}{i}{0} = \frac{\bar{\omega}x^i}{2\omega \rho}\\
\crs{i}{0}{0} &= \frac{\bar{\omega}x^i}{2\delta \rho}\\
\crs{i}{i}{j} &= \crs{i}{j}{i} = \frac{\bar{\delta}x^j}{2\delta \rho}\\
\crs{i}{j}{j} &= -\frac{\bar{\delta}x^i}{2\delta \rho} ~\forall~ i \neq j
\end{align}
wobei der Balken die Ableitung nach der zugehörigen Koordinate (hier $\rho$) bezeichnet. \vspace{12pt}\\
Mit Hilfe von \eqref{GroGA} werden nun die $\Gamma_\mu$ berechnet.
\begin{align}
  \Gamma_0 &=  \frac{\bar{\omega}}{4\omega\rho} \gamma_0 \sum_{i=1}^3 x^i \gamma^i\\
 \Gamma_i &= \frac{\bar{\alpha}}{4\alpha \rho}\gamma_i \sum_{\substack{j=1\\j\neq i}}^3 x^j \gamma^j
\end{align}
Daraus folgt
\begin{align}
  \gamma^\mu \Gamma_\mu &= \frac{1}{4\rho} \left ( \frac{\bar{\gamma}}{\gamma} + \frac{2\bar{\alpha}}{\alpha} \right ) \sum_{i=1}^3 x^i \gamma^i
\end{align}
Dementsprechend ist die Dirac-Gleichung gegeben durch
\begin{align}
 i\hbar \gamma^0 \fracpd{\Psi}{x^0} &= -i\hbar \gamma^i \fracpd{}{x^i} \Psi +  mc \Psi -i\hbar \frac{1}{4\rho} \left ( \frac{\bar{\omega}}{\omega} + \frac{2\bar{\delta}}{\delta} \right ) \sum_{i=1}^3 x^i \gamma^i\Psi\\
\Leftrightarrow \frac{i\hbar}{\sqrt{\omega}} \fracpd{\Psi}{t} &= c\vec{\tilde{\delta}} \left [\vec{p} -i\hbar \frac{1}{4\sqrt{\delta}} \left ( \frac{\bar{\omega}}{\omega} + \frac{2\bar{\delta}}{\delta} \right ) \vec{e}_\rho  \right ]\Psi + \beta mc^2\Psi
\end{align}
Ein Vergleich der beiden Linienelemente führt zu den Gleichungen
\begin{align}
 \omega &= e^\nu\\
 r &= \sqrt{\delta} \rho\\
\frac{dr}{d\rho} &= \sqrt{\delta} e^{-\frac{\lambda}{2}}
\end{align}
und daraus folgt
\begin{align}
 \bar{\delta} &= -\frac{2r^2}{\rho^3} + \frac{2r}{\rho^2} \frac{dr}{d\rho} = -\frac{2}{r} \delta^{\frac{3}{2}} + \frac{2}{r} e^{-\frac{\lambda}{2}} \delta^{\frac{3}{2}}\\  
\bar{\omega} &= \frac{de^\nu}{dr} \frac{dr}{d\rho} = \bar{\nu}(r) \omega  \sqrt{\delta} e^{-\frac{\lambda}{2}} 
\end{align}
Da zudem die Einheitsvektoren in r- und $\rho$-Richtung gleich sind, transformiert sich die Dirac-Gleichung zu
\begin{align}
 i\hbar e^{-\frac{\nu}{2}} \fracpd{\Psi}{t} &= c\vec{\alpha} \left [\vec{p} -i\hbar \left ( \frac{\bar{\nu}}{4}e^{-\frac{\lambda}{2}} + \frac{1}{r} \left ( e^{-\frac{\lambda}{2}} - 1 \right )  \right ) \vec{e}_r  \right ]\Psi + \beta mc^2\Psi
\end{align}
wobei $\vec{\alpha}$ der aus dem Minkowskiraum bekannte Vektor der $\alpha$-Matrizen ist.\vspace{12pt}\\
Somit liegt die Dirac-Gleichung mitsamt den Verschiebungen nun in den gewöhnlichen Schwarzschildkoordinaten vor, und es ist möglich Lösungen, die aus den Rechnungen dieses und des vorherigen Kapitels hervorgehen, direkt einzusetzen.\vspace{12pt}\\
Damit die Gleichungen übersichtlicher werden, wird 
\begin{align}
 \vec{\tilde{\Gamma}} := -i\hbar \left ( \frac{\bar{\nu}}{4}e^{-\frac{\lambda}{2}} + \frac{1}{r} \left ( e^{-\frac{\lambda}{2}} - 1 \right )  \right ) \vec{e}_r \label{Gammatilde}
\end{align}
definiert.\vspace{12pt}\\
An dieser Stelle kommt der in \cite{GMRbuch} gegebene Ansatz zum Einsatz
\begin{align}
 \Psi =  e^{-\frac{\lambda}{4}} \frac{1}{r} \begin{pmatrix} \Phi_1(r,t) \chi_\kappa^\mu(\vartheta,\varphi)\\ i\Phi_2(r,t) \chi_{-\kappa}^\mu(\vartheta,\varphi) \end{pmatrix}
\end{align}
wobei die $\chi_{\pm \kappa}^\mu(\vartheta,\varphi)$ die aus dem Minkwoskiraum bekannten Zweierspinoren für den Winkelanteil sind.\vspace{12pt}
Damit folgen zwei Gleichungen
\begin{align}
 &(1)~ i\hbar\fracpd{\Phi_1}{t} \chi_\kappa^\mu = mc^2 e^{\frac{\nu}{2}} \Phi_1 \chi_\kappa^\mu + ic r e^{\frac{\nu}{2}}e^{\frac{\lambda}{4}} \vec{\sigma} \left ( \vec{p}+\vec{\tilde{\Gamma}}\right ) e^{-\frac{\lambda}{4}} \frac{1}{r} \Phi_2 \chi_{-\kappa}^\mu\\
&(2)~ i\hbar\fracpd{\Phi_2}{t} \chi_{-\kappa}^\mu = - mc^2 e^{\frac{\nu}{2}} \Phi_2 \chi_{-\kappa}^\mu - ic e^{\frac{\nu}{2}} r e^{\frac{\lambda}{4}} \vec{\sigma} \left ( \vec{p}+\vec{\tilde{\Gamma}}\right ) e^{-\frac{\lambda}{4}} \frac{1}{r} \Phi_1 \chi_\kappa^\mu    
\end{align}
wobei $\vec{\sigma}$ der Vektor aus Pauli-Matrizen ist, der aufgrund der Struktur der $\alpha$-Matrizen vorkommt.\vspace{12pt}\\
Die Paulimatrizen besitzen einige nützliche Eigenschaften, die beispielsweise in \cite{Schwabl2005} nachgelesen werden können. Unter anderem gilt
\begin{align}
 (\vec{\sigma}\vec{a}) (\vec{\sigma}\vec{b}) = \vec{a}\vec{b} + i \vec{\sigma} \left ( \vec{a} \times \vec{b}\right )
\end{align}
und daraus folgt
\begin{align}
 \vec{\sigma} \left ( \vec{p}+\vec{\tilde{\Gamma}}\right ) &= (\vec{\sigma}\vec{e}_r) \left ( -i\hbar e^{-\frac{\lambda}{2}} \fracpd{}{r} -i\hbar \left ( \frac{\bar{\nu}}{4}e^{-\frac{\lambda}{2}} + \frac{1}{r} \left ( e^{-\frac{\lambda}{2}} - 1  \right ) \right ) + \frac{i}{r} \vec{\sigma} \vec{L} \right )
\end{align}
Zudem erfüllt der Vektor aus den Zweierspinoren für beliebige Funktionen f und g die Eigenwertgleichung
\begin{align}
 \hat{K} \begin{pmatrix} f(r,t) \chi_\kappa^\mu(\vartheta,\varphi)\\ g(r,t)  \chi_{-\kappa}^\mu(\vartheta,\varphi) \end{pmatrix} &= \beta (\vec{\Sigma}\vec{L} + \hbar) \begin{pmatrix} f(r,t) \chi_\kappa^\mu(\vartheta,\varphi)\\ g(r,t)  \chi_{-\kappa}^\mu(\vartheta,\varphi) \end{pmatrix}\\ 
&= -\hbar \kappa \begin{pmatrix} f(r,t) \chi_\kappa^\mu(\vartheta,\varphi)\\ ig(r,t)  \chi_{-\kappa}^\mu(\vartheta,\varphi) \end{pmatrix}
\end{align} 
mit
\begin{align}
 \kappa = \begin{cases} - \left (j+\frac{1}{2} \right ) ~\mathrm{ falls }~ j= l + \frac{1}{2}\\
           j + \frac{1}{2} ~\mathrm{ falls }~ j = l - \frac{1}{2}
          \end{cases}
\end{align}
sodass sie Eigenvektoren zum Operator $\vec{\sigma}\vec{L}$ und der Eigenwert aus der Relation in Abhängigkeit von der Drehimpuls- und der Bahndrehimpulsquantenzahl abgelesen werden kann.\vspace{12pt}\\
Weiterhin erfüllen sie die Relation
\begin{align} 
\vec{\sigma}\vec{e}_r \chi_{\pm \kappa} &= \chi_{\mp \kappa}
\end{align}
und somit ergibt sich für $\Phi = \begin{pmatrix} \Phi_1 \\ \Phi_2 \end{pmatrix}$ die Gleichung
\begin{align}
 i\hbar \fracpd{\Phi}{t} = \left [ i\hbar c\alpha_r e^{\frac{\nu-\lambda}{4}} \fracpd{}{r} e^{\frac{\nu-\lambda}{4}} - i\hbar c\beta_r\alpha_r e^{\frac{\nu}{2}} \frac{\kappa}{r} + \beta e^{\frac{\nu}{2}} mc^2 \right ]\Phi
\end{align}
wobei hier
\begin{align}
 \beta := \begin{pmatrix} 1 & 0 \\ 0 & -1 \end{pmatrix} && \alpha_r := \begin{pmatrix} 0 & -i \\ i & 0 \end{pmatrix} 
\end{align}
definiert wurden. Diese Gleichung entspricht bis auf ein Vorzeichen bei dem Impuls- und dem Drehimpulsterm der Formel (21.21) in \cite{GMRbuch} für $V = 0$. Somit ergeben sich auch im Folgenden Unterschiede. \vspace{12pt}\\
Sofern keine weiteren Einschränkungen vorgenommen werden, ist dies das abschließende Ergebnis. Allerdings können für eine stationäre Metrik die stationären Zustände gesucht werden. In diesem Fall kann ein erneuter Ansatz gemacht werden
\begin{align}
 \Phi(r,t) = e^{\frac{\lambda - \nu}{4}} \begin{pmatrix} f(r) \\ g(r) \end{pmatrix} e^{-i\frac{E}{\hbar}t}
\end{align}
und erst ein erneutes Umstellen der Gleichungen nach
\begin{align}
 \frac{d}{dr} \begin{pmatrix} f\\ g \end{pmatrix} = e^{\frac{\lambda}{2}} \begin{pmatrix} -\frac{\kappa}{r} & - \left ( \frac{mc}{\hbar} + \frac{E}{\hbar c}e^{-\frac{\nu}{2}}  \right ) \\ -\left ( \frac{mc}{\hbar} - \frac{E}{\hbar c}e^{-\frac{\nu}{2}}  \right ) &  \frac{\kappa}{r} \end{pmatrix}\begin{pmatrix} f\\ g \end{pmatrix} \label{radDira}
\end{align}
führt zum abschließenden Ergebnis.\vspace{12pt}\\
Auf dieser Grundlage können nun die stationären Zustände für jedes durch die Dirac-Gleichung beschreibbare Problem in einer beliebigen Metrik bestimmt und dadurch experimentell testbare Vorhersagen gemacht werden. 
\subsection{Atomare Energieniveaus}
\begin{figure}[hbt]
\centering
\includegraphics[width = \textwidth]{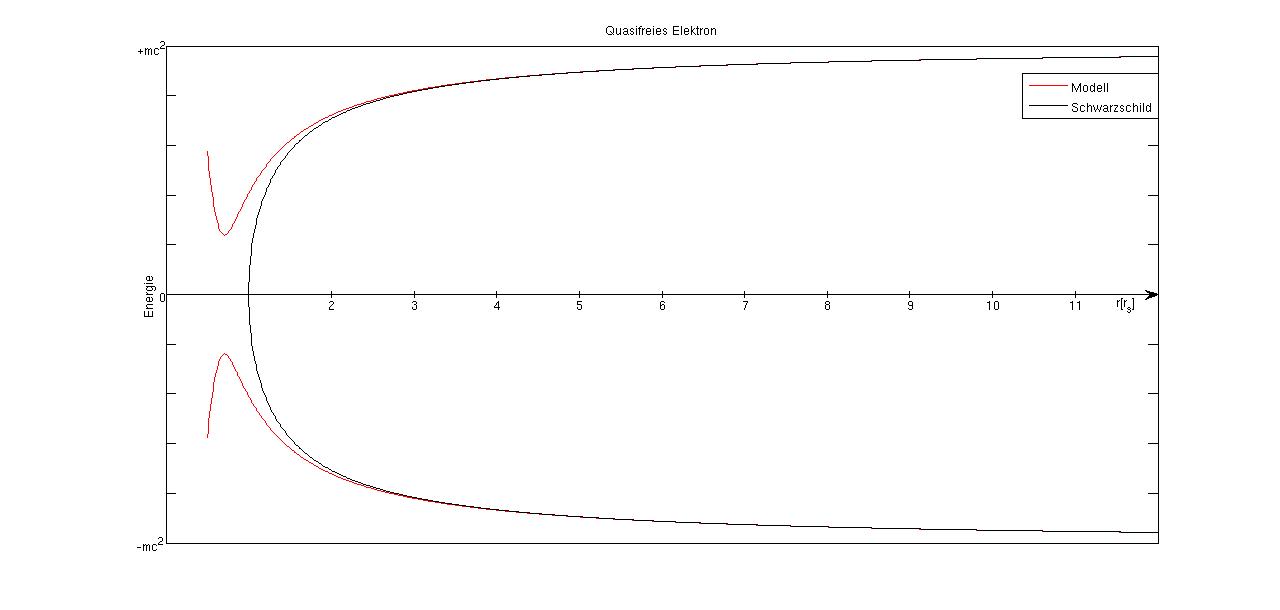}
\caption{Die Ruheenergie eines lokalisierten Elektrons im Gravitationsfeld ist in Abhängigkeit von der Position r in Einheiten des Schwarzschildradius r$_s$ dargestellt. }
\label{fig:ru} 
\end{figure}

\begin{figure}[hbt]
\centering
\includegraphics[width = \textwidth]{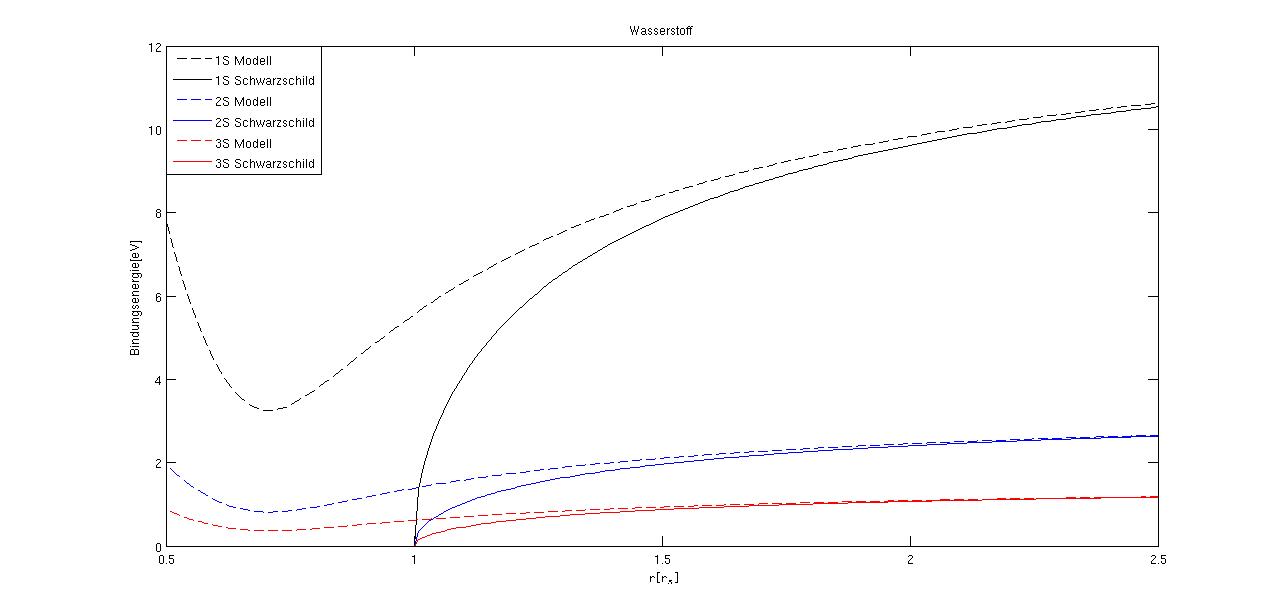}
\caption{Die Vorhersagen der ART und dem Modell werden für die ersten S-Energieniveaus von Wasserstoff im Gravitationsfeld verglichen, wobei die x-Achse den Abstand zum Zentrum in Einheiten des Schwarschildradius angibt.}
\label{fig:wa} 
\end{figure}

\begin{figure}[hbt]
\centering
\includegraphics[width = \textwidth]{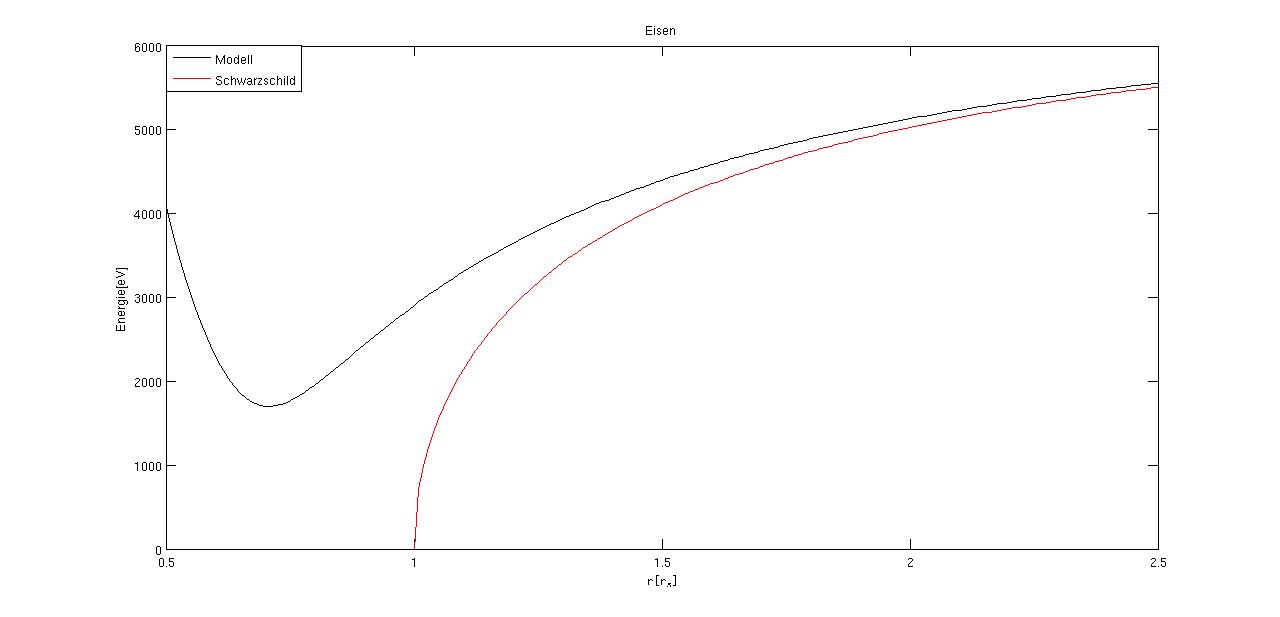}
\caption{Die Vorhersagen der ART und dem Modell werden für das 1S-Niveau von Eisen im Gravitationsfeld verglichen, wobei die x-Achse den Abstand zum Zentrum in Einheiten des Schwarschildradius angibt. }
\label{fig:fe} 
\end{figure}

Ein Elektron in einem Atom stellt im effektiven Potential des Restsystems ein solches Problem dar. Allerdings wurde schon 1956 die Verschiebung atomarer Energieniveaus von Papapetrou \cite{Papapetrou:1956} im Gravitationsfeld näherungsweise berechnet. Es stellt sich heraus, dass die Näherung für übliche sphärisch symmetrische Massenverteilungen so gut ist, dass ein numerisches Lösen der exakten Gleichungen \eqref{radDira} nicht nötig ist. Von daher werden in dieser Sektion seine Überlegungen ausgebaut und die Auswirkungen einer veränderten Schwarzschild-Metrik graphisch dargestellt.\vspace{12pt}\\
Mit einem minimal gekoppelten elektromagnetischen Feld ist die Dirac-Gleichung in einem gekrümmten Raum durch
\begin{align}
 i\hbar \gamma^\mu \left (\fracpd{}{x^\mu} + \Gamma_\mu + \frac{ie}{\hbar}A_\mu \right ) \Psi - mc \Psi = 0
\end{align}
gegeben.\vspace{12pt}\\
Erneut wird eine sphärisch symmetrische Metrik betrachtet und somit können die $\Gamma_\mu$ aus dem Vorherigen übernommen werden, sodass für die Dirac-Gleichung folgt
\begin{align}
 i\hbar e^{-\frac{\nu}{2}} \fracpd{\Psi}{t} =  c\vec{\tilde{\alpha}} \left [\vec{p} - e\vec{A} + \vec{\tilde{\Gamma}}  \right ]\Psi + e\Phi\Psi + \beta mc^2\Psi
\end{align}
Für die Bewegung eines gebundenen Elektrons in einem Atom ist offensichtlich, dass die Wellenfunktion außerhalb des atomaren Radius äußerst stark abfällt, sodass in sehr guter Näherung angenommen werden kann, dass die metrikabhängigen Funktionen konstant sind, solange das Atom insgesamt ruht. Da jedoch der rein gravitative Effekt auf die atomaren Energieniveaus betrachtet werden soll, kann dies problemlos angenommen werden. Außerdem weist Papapetrou unter Betrachtung der isotropen Form darauf hin, dass $\vec{\tilde{\Gamma}}$ vernachlässigbar ist, da es für hinreichend großen Abstand von der Massenansammlung proportional zum Inversen der Radialkomponente ist und somit mit $\frac{r_{Atom}}{\rho}$ im Vergleich zu den anderen Termen unterdrückt ist. Dies wird beispielsweise durch die Heisenbergsche Unschärferelation ersichtlich, weil die Ortsunsicherheit durch den Durchmesser des Atoms gegeben ist und und somit $\langle \Delta p\rangle \geq \frac{\hbar}{2r_{Atom}}$ gilt.\vspace{12pt}\\ 
Selbstverständlich dürfen diese Näherungen nicht für Bereiche genutzt werden, in denen ein Metrikkoeffizient verschwindend klein oder eine Steigung extrem hoch ist. Allerdings ermöglicht der Betrag von \eqref{Gammatilde} die Identifikation dieser Bereiche, die, wie später zu sehen sein wird, im Allgemeinen äußerst klein sind. \vspace{12pt}\\
Üblicherweise wird zudem die Kernbewegung vernachlässigt (beziehungsweise die des restlichen Atoms). Dies führt zu einer Korrektur proportional zum Quotienten der Elektronenmasse durch die Kernmasse, das heißt maximal grob $\frac{1}{2000}$. Dies ist einerseits eine weitere Vereinfachung der Gleichungen, da somit das Vektorpotential verschwindet und $\Phi$ ein reines Coulombpotential ist, und andererseits gibt es ungefähr eine Grenze ab welcher Größe die vorherigen Näherungen nicht mehr gerechtfertigt sind.\vspace{12pt}\\
Alle Näherungen zusammen überführen die Dirac-Gleichung nach
\begin{align}
 i\hbar e^{-\frac{\nu(r_{Kern})}{2}} \fracpd{\Psi}{t} =  c\vec{\tilde{\alpha}}\vec{p}\Psi + e\Phi\Psi + \beta mc^2\Psi
\end{align}
wobei r$_{Kern}$ den Ort des Atomkerns (beziehungsweise des Schwerpunkts) bezeichnet.\vspace{12pt}\\
Für die Eigenzustände kann der Ansatz $\Psi = \tilde{\Psi} e^{-i\frac{\tilde{E}}{\hbar}e^{\frac{\nu(r_{Kern})}{2}}t}$ gewählt werden, sodass die Dirac-Gleichung nach
\begin{align}
 \tilde{E}\tilde{\Psi} =  c\vec{\tilde{\alpha}}\vec{p}\tilde{\Psi} + e\Phi\tilde{\Psi} + \beta mc^2\tilde{\Psi}
\end{align}
übergeht\vspace{12pt}\\
Dies ist jedoch die gewöhnliche stationäre Dirac-Gleichung, sodass die Werte $\tilde{E}$ die Energien im ebenen Raum sind und somit folgt für die Gesamtenergie E
\begin{align}
 E = \tilde{E} e^{\frac{\nu(r_{Kern})}{2}}
\end{align}
Dies bedeutet, dass im größten Teil des Raums die Energieniveaus bis auf winzige Korrekturen nur gravitativ rotverschoben werden und somit praktisch der klassischen Erwartung entsprechen, da für einen entfernten Beobachter das ausgesandte Licht rotverschoben wird und folglich aus seinem Blickwinkel dies auch mit den atomaren Energieniveaus geschieht.\vspace{12pt}\\   
In \cite{GMRbuch} wurden die Auswirkungen für die Schwarschild-Metrik betrachtet und in \ref{testorig} wurde gezeigt, dass eine quadratische Korrektur nur klein sein kann, sodass im Inneren Teil die atomaren Energieniveaus nicht beobachtbar sein werden und das Verhalten im Außenbereich dem der Schwarzschild-Metrik stark ähnelt. Deshalb wird nun von einer Korrektur dritter Ordnung ohne Nullstellen ausgegangen
\begin{align}
 ds^2 = 1 - \frac{r_s}{r} + \frac{r_s^3}{6r^3} c^2dt^2 - \frac{1}{1 - \frac{r_s}{r} + \frac{r_s^3}{6r^3}} dr^2 -r^2 d\vartheta^2 - r^2\sin^2(\vartheta) d\varphi^2
\end{align}
Der Betrag von $\vec{\tilde{\Gamma}}$ ist dann
\begin{align}
 |\vec{\tilde{\Gamma}}| = \hbar \left | \frac{ r_s - \frac{r_s^3}{2r^2}}{{\sqrt{r^4 - r_sr^3 + \frac{r_s^3r}{6}}}} + \frac{1}{r^3} \left ( \sqrt{r^4 - r_sr^3 + \frac{r_s^3r}{6}} - r^2  \right ) \right |
\end{align}
Offensichtlich steigt er für kleine r stark an, sodass mit $r = \epsilon r_s$ angesetzt und nach der niedrigsten nichtverschwindenden Ordnung in $\epsilon$ entwickelt wird. Daraus erhält man
\begin{align}
 |\vec{\tilde{\Gamma}}| \approx \frac{\hbar}{r_s} \frac{\sqrt{6} - 2}{2\sqrt{6}\epsilon^{\frac{5}{2}}}
\end{align}
Aufgrund der vorherigen Überlegungen kann ungefähr $\frac{\hbar}{1000r_{Atom}}$ als Grenze genommen werden, ab der der Term relevant wird, das heißt
\begin{align}
&\frac{\hbar}{r_s} \frac{\sqrt{6} - 2}{2\sqrt{6} \epsilon^{\frac{5}{2}}} \gtrapprox \frac{\hbar}{1000r_{Atom}}\\
&\epsilon \lessapprox 6 \frac{r_{Atom}^{\frac{2}{5}}}{r_s^{\frac{2}{5}}}
\end{align}
Der atomare Radius ist ungefähr $10^{-10}$ Meter und der Schwarzschildradius für ein Objekt mit einer Sonnenmasse grob 3 Kilometer. Somit folgt
\begin{align}
 \epsilon \lessapprox 2,5 \cdot 10^{-5} \left ( \frac{M_S}{M}\right )^{\frac{2}{5}} 
\end{align}
Folglich ist klar, dass die Vernachlässigung für alle astronomischen Massen außer für äußerst kleine Radien im Vergleich zum Schwarzschildradius gerechtfertigt ist und daher taucht im Folgenden $\vec{\tilde{\Gamma}}$ nicht mehr auf.\vspace{12pt}\\
Bei den meisten Atomen ist die Bindungsenergie eines Elektrons viel kleiner als seine Ruhemasse, sodass für einen Graphen, der die Gesamtenergie darstellt alle Linien zusammenfallen. Somit spielt ihre Bindungsenergie für eine solche graphische Darstellung keine Rolle und sie sind quasifrei. Deshalb zeigt der Graph \ref{fig:ru} den Verlauf der Ruheenergie eines lokalisierten Elektrons in Abhängigkeit von seiner Position beziehungsweise der Position des zugehörigen Atoms. Dabei treten erst zwischen zwei und drei Schwarzschildradien sichtbare Unterschiede zwischen der Schwarzschildmetrik und dem Modell auf. Folglich wird der Bereich weiter außen in den weiteren Graphen nicht mehr beachtet und sie stellen nur die Bindung- statt der Gesamtenergie dar.\vspace{12pt}\\   
\begin{flushleft} Die Energieniveaus für Wasserstoff im Minkowskiraum können beispielsweise in \cite{Schwabl2005} nachgeschlagen werden \end{flushleft}
\begin{align}
 \tilde{E}_{n,j} = mc^2 \left [ 1 + \left ( \frac{\alpha}{n-(j+\frac{1}{2}) + \sqrt{(j+\frac{1}{2})^2 - \alpha^2}} \right )^2  \right ]^{-\frac{1}{2}} 
\end{align}
Damit ergibt sich der Graph \ref{fig:wa}. Er zeigt anschaulich die Vorhersage, dass die totale Abweichung zwischen den Voraussagen des Modells und der Schwarzschild-Metrik proportional zur Energie ist.
\begin{flushleft}\end{flushleft}       
Für komplizierte Atome ist eine analytische Herleitung nicht so einfach möglich, da es sich um ein Vielteilchensystem handelt. Jedoch werden die Energieniveaus näherungsweise ebenfalls nur rotverschoben, da die Dirac-Gleichung für jedes einzelne Elektron mit einer effektiven Ladungsverteilung nach wie vor anwendbar ist und somit der Einfluss des Gravitationsfelds gleich bleibt. Im Minkowskiraum sind die niedrigsten Energieniveaus für Eisen \cite{Bearden1967}\\
\begin{center}
 \begin{tabular}{|c|c|}
  \hline
  Niveau & Bindungsenergie [eV] \\
  \hline
  1s$_{\frac{1}{2}}$ & 7112,0 \\
  \hline
  2s$_{\frac{1}{2}}$ & 846,1 \\
  \hline
  2p$_{\frac{1}{2}}$ & 721,1 \\
  \hline
  2p$_{\frac{3}{2}}$ & 708,1 \\
  \hline
  3s$_\frac{1}{2}$ & 92,9 \\
  \hline
 \end{tabular}
\vspace{12pt}
\end{center}
Damit ergeben sich für die Energieniveaus im Gravitationsfeld in Abhängigkeit vom Abstand zum Zentrum und der Theorie auf die erste Nachkommastelle gerundet\\
\begin{center}
 \begin{tabular}{c|c|c|c|c|c|}
  \cline{2-6}
  & 1s$_{\frac{1}{2}}$ & 2s$_{\frac{1}{2}}$ & 2p$_{\frac{1}{2}}$ & 2p$_{\frac{3}{2}}$ & 3s$_\frac{1}{2}$ \\
  \hline
  \multicolumn{1}{|c|}{Energie ART bei $r=r_s$ [eV]} & 0 & 0 & 0 & 0 & 0\\
  \hline
  \multicolumn{1}{|c|}{Energie Modell bei $r=r_s$ [eV]} & 2903,5 & 345,4 & 294,4 & 289,1 & 37,9 \\
  \hline
  \multicolumn{1}{|c|}{Energie ART bei $r=1,5r_s$ [eV]} & 4106,1 & 488,5 & 416,3 & 408,8 & 53,6 \\
  \hline
  \multicolumn{1}{|c|}{Energie Modell bei $r=1,5r_s$ [eV]} & 4399,8 & 523,4 & 446,1 & 438,0 & 57,5 \\
  \hline
  \multicolumn{1}{|c|}{Energie ART bei $r=2r_s$ [eV]} & 5028,9 & 598,3 & 509,9 & 500,7 & 65,7 \\
  \hline
  \multicolumn{1}{|c|} {Energie Modell bei $r=2r_s$ [eV]} & 5132,6 & 610,6 & 520,4 & 511,0 & 67,0\\
  \hline
 \end{tabular}
\vspace{12pt}
\end{center}
Aufgrund der Proportionalität zur Energie zeigt \ref{fig:fe} nur das 1S-Niveau.\vspace{12pt}\\
Allgemein ist zu sehen, dass die Abweichungen zwar im Bereich vom Schwarzschildradius zunehmen aber noch relativ klein sind. Von daher sind sehr präzise Messungen und in der Nähe des Schwarzschildradius nötig, um eine eventuelle Abweichung zur ART zu verifizieren. 
 
\chapter{Zusammenfassung und Ausblick}
Da diese Arbeit eine Theorie der Gravitation behandelt, führt das erste Kapitel beginnend mit der historischen Entwicklung des menschlichen Verständnisses über sie in den aktuellen Stand der Wissenschaft ein. Die Grundüberlegungen zur Herleitung der Feldgleichungen der ART und die für diese Arbeit relevanten Lösungen werden vorgestellt. Das Kapitel wird mit ausgewählten Tests der Theorie abgeschlossen. Damit werden einerseits die Effekte der Theorie und ihre Größe verdeutlicht und andererseits aufgezeigt wie eine neue Theorie unter anderem getestet werden kann.\vspace{12pt}\\
Das zweite Kapitel beschäftigt sich mit dem Pseudokomplexen Zahlensystem, da die neue Theorie auf ihm basiert. Zu Anfang werden die grundlegenden Postulate behandelt und damit die algebraischen Eigenschaften des Zahlensystems bestimmt. Im nächsten Teil über die metrischen Eigenschaften wird gezeigt wie auf diesem System das Inverse und eine Norm definiert werden kann. Dadurch wird der Umgang mit dem Zahlensystem und die Messung von Abständen verständlich. Im Weiteren wird die Differentation und Integration behandelt. Dabei wird herausgestellt, wie eine holomorphe Funktion definiert wird und dass für sie ein Analogon zum Cauchyschen Integralsatz gilt.\vspace{12pt}\\
In der Pseudokomplexen Allgemeinen Relativitätstheorie wird die Algebra der Raumzeitkoordinaten auf die Pseudokomplexe erweitert. Der daraus resultierende Formalismus wird im dritten Kapitel vorgestellt, indem die Eigenschaften der Metrik und der kovarianten Ableitung betrachtet werden. Zudem setzt es sich im ersten Teil mit der Realprojektion und den neuen Feldgleichungen auseinander. Im Unterschied zu den Feldgleichungen der ART wird im leeren Raum eine Quelle proportional zu einem der Nullteiler angenommen, sodass deren Bestimmung ein zentrales Problem der Theorie ist. Der zweite Teil zeigt, dass diese Problematik noch nicht abschließend geklärt ist, jedoch derzeit zwei Ansätze existieren, deren Untersuchung vielversprechend ist. Im Folgenden wird die Schwarzschildlösung in beiden Ansätzen behandelt und gezeigt, dass für den Originalansatz schon eine bis zu einem gewissen Grad testbare Vorhersage existiert.\vspace{12pt}\\
Im vierten Kapitel werden die eigentlichen Resultate dieser Arbeit vorgestellt, soweit sie nicht unter die gemeinschaftliche Arbeit im dritten Kapitel fallen. Es beginnt mit der Herleitung der Reissner-Nordström-Lösung für die Psudokomplexe Allgemeine Relativitätstheorie. Dabei wird zwar aufgrund der mangelnden Kenntnisse über die Quelle keine abschließende Lösung gewonnen, jedoch kann die Lösung in Abhängigkeit von der Schwarzschildlösung und den Quellfunktionen allgemein dargestellt werden, sodass eine kurze Rechnung zu einer Vorhersage führt, sobald die Quellfunktionen bestimmt werden konnten. Somit stellt das unter den gegebenen Umständen die optimale Lösung dar. \vspace{12pt}\\
Im nächsten Teil wird die Vorhersage der Originallösung mit aktuellen experimentellen Daten verglichen und das daraus resultierende Ergebnis diskutiert. Es wird klar, dass die Interpretation bis auf das Fehlen der Singularität anlog zu der  der Schwarzschild-Metrik ist und folglich die Lösung zwar nicht auszuschließen ist, aber aus philospohischer Sicht keinen essentiellen Fortschritt zur Schwarzschild-Metrik darstellt. \vspace{12pt}\\ 
Im dritten und letzten Teil des vierten Kapitels werden die Dirac-Gleichung im gekrümmten Raum und die Auswirkungen auf von ihr beschriebene Objekte diskutiert. Dafür werden die radialen Gleichungen für sphärisch symmetrische Problemstellungen hergeleitet und mit vorhandener Literatur verglichen. Zudem werden atomare Energieniveaus von durch die Dirac-Gleichung beschriebenen Elektronen in Näherung betrachtet. Es handelt sich dabei um schon bekannte Ergebnisse, jedoch wird die Rechtfertigung der Näherung genauer diskutiert und für eine angenommene Metrik vorgerechnet. Zudem werden Energieniveaus von Wasserstoff und Eisen in dieser Metrik graphisch dargestellt. Die Änderung der atomaren Energieniveaus eignet sich äußerst gut für einen Test der neuen Theorie, da elektromagnetische Spektren recht genau vermessen werden können, und sobald die Quellfunktionen bekannt sind, gibt ein analoges Vorgehen eine testbare Vorhersage bezüglich der Rotverschiedung beobachteter Spektren. Zudem kann die angenommene Metrik als phänomenologisches Modell verstanden werden, sodass auch diese Graphen genutzt werden können.\vspace{12pt}\\
Wie schon im Vorherigen angesprochen existieren noch viele Bereiche in der Pseudokomplexen Allgemeinen Relativitätstheorie, in denen noch gearbeitet werden muss. Das größte Problem ist eindeutig die mangelnde Kenntnis über die Quellfunktionen, sodass weitere Arbeit in diese Richtung notwendig und wenn erfolgreich vielversprechend ist. Zudem gibt es noch viele weitere Lösungen der Einsteingleichung, die für die Feldgleichungen der Pseudokomplexen Allgemeinen Relativitätstheorie noch nicht existieren (beispielsweise die innere Schwarzschildlösung) und deren Herleitung und Interpretation von daher noch ausstehen. Weiterhin gibt es auch im Bereich der Semiklassischen Beschreibung noch viele offene Fragen, so ist beispielsweise noch nicht geklärt, wie sich die veränderten Metriken auf den Hawking-Effekt auswirken.\vspace{12pt}\\
Von daher ist die Pseudokomplexe Allgemeine Relativitätstheorie noch ein weitestgehend unerforschtes Gebiet, und es ist viel weitere Arbeit nötig bis sie wirklich verstanden und geprüft ist. Jedoch trägt jeder Schritt die Hoffnung unserem Jahrtausende altem Wunsch nach Verständnis (der Gestirne) näher zu kommen.       

\chapter*{Danksagung}
Zuerst möchte ich den drei Menschen danken ohne die die Masterarbeit in dieser Form nicht möglich gewesen wäre.\vspace{12pt}\\
Den beiden Betreuern danke ich dafür, dass sie uns jederzeit Ernst genommen und sich auf Diskussionen eingelassen haben, obwohl dies bedauerlicherweise keine Selbstverständlichkeit ist. Sie haben es dadurch geschafft, dass ich mich als Mitarbeiter und nicht als Belast gefühlt habe und mir das Erstellen dieser Arbeit, so anstrengend es auch war, Spaß gemacht hat. \vspace{12pt}\\
Insbesondere möchte ich mich bei Herrn Prof. Dr. Dr. h.c. mult Walter Greiner dafür bedanken, dass er mich auf das Thema aufmerksam und für es begeistert hat, mir bei jeglicher Bürokratie geholfen hat und mir mit guten Ideen, Ratschlägen und Arbeitsvorschlägen zur Seite gestanden hat.\vspace{12pt}\\
Zugegebenermaßen war ich anfangs skeptisch einen Betreuer zu haben, der den größten Teil der Zeit nicht vor Ort ist. Dies wurde jeodch schon in der Einarbeitungsphase von Herrn Prof. Dr. Peter Otto Hess zerstreut als er sich bei Problemen die Mühe machte Gleichungen im Lehrbuch selbst erneut nachzuvollziehen und für Thomas und mich aufzuschreiben. Dieser Einsatz ist bezeichnend für das Betreuungsverhältnis und dafür danke ich ihm. Zudem möchte ich mich für das Vertrauen bedanken, dass er uns entgegenbrachte, indem er uns schon recht früh seine aktuellen Rechnungen nachprüfen ließ.\vspace{12pt}\\
Meinem Kommilitonen und guten Freund Thomas Schönenbach danke ich für die tolle Zusammenarbeit im gesamten Studium. Insbesondere beim Einarbeiten für die Masterarbeit und beim Ausarbeiten der grundlegenden Rechnungen, deren Ergebnisse sich auch im dritten Kapitel wiederfinden. Außerdem danke ich ihm dafür, dass er so oft davon zu überzeugen war, dass eine bestimmte Veranstaltung  wahrscheinlich interessant oder nützlich ist, sodass wir von daher und aufgrund ähnlicher Interessen sehr häufig einander helfen und uns gegenseitig motivieren konnten.\vspace{12pt}\\
Ein besonderer Dank gebührt Herrn Dr. Joachim Reinhardt für seinen Hinweis, dass die Messung der PPN- Parameter die möglichen Korrekturen der Metriken einschränken.\vspace{12pt}\\
Weiterhin möchte ich meiner Familie für die Unterstützung, während meines Studiums danken. Insbesondere meiner Mutter, die sich währendessen viel Mühe gemacht und mir zudem häufiger durch eine orthografische Überprüfung von Arbeiten und Vorträgen geholfen hat.\vspace{12pt}\\
Ich danke Dominik für seine Bereitschaft mir beim Retten der Daten auf meiner Festplatte zu helfen, nachdem mein Rechner "repariert" wurde, auch wenn sie vorher zerstört wurde.\vspace{12pt}\\
Zudem möchte ich mich bei meinen Freunden Alex, Dominik, Mike, Robert und Thomas für die schöne Zeit im und neben dem Studium bedanken.

\appendix

\addchap{Anhang}
\section{Herleitung der sphärisch symmetrischen Diracgleichung im gekrümmten Raum}
Der Zweck dieses Kapitels ist es die Dirac-Gleichung für sphärisch symmetrische Probleme in einer sphärisch symmetrischen Metrik in den gewöhnlichen Schwarzschildkoordinaten aufzustellen. Da die Berechnung des Einflusses des Gravitationsfelds in isotropen Koordinaten einfacher ist, werden diese anfangs genutzt und die Dirac-Gleichung dann in das gewünschte Koordinatensystem transformiert. Dann wird die sphärische Symmetrie ausgenutzt, um die Gleichungen zu vereinfachen, sodass eine rein radiale Gleichung zu lösen ist. Abschließend werden stationäre Zustände und die daraus folgenden weiteren Vereinfachungen der Gleichungen diskutiert, sodass letztendlich die weitestmöglichst vereinfachte Dirac-Gleichung erhalten wird.\vspace{12pt}\\  
Die Dirac-Gleichung im gekrümmten Raum ist gegeben durch \cite{GMRbuch}
\begin{align}
 \left [ i\hbar \gamma^\mu \left ( \fracpd{}{x^\mu} + \Gamma_\mu \right ) - mc \right ] \Psi= 0
\end{align}
wobei $\Gamma_\mu$ über
\begin{align}
 \Gamma_\mu = \frac{1}{4} \gamma_\nu \left ( \fracpd{\gamma^\nu}{x^\mu} + \crs{\nu}{\lambda}{\mu} \gamma^\lambda \right ) \label{GroGA2}
\end{align}
berechnet werden kann und die Antikommutationsrelation der Dirac-Matrizen zu
\begin{align}
 \left \{ \gamma^\mu , \gamma^\nu \right \} = 2g^{\mu\nu}
\end{align}
abgeändert werden muss.\vspace{12pt}\\
Es wird ein sphärisch symmetrisches Problem betrachtet und aus rechentechnischen Gründen werden isotrope Koordinaten benutzt. Das Linienelement ist daher durch  
\begin{align}
 ds^2 = \omega(\rho) c^2dt^2 - \delta(\rho) \left ( (dx^1)^2 + (dx^2)^2 + (dx^3)^2 \right )
\end{align}
gegeben.\vspace{12pt}\\
Somit ist die Metrik diagonal und es ist möglich die Dirac-Matrizen als Produkt der jeweiligen alten Matrix und einer vom zugehörigen Metrikkoeffizient abhängigen Funktion auszudrücken
\begin{align}
\gamma_0 &= \sqrt{\omega} \tilde{\gamma}_0 & \gamma^0 &= \frac{1}{\sqrt{\omega}} \tilde{\gamma}^0\notag\\
\gamma_i &= \sqrt{\delta} \tilde{\gamma}_i & \gamma^i &= \frac{1}{\sqrt{\delta}} \tilde{\gamma}^i
\end{align}
wobei $\tilde{\gamma_\mu}$ wie in \cite{GMRbuch} die jeweilige Dirac-Matirx im Minkowskiraum bezeichnet. Dabei ist zu beachten, dass $\tilde{\gamma}_0 = \tilde{\gamma}^0$ und $\tilde{\gamma}_i = - \tilde{\gamma}^i$ gilt.\vspace{12pt}\\
Die nichtverschwindenden Christoffelsymbole für die isotrope Metrik können mittels der Geodätengleichung berechnet werden. Die zu variierende Lagrangefunktion ist
\begin{align}
 L = \omega(\rho) c^2 t'^2- \delta(\rho) \left ( (x'^1)^2 + (x'^2)^2 + (x'^3)^2 \right )
\end{align}
Somit folgt für t
\begin{align}
0 &= \frac{d}{ds} \fracpd{L}{ct'} = \frac{d}{ds} (2\omega c t')\\
  &=  2t'' c \omega + 2\frac{d\omega}{d\rho} \frac{d\rho}{ds}c t'
\end{align}
Die Ableitung $\frac{d\rho}{ds} = \frac{d\sqrt{(x^1)^2+(x^2)^2 + (x^3)^2}}{ds}$ kann umgeschrieben werden nach $\frac{d\rho}{ds} = \sum_{i=1}^3 \frac{d\rho}{dx^i} \frac{dx^i}{ds} = \sum_{i=1}^3 \frac{x^i}{\rho} x'^i$ und damit folgt
\begin{align}
0 &=  ct'' + \frac{\bar{\omega}}{\omega\rho} \sum_{i=1}^3 x^i x'^i ct'
\end{align}
und analog für die räumlichen Koordinaten
\begin{align}
 0 &=  \frac{d}{ds} \fracpd{L}{x'^i} - \fracpd{L}{x^i} = \frac{d}{ds} (-2\delta x'^i) - \fracpd{\omega}{x^i} c^2 t'^2 - \fracpd{\delta}{x^i} \sum_{j=1}^3 (x'^j)^2\\
&= - 2 \delta x''^i - 2 \frac{\bar{\delta}}{\rho} \sum_{j=1}^3 x^j x'^j x'^i - \frac{\bar{\omega}x^i}{\rho} c^2t'^2 + \frac{\bar{\delta}x^i}{\rho} \sum_{j=1}^3 (x'^j)^2\\
0 &= x''^i + \frac{\bar{\delta}}{\delta\rho} \sum_{j=1}^3 x^j x'^j x'^i + \frac{\bar{\omega}x^i}{2\delta\rho} c^2 t'^2 - \frac{\bar{\delta}x^i}{2\delta\rho} \sum_{j=1}^3 (x'^j)^2\\
&= x''^i + \frac{\bar{\delta}x^i}{2\delta\rho} (x'^i)^2  + \frac{\bar{\delta}}{\delta\rho} \sum_{\substack{j=1\\j\neq i}}^3 x^j x'^j x'^i + \frac{\bar{\omega}x^i}{2\delta\rho} c^2 t'^2 - \frac{\bar{\delta}x^i}{2\delta\rho} \sum_{\substack{j=1\\j\neq i}}^3 (x'^j)^2
\end{align}
wobei $x'^\mu$ die Ableitung der Koordinate nach s und $\bar{f}$ die Ableitung der Funktion nach ihrer jeweiligen Koordinate (hier $\rho$) bezeichnet.\vspace{12pt}\\
Daraus lassen sich die Christoffelsymbole einfach durch Vergleich mit \eqref{Geoglg} ablesen. 
\begin{align}
 \crs{0}{0}{i} &= \crs{0}{i}{0} = \frac{\bar{\omega}x^i}{2\omega \rho}\\
\crs{i}{0}{0} &= \frac{\bar{\omega}x^i}{2\delta \rho}\\
\crs{i}{i}{j} &= \crs{i}{j}{i} = \frac{\bar{\delta}x^j}{2\delta \rho}\\
\crs{i}{j}{j} &= -\frac{\bar{\delta}x^i}{2\delta \rho} ~\forall~ i \neq j
\end{align}
Dieses Ergebnis ist konsistent mit dem von \cite{Papapetrou:1956}\vspace{12pt}\\
Mit Hilfe von \eqref{GroGA2} werden nun die $\Gamma_\mu$ berechnet. Dabei wird mit $\Gamma_0$ angefangen
\begin{align}
  \Gamma_0 &= \frac{1}{4} \gamma_\nu \crs{\nu}{\lambda}{0} \gamma^\lambda\\
&= \frac{1}{4} \sum_{i=1}^3 \left [ \gamma_0 \crs{0}{i}{0} \gamma^i + \gamma_i \crs{i}{0}{0} \gamma^0 \right]\\
&= \frac{1}{4} \sum_{i=1}^3 \left [ \sqrt{\omega}\tilde{\gamma}_0 \frac{\bar{\omega}x^i}{2\omega \rho} \frac{1}{\sqrt{\delta}}\tilde{\gamma}^i + \sqrt{\delta}\tilde{\gamma}_i \frac{\bar{\omega}x^i}{2\delta\rho} \frac{1}{\sqrt{\omega}} \gamma^0 \right]\\
&= \frac{\bar{\omega}}{8\sqrt{\omega\delta}\rho} \sum_{i=1}^3 x^i \left ( \tilde{\gamma}_0 \tilde{\gamma}^i+ \tilde{\gamma}_i \tilde{\gamma}^0  \right ) =  \frac{\bar{\omega}}{4\sqrt{\omega\delta}\rho} \sum_{i=1}^3 x^i \tilde{\gamma}_0 \tilde{\gamma}^i\\
&= \frac{\bar{\omega}}{4\omega\rho} \gamma_0 \sum_{i=1}^3 x^i \gamma^i
\end{align}
und mit den räumlichen $\Gamma_i$ fortgefahren
\begin{align}
 \Gamma_i &= \frac{1}{4} \gamma_\nu \left ( \fracpd{\gamma^\nu}{x^i} + \crs{\nu}{\lambda}{i} \gamma^\lambda \right )\\
\gamma_\nu \fracpd{\gamma^\nu}{x^i} &= \sqrt{\omega} \tilde{\gamma}_0 \fracpd{\frac{1}{\sqrt{\omega}}}{x^i} \tilde{\gamma}_0 + \sum_i \sqrt{\delta} \tilde{\gamma}_i \fracpd{\frac{1}{\sqrt{\delta}}}{x^i} \tilde{\gamma}^i\\
&= -\frac{x^i}{2\rho} \left ( \frac{\bar{\omega}}{\omega} + \frac{3\bar{\delta}}{\delta} \right )\\
\gamma_\nu \crs{\nu}{\lambda}{i} \gamma^\lambda &= \gamma_0 \crs{0}{0}{i} \gamma^0 + \gamma_i \sum_{\substack{j=1\\j\neq i}}^3 \crs{i}{j}{i} \gamma^j + \sum_{\substack{j=1\\j\neq i}}^3 \gamma_j \crs{j}{i}{i} \gamma^i + \sum_{j=1}^3 \gamma_j \crs{j}{j}{i} \gamma^j\\
&= \crs{0}{0}{i} + \sum_{\substack{j=1\\j\neq i}}^3 \gamma_i \crs{i}{j}{i} \gamma^j + \sum_{\substack{j=1\\j\neq i}}^3 \gamma_j \crs{j}{i}{i} \gamma^i + \sum_{j=1}^3 \crs{j}{j}{i}\\
&= \frac{\bar{\omega}x^i}{2\omega \rho} + \frac{\bar{\delta}}{2\delta \rho} \sum_{\substack{j=1\\j\neq i}}^3 x^j \left ( \gamma_i \gamma^j - \gamma_j \gamma^i \right ) + \frac{3\bar{\delta}x^i}{2\delta \rho}\\
&= \frac{\bar{\omega}x^i}{2\omega \rho} + \frac{3\bar{\delta}x^i}{2\delta \rho} + \frac{\bar{\delta}}{\delta \rho} \sum_{\substack{j=1\\j\neq i}}^3 x^j \gamma_i \gamma^j\\
\Rightarrow \Gamma_i &= \frac{\bar{\delta}}{4\delta \rho}\gamma_i \sum_{\substack{j=1\\j\neq i}}^3 x^j \gamma^j
\end{align}
wobei die Einschränkung der Summen die Mehrfachzählung des Terms $\gamma_i \crs{i}{i}{i} \gamma^i $ vermeidet und die allgemeine Eigenschaft $\gamma_\mu \gamma^\mu = \gamma^\mu \gamma_\mu = \mathds{1}$ (ohne Summenkonvention) ausgenutzt wurde.\vspace{12pt}\\
Daraus folgt
\begin{align}
  \gamma^\mu \Gamma_\mu &= \frac{\bar{\omega}}{4\omega \rho} \sum_{i=1}^3 x^i \gamma^i +  \frac{\bar{\delta}}{2\delta \rho} \sum_{i=1}^3 x^i \gamma^i\\
&= \frac{1}{4\rho} \left ( \frac{\bar{\omega}}{\omega} + \frac{2\bar{\delta}}{\delta} \right ) \sum_{i=1}^3 x^i \gamma^i
\end{align}
Dabei wurde ausgenutzt, dass das Innere der Summe in $\Gamma_i$ von i unabhängig ist und somit $\sum_{i=1}^3 \sum_{\substack{j=1\\j\neq i}}^3 a_j = 2 \sum_{j=1}^3 a_j $ angewendet werden kann.\vspace{12pt}\\ 
Dementsprechend ist die Dirac-Gleichung gegeben durch
\begin{align}
0 &= \left [ i\hbar\gamma^\mu \fracpd{}{x^\mu}  + \frac{i\hbar}{4\rho} \left (  \frac{\bar{\omega}}{\omega} + \frac{2\bar{\delta}}{\delta}\right )\sum_{j=1}^3 x^j \gamma ^j - mc \right ] \Psi\\
\Leftrightarrow i\hbar \gamma^0 \fracpd{\Psi}{x^0} &= -i\hbar \sum_{j=1}^3 \gamma^j \fracpd{}{x^j} \Psi  -i\hbar \frac{1}{4\rho} \left ( \frac{\bar{\omega}}{\omega} + \frac{2\bar{\delta}}{\delta} \right ) \sum_{j=1}^3 x^j \gamma^j \Psi+  mc \Psi\\
\Leftrightarrow \frac{i\hbar}{\sqrt{\omega}} \fracpd{\Psi}{t} &= -i\hbar c \sum_{j=1}^3 \alpha_j \frac{1}{\sqrt{\delta}} \fracpd{}{x^j} \Psi - i\hbar c \frac{1}{4\sqrt{\delta}\rho} \left ( \frac{\bar{\omega}}{\omega} + \frac{2\bar{\delta}}{\delta} \right ) \sum_{j=1}^3 x^j \alpha_j \Psi + \beta mc^2\Psi
\end{align}
wobei die $\alpha_j = \beta\tilde{\gamma}^j$ die bekannten Matrizen aus dem Minkowskiraum sind.\vspace{12pt}\\
Der Impulsoperator ist in Ortsdarstellung über $\vec{p} = \frac{\hbar}{i} \nabla$ definiert und somit gilt für die isotrope sphärisch symmetrische Metrik $p_j = \frac{\hbar}{i\sqrt{\delta}} \fracpd{}{x^j}$. Eingesetzt folgt daraus
\begin{align}
\frac{i\hbar}{\sqrt{\omega}} \fracpd{\Psi}{t} &= c\vec{\alpha}\vec{p}\Psi -i\hbar c \frac{1}{4\sqrt{\delta}\rho} \left ( \frac{\bar{\omega}}{\omega} + \frac{2\bar{\delta}}{\delta}\right ) \vec{\alpha}\vec{x}  \Psi + \beta mc^2\Psi\label{Papape}\\
\Leftrightarrow \frac{i\hbar}{\sqrt{\omega}} \fracpd{\Psi}{t} &= c\vec{\alpha} \left [\vec{p} -i\hbar \frac{1}{4\sqrt{\delta}} \left ( \frac{\bar{\omega}}{\omega} + \frac{2\bar{\delta}}{\delta} \right ) \vec{e}_\rho  \right ]\Psi + \beta mc^2\Psi
\end{align}
Die Gleichung \eqref{Papape} ist analog zum Resultat von \cite{Papapetrou:1956}.\vspace{12pt}\\
Um die Dirac-Gleichung in die gewöhnlichen Schwarzschildkoordinaten umzuschreiben, müssen die Linienelemente verglichen werden
\begin{align}
 ds^2 &= \omega(\rho) c^2dt^2 - \delta(\rho) \left ( (dx^1)^2 + (dx^2)^2 + (dx^3)^2 \right )\\
&= \omega(\rho) c^2dt^2  - \delta(\rho) d\rho^2 - \delta(\rho) \rho^2 d\Omega^2\\
&= e^\nu c^2 dt^2 - e^\lambda dr^2 - r^2 d\Omega^2
\end{align}
Daraus folgen die Gleichungen
\begin{align}
  \omega &= e^\nu\\
  r^2 &= \delta \rho^2 ~\Rightarrow~ r = \sqrt{\delta} \rho\\
  \delta d\rho^2 &= e^\lambda dr^2 ~\Rightarrow~ \frac{dr}{d\rho} = \sqrt{\delta} e^{-\frac{\lambda}{2}} \label{Isoschw}
\end{align}
Dabei kann bei den letzten beiden Gleichungen das Vorzeichen frei gewählt werden. Jedoch ist jeweils nur die positive Variante sinnvoll, da beide Koordinaten radiale Koordinaten darstellen und somit stets positiv und nach außen ansteigend sind. Daraus folgt auch, dass die Einheitsvektoren von $\rho$ und r gleich sind.\vspace{12pt} \\
Damit können die in der Dirac-Gleichung auftretenden Terme in den gewöhnlichen Koordinaten bestimmt werden
\begin{align}
 \bar{\delta} &= \frac{d\delta}{d\rho} = \frac{d}{d\rho} \frac{r^2}{\rho^2} = -\frac{2r^2}{\rho^3} + \frac{2r}{\rho^2} \frac{dr}{d\rho}\\
&= -\frac{2r^2}{\left (\frac{r}{\sqrt{\delta}} \right )^3} + \frac{2r}{\left (\frac{r}{\sqrt{\delta}} \right )^2} \sqrt{\delta} e^{-\frac{\lambda}{2}}\\
&= -\frac{2}{r} \delta^{\frac{3}{2}} + \frac{2}{r} e^{-\frac{\lambda}{2}} \delta^{\frac{3}{2}}\\
\Rightarrow \frac{\bar{\delta}}{\delta^{\frac{3}{2}}} &= \frac{2}{r} \left ( e^{-\frac{\lambda}{2}} - 1 \right )\\
\bar{\omega} &= \frac{d\omega}{d\rho} = \frac{de^\nu}{dr} \frac{dr}{d\rho} = \bar{\nu} e^\nu \sqrt{\delta} e^{-\frac{\lambda}{2}}\\
\Rightarrow \frac{\bar{\omega}}{\omega \sqrt{\delta}} &= \bar{\nu} e^{-\frac{\lambda}{2}}
\end{align}
Diese beiden Erkenntnisse in die Dirac-Gleichung eingesetzt führen zu
\begin{align}
 i\hbar e^{-\frac{\nu}{2}} \fracpd{\Psi}{t} &= c\vec{\alpha} \left [\vec{p} -i\hbar \left ( \frac{\bar{\nu}}{4}e^{-\frac{\lambda}{2}} + \frac{1}{r} \left ( e^{-\frac{\lambda}{2}} - 1 \right )  \right ) \vec{e}_r  \right ]\Psi + \beta mc^2\Psi
\end{align}
Der Übersichtlichkeit halber wird
\begin{align}
 \vec{\tilde{\Gamma}} := -i\hbar \left ( \frac{\bar{\nu}}{4}e^{-\frac{\lambda}{2}} + \frac{1}{r} \left ( e^{-\frac{\lambda}{2}} - 1 \right )  \right ) \vec{e}_r
\end{align}
definiert.\vspace{12pt}\\
Analog zu \cite{GMRbuch} wird der Ansatz
\begin{align}
  \Psi =  e^{-\frac{\lambda}{4}} \frac{1}{r} \begin{pmatrix} \Phi_1(r,t) \chi_\kappa^\mu(\vartheta,\varphi)\\ i\Phi_2(r,t)  \chi_{-\kappa}^\mu(\vartheta,\varphi) \end{pmatrix}
\end{align}
benutzt. Dabei sind die Winkelanteile wie im Minkowskiraum und der Spinor erfüllt für beliebige Funktionen f(r,t) und g(r,t) die Eigenwertgleichung \cite{Greiner1987}
\begin{align}
 \hat{K} \begin{pmatrix} f(r,t) \chi_\kappa^\mu(\vartheta,\varphi)\\ g(r,t)  \chi_{-\kappa}^\mu(\vartheta,\varphi) \end{pmatrix} &= \beta (\vec{\Sigma}\vec{L} + \hbar) \begin{pmatrix} f(r,t) \chi_\kappa^\mu(\vartheta,\varphi)\\ g(r,t)  \chi_{-\kappa}^\mu(\vartheta,\varphi) \end{pmatrix}\\ 
&= -\hbar \kappa \begin{pmatrix} f(r,t) \chi_\kappa^\mu(\vartheta,\varphi)\\ ig(r,t)  \chi_{-\kappa}^\mu(\vartheta,\varphi) \end{pmatrix} \label{Keigen}
\end{align}
wobei
\begin{align}
 \vec{\Sigma} = \begin{pmatrix} \vec{\sigma} & 0 \\ 0 & \vec{\sigma} \end{pmatrix}
\end{align}
gilt und $\kappa$ durch die Drehimpulsquantenzahl und die Bahndrehimpulsquantenzahl wie folgt festgelegt wird
\begin{align}
 \kappa = \begin{cases} - \left (j+\frac{1}{2} \right ) ~\mathrm{ falls }~ j= l + \frac{1}{2}\\
           j + \frac{1}{2} ~\mathrm{ falls }~ j = l - \frac{1}{2}
          \end{cases}
\end{align}
In die Dirac-Gleichung eingesetzt ergibt sich damit
\begin{align}
 i\hbar \frac{e^{-\frac{2\nu + \lambda}{4}}}{r} \fracpd{}{t} \begin{pmatrix} \Phi_1 \chi_\kappa^\mu\\ i\Phi_2  \chi_{-\kappa}^\mu \end{pmatrix}  &= \left [c\vec{\alpha}(\vec{p} + \vec{\tilde{\Gamma}}) + \beta m c^2 \right ] \frac{e^{-\frac{\lambda}{4}}}{r} \begin{pmatrix} \Phi_1 \chi_\kappa^\mu\\ i\Phi_2 \chi_{-\kappa}^\mu \end{pmatrix}\\
\Leftrightarrow  i\hbar \fracpd{}{t} \begin{pmatrix} \Phi_1 \chi_\kappa^\mu\\ i\Phi_2 \chi_{-\kappa}^\mu \end{pmatrix} &= \left [ ce^{\frac{2\nu + \lambda}{4}}r \begin{pmatrix} 0 & \vec{\sigma} \\ \vec{\sigma} & 0 \end{pmatrix} (\vec{p} + \vec{\tilde{\Gamma}}) \frac{e^{-\frac{\lambda}{4}}}{r} + e^{\frac{\nu}{2}}\beta mc^2 \right ] \begin{pmatrix} \Phi_1 \chi_\kappa^\mu\\ i\Phi_2 \chi_{-\kappa}^\mu \end{pmatrix}
\end{align}
In die Zweierspinoren umgeschrieben folgen die Gleichungen für die obere Komponente 
\begin{align}
 i\hbar \fracpd{\Phi_1}{t} \chi_\kappa^\mu &= icre^{\frac{2\nu + \lambda}{4}} \vec{\sigma} (\vec{p} + \vec{\tilde{\Gamma}}) \frac{e^{-\frac{\lambda}{4}}}{r} \Phi_2 \chi_{-\kappa}^\mu + mc^2 e^{\frac{\nu}{2}} \Phi_1 \chi_\kappa^\mu
\end{align}
und für die untere Komponente
\begin{align}
 -\hbar \fracpd{\Phi_2}{t} \chi_{-\kappa}^\mu &= cre^{\frac{2\nu + \lambda}{4}}(\vec{p} + \vec{\tilde{\Gamma}}) \frac{e^{-\frac{\lambda}{4}}}{r} \Phi_1 \chi_{\kappa}^\mu - mc^2 e^{\frac{\nu}{2}} i\Phi_2 \chi_{-\kappa}^\mu\\
\Leftrightarrow i\hbar \fracpd{\Phi_2}{t} \chi_{-\kappa}^\mu &= -icre^{\frac{2\nu + \lambda}{4}}(\vec{p} + \vec{\tilde{\Gamma}}) \frac{e^{-\frac{\lambda}{4}}}{r} \Phi_1 \chi_{\kappa}^\mu - mc^2 e^{\frac{\nu}{2}}\Phi_2 \chi_{-\kappa}^\mu
\end{align}
sodass sich das Gleichungssystem
\begin{align}
 &(1)~ i\hbar\fracpd{\Phi_1}{t} \chi_\kappa^\mu = ic r e^{\frac{\nu}{2}}e^{\frac{\lambda}{4}} \vec{\sigma} \left ( \vec{p}+\vec{\tilde{\Gamma}}\right ) e^{-\frac{\lambda}{4}} \frac{1}{r} \Phi_2 \chi_{-\kappa}^\mu + mc^2 e^{\frac{\nu}{2}} \Phi_1 \chi_\kappa^\mu\notag\\
&(2)~ i\hbar\fracpd{\Phi_2}{t} \chi_{-\kappa}^\mu =  - ic e^{\frac{\nu}{2}} r e^{\frac{\lambda}{4}} \vec{\sigma} \left ( \vec{p}+\vec{\tilde{\Gamma}}\right ) e^{-\frac{\lambda}{4}} \frac{1}{r} \Phi_1 \chi_\kappa^\mu  - mc^2 e^{\frac{\nu}{2}} \Phi_2 \chi_{-\kappa}^\mu \label{Phiglgsys}
\end{align}
ergibt.\vspace{12pt}\\
Für beliebige Vektoren $\vec{a}$ und $\vec{b}$ gilt die Relation
\begin{align}
 (\vec{\sigma}\vec{a}) (\vec{\sigma}\vec{b}) = \vec{a}\vec{b} + i \vec{\sigma} \left ( \vec{a} \times \vec{b}\right )
\end{align}
und von daher ist
\begin{align}
 (\vec{\sigma}\vec{e}_r) (\vec{\sigma}\vec{e}_r) = 1
\end{align}
Diese Relation wird zur Evaluation von $\vec{\sigma} \left ( \vec{p}+\vec{\tilde{\Gamma}}\right )$ verwendet:
\begin{align}
 \vec{\sigma} \left ( \vec{p}+\vec{\tilde{\Gamma}}\right ) &= (\vec{\sigma}\vec{e}_r)(\vec{\sigma}\vec{e}_r)\left [ \vec{\sigma} \left ( \vec{p}+\vec{\tilde{\Gamma}}\right ) \right ]\\
&= (\vec{\sigma}\vec{e}_r) \left [\vec{e}_r \left ( \vec{p}+\vec{\tilde{\Gamma}}\right ) + i \vec{\sigma} \left ( \vec{e}_r  \times \left ( \vec{p}+\vec{\tilde{\Gamma}}\right ) \right )  \right ]\\
&= (\vec{\sigma}\vec{e}_r) \left [p_r + \tilde{\Gamma}_r + \frac{i}{r} \vec{\sigma} \left ( \vec{r}  \times \vec{p} \right )  \right ]\\
\end{align}
Dabei wurden $p_r := \vec{e}_r \cdot \vec{p}$ und $\tilde{\Gamma}_r := \vec{e}_r \cdot \vec{\tilde{\Gamma}}$ definiert.\vspace{12pt}\\
Durch explizites Einsetzen der beiden Ausdrücke und der Definition des Bahndrehimpulses $\vec{L} = \vec{r}\times\vec{p}$ ergibt sich
\begin{align}
&= (\vec{\sigma}\vec{e}_r) \left ( -i\hbar e^{-\frac{\lambda}{2}} \fracpd{}{r} -i\hbar \left ( \frac{\bar{\nu}}{4}e^{-\frac{\lambda}{2}} + \frac{1}{r} \left ( e^{-\frac{\lambda}{2}} - 1  \right ) \right ) + \frac{i}{r} \vec{\sigma} \vec{L} \right )
\end{align}
Zudem gilt (siehe beispielsweise \cite{Schwabl2005})
\begin{align}
 (\vec{\sigma}\vec{e}_r)\chi^\mu_{\pm \kappa} = \chi^\mu_{\mp \kappa}
\end{align}
Über \eqref{Keigen} kann die Wirkung von $\vec{\sigma}\vec{L}$ auf die Zweierspinoren bestimmt werden
\begin{align}
 (\vec{\sigma}\vec{L} + \hbar) \chi_\kappa^\mu &= -\hbar \kappa \chi_\kappa^\mu\\
\Leftrightarrow \vec{\sigma}\vec{L} \chi_\kappa^\mu &= -\hbar (\kappa + 1) \chi_\kappa^\mu\\
-(\vec{\sigma}\vec{L} + \hbar) \chi_{-\kappa}^\mu &= -\hbar \kappa \chi_{-\kappa}^\mu\\
\Rightarrow \vec{\sigma}\vec{L} \chi_{-\kappa}^\mu &= \hbar (\kappa - 1) \chi_{-\kappa}^\mu
\end{align}
Nachdem der Winkelanteil herausgekürzt ist kann \eqref{Phiglgsys} dementsprechend zu
\begin{align}
 &(1)~ i\hbar\fracpd{\Phi_1}{t} = i\hbar c r e^{\frac{\nu}{2}}e^{\frac{\lambda}{4}} \left [ -i e^{-\frac{\lambda}{2}} \fracpd{}{r} - i\left ( \frac{\bar{\nu}}{4} e^{-\frac{\lambda}{2}} + \frac{1}{r} \left (e^{-\frac{\lambda}{2}} - 1 \right )   \right ) + i\frac{\kappa-1}{r} \right ] e^{-\frac{\lambda}{4}} \frac{1}{r} \Phi_2 + mc^2 e^{\frac{\nu}{2}} \Phi_1 \notag\\
&(2)~ i\hbar\fracpd{\Phi_2}{t} =  - i\hbar cr e^{\frac{\nu}{2}}  e^{\frac{\lambda}{4}} \left [-i e^{-\frac{\lambda}{2}} \fracpd{}{r} - i\left ( \frac{\bar{\nu}}{4} e^{-\frac{\lambda}{2}} + \frac{1}{r} \left (e^{-\frac{\lambda}{2}} - 1 \right )   \right ) - i\frac{\kappa+1}{r}  \right ] e^{-\frac{\lambda}{4}} \frac{1}{r} \Phi_1   - mc^2 e^{\frac{\nu}{2}} \Phi_2 
\end{align}
umgeformt werden.\vspace{12pt}\\
Die Terme $\pm \frac{i}{r}$ heben sich in der Klammer weg und so folgt
\begin{align}
 &(1)~ i\hbar\fracpd{\Phi_1}{t} = i\hbar c r e^{\frac{\nu}{2}}e^{\frac{\lambda}{4}} \left [ -i e^{-\frac{\lambda}{2}} \fracpd{}{r} - i\left ( \frac{\bar{\nu}}{4} e^{-\frac{\lambda}{2}} + \frac{e^{-\frac{\lambda}{2}}}{r}   \right ) + i\frac{\kappa}{r} \right ] e^{-\frac{\lambda}{4}} \frac{1}{r} \Phi_2 + mc^2 e^{\frac{\nu}{2}} \Phi_1 \notag\\
&(2)~ i\hbar\fracpd{\Phi_2}{t} =  - i\hbar cr e^{\frac{\nu}{2}}  e^{\frac{\lambda}{4}} \left [-i e^{-\frac{\lambda}{2}} \fracpd{}{r} - i\left ( \frac{\bar{\nu}}{4} e^{-\frac{\lambda}{2}} + \frac{e^{-\frac{\lambda}{2}}}{r}   \right ) - i\frac{\kappa}{r}  \right ] e^{-\frac{\lambda}{4}} \frac{1}{r} \Phi_1   - mc^2 e^{\frac{\nu}{2}} \Phi_2 
\end{align}
Durch die Ableitung in der Klammer führt das Verschieben von r abhängigen Funktionen von der rechten Seite zur linken Seite der Klammer zu zusätzlichen Termen. Verschiebt man den Faktor $\frac{1}{r}$ auf die linke Seite ergibt sich
\begin{align}
&(1)~ i\hbar\fracpd{\Phi_1}{t} = i\hbar c e^{\frac{\nu}{2}}e^{\frac{\lambda}{4}} \left [ -i e^{-\frac{\lambda}{2}} \fracpd{}{r} - i \frac{\bar{\nu}}{4} e^{-\frac{\lambda}{2}} + i\frac{\kappa}{r} \right ] e^{-\frac{\lambda}{4}} \Phi_2 + mc^2 e^{\frac{\nu}{2}} \Phi_1 \notag\\
&(2)~ i\hbar\fracpd{\Phi_2}{t} =  - i\hbar c e^{\frac{\nu}{2}}  e^{\frac{\lambda}{4}} \left [-i e^{-\frac{\lambda}{2}} \fracpd{}{r} - i\frac{\bar{\nu}}{4} e^{-\frac{\lambda}{2}} - i\frac{\kappa}{r}  \right ] e^{-\frac{\lambda}{4}} \Phi_1   - mc^2 e^{\frac{\nu}{2}} \Phi_2 
\end{align}
Im nächsten Schritt wird der Faktor $e^{\frac{\nu}{4}}$ von links nach rechts gebracht und dadurch folgt
\begin{align}
 &(1)~ i\hbar\fracpd{\Phi_1}{t} = i\hbar c  e^{\frac{\nu+\lambda}{4}} \left [ -i e^{-\frac{\lambda}{2}} \fracpd{}{r}  + i\frac{\kappa}{r} \right ] e^{\frac{\nu-\lambda}{4}} \Phi_2 + mc^2 e^{\frac{\nu}{2}} \Phi_1 \notag\\
&(2)~ i\hbar\fracpd{\Phi_2}{t} =   i\hbar c e^{\frac{\nu+\lambda}{4}}  \left [i e^{-\frac{\lambda}{2}} \fracpd{}{r}  + i\frac{\kappa}{r}  \right ] e^{\frac{\nu-\lambda}{4}} \Phi_1   - mc^2 e^{\frac{\nu}{2}} \Phi_2 
\end{align}
Somit gilt für den Vektor $\Phi = \begin{pmatrix} \Phi_1 \\ \Phi_2 \end{pmatrix}$
\begin{align}
 i\hbar \fracpd{\Phi}{t} &=  i\hbar c\begin{pmatrix} \left (-i e^{\frac{\nu-\lambda}{4}} \fracpd{}{r} e^{\frac{\nu-\lambda}{4}}  + i e^{\frac{\nu}{2}}\frac{\kappa}{r}\right ) \Phi_2 \\ \left (i e^{\frac{\nu-\lambda}{4}} \fracpd{}{r} e^{\frac{\nu-\lambda}{4}}  + i e^{\frac{\nu}{2}} \frac{\kappa}{r}\right ) \Phi_1   \end{pmatrix} + mc^2 e^{\frac{\nu}{2}} \begin{pmatrix} \Phi_1 \\ -\Phi_2 \end{pmatrix}
\end{align}
Dies kann umgeschrieben werden zu
\begin{align}
  i\hbar \fracpd{\Phi}{t} &=  i\hbar c\begin{pmatrix} 0 & -i \\ i & 0  \end{pmatrix} e^{\frac{\nu-\lambda}{4}} \fracpd{}{r} e^{\frac{\nu-\lambda}{4}} \Phi + i\hbar c \begin{pmatrix} 0 & i \\ i & 0  \end{pmatrix}e^{\frac{\nu}{2}} \frac{\kappa}{r} \Phi + mc^2 e^{\frac{\nu}{2}} \begin{pmatrix} 1 & 0 \\ 0 & -1 \end{pmatrix} \Phi
\end{align}
Analog zu \cite{GMRbuch} werden die Matrizen
\begin{align}
 \alpha_r = \begin{pmatrix} 0 & -i \\ i & 0\end{pmatrix} && \beta_r = \begin{pmatrix} 1 & 0 \\ 0 & -1 \end{pmatrix}
\end{align}
definiert und eingesetzt ergibt sich
\begin{align}
 i\hbar \fracpd{\Phi}{t} &= \left [ i\hbar c\alpha_r  e^{\frac{\nu-\lambda}{4}} \fracpd{}{r} e^{\frac{\nu-\lambda}{4}} - i\hbar c\beta_r \alpha_r  e^{\frac{\nu}{2}} \frac{\kappa}{r} + \beta_r e^{\frac{\nu}{2}}mc^2 \right ] \Phi
\end{align}
Somit tritt im Impuls- und im Drehimpulsterm ein Vorzeichenunterschied zu \cite{GMRbuch} auf, der dementsprechend auch die abschließenden Gleichungen verändert.\vspace{12pt}\\
Sobald ein statisches Gravitationsfeld vorliegt, existieren statische Lösungen, für die der Ansatz 
\begin{align}
 \Phi(r,t) = e^{\frac{\lambda - \nu}{4}} \begin{pmatrix} f(r) \\ g(r) \end{pmatrix} e^{-i\frac{E}{\hbar}t}
\end{align}
gewählt wird.\\
Eingesetzt bedeutet das 
\begin{align}
 E e^{\frac{\lambda-\nu}{4}} \begin{pmatrix} f\\ g \end{pmatrix} &= \left [ i\hbar c \alpha_r e^{\frac{\nu-\lambda}{4}} \fracpd{}{r} - i\hbar c \beta_r \alpha_r  e^{\frac{\lambda+\nu}{4}} \frac{\kappa}{r}  + \beta_r e^{\frac{\lambda+\nu}{4}} mc^2 \right ] \begin{pmatrix} f\\ g \end{pmatrix}\\
\end{align}
und nachdem die Gleichung mit $e^{\frac{\lambda-\nu}{4}}$ multipliziert wurde, ergibt sich 
\begin{align}
\frac{E}{\hbar c} e^{\frac{\lambda-\nu}{2}} \begin{pmatrix} f\\ g \end{pmatrix} &= \left [ i\alpha_r \fracpd{}{r} - i \beta_r \alpha_r  e^{\frac{\lambda}{2}} \frac{\kappa}{r} + \beta_r e^{\frac{\lambda}{2}} \frac{mc}{\hbar} \right ] \begin{pmatrix} f\\ g \end{pmatrix} 
\end{align}
Somit folgt für die erste Komponente
\begin{align}
 \frac{E}{\hbar c} e^{\frac{\lambda-\nu}{2}} f &= \frac{dg}{dr} - e^{\frac{\lambda}{2}} \frac{\kappa}{r} g + e^{\frac{\lambda}{2}} \frac{mc}{\hbar} f\\
\Leftrightarrow \frac{dg}{dr} &= e^{\frac{\lambda}{2}} \frac{-mc^2 + Ee^{-\frac{\nu}{2}}}{\hbar c} f +  e^{\frac{\lambda}{2}} \frac{\kappa}{r} g 
\end{align}
und für die zweite Komponente
\begin{align}
 \frac{E}{\hbar c} e^{\frac{\lambda-\nu}{2}} g &= -\frac{df}{dr} - e^{\frac{\lambda}{2}} \frac{\kappa}{r} f - e^{\frac{\lambda}{2}} \frac{mc}{\hbar} g\\
\Leftrightarrow \frac{df}{dr} &= -e^{\frac{\lambda}{2}} \frac{\kappa}{r} f -  e^{\frac{\lambda}{2}} \frac{mc^2 + Ee^{-\frac{\nu}{2}}}{\hbar c} g
\end{align}
Damit folgen die abschließenden radialen Gleichungen
\begin{align}
 \frac{d}{dr} \begin{pmatrix} f\\ g \end{pmatrix} = e^{\frac{\lambda}{2}} \begin{pmatrix} -\frac{\kappa}{r} & - \left ( \frac{mc}{\hbar} + \frac{E}{\hbar c}e^{-\frac{\nu}{2}}  \right ) \\ -\left ( \frac{mc}{\hbar} - \frac{E}{\hbar c}e^{-\frac{\nu}{2}}  \right ) &  \frac{\kappa}{r} \end{pmatrix}\begin{pmatrix} f\\ g \end{pmatrix}
\end{align}
\bibliographystyle{unsrt}
\bibliography{Literatur.bib}
%
\newpage
\begin{center} \huge Erklärung \end{center}
Ich versichere hiermit, dass ich diese Masterarbeit selbständig verfasst und keine anderen als die angegebenen Quellen und Hilfsmittel genutzt habe.\vspace{12pt}\\
\bigskip\\
Frankfurt am Main, den 24. März 2011
\flushright Gunther Caspar

\end{document}